\PassOptionsToPackage{sort&compress}{natbib}
\documentclass[cmc,review,submit,moreauthors,pdftex]{Definitions/tsp}
\let\linenumbers\relax
\usepackage[normalem]{ulem} % For using the command \uuline{} in the \abstract 
\usepackage{amsfonts} % Use for some special math mark
\usepackage{mathcomp} % For permille mark
\usepackage{CJKutf8} % For Chinese font
\usepackage{pifont} % For some special mark 
\usepackage{bm} % Forr math enviroment bold format
\usepackage{bbm} % For some math fancy characters
\graphicspath{{./Definitions/}} % Use for import path the figures paper used

\makeatletter
\def\T@n@@nc@d@ngM@cr@M@d{}
\def\LY@n@@nc@d@ngM@cr@M@d{}
\makeatother

\let\orignewcommand\newcommand  % Store the original \newcommand
\let\newcommand\providecommand  % Make \newcommand behave like \providecommand
\usepackage{verse}
\let\newcommand\orignewcommand  % Use the original `\newcommand` in future

\newsavebox\foobox

\newcommand{\mmathit}[1]{
  \ifthenelse{\equal{#1}{\ln}}{\mathit{ln}}{
    \ifthenelse{\equal{#1}{\max}}{\mathit{max}}{\mathit{#1}}
  }
}
\makeatother
\robustify{\footnote}

\DeclareUnicodeCharacter{1E45}{\.{n}}
\DeclareUnicodeCharacter{1E41}{\.{m}}
\DeclareUnicodeCharacter{2003}{\quad}
\DeclareUnicodeCharacter{2009}{\thinspace}
\DeclareUnicodeCharacter{2002}{\enspace{}}
\DeclareUnicodeCharacter{2005}{\thinspace}
\DeclareUnicodeCharacter{0263}{\textipa{G}}
\DeclareUnicodeCharacter{A0}{~}
\DeclareUnicodeCharacter{2460}{\textcircled{\scriptsize{1}}}
\DeclareUnicodeCharacter{2461}{\textcircled{\scriptsize{2}}}
\DeclareUnicodeCharacter{2462}{\textcircled{\scriptsize{3}}}
\DeclareUnicodeCharacter{2463}{\textcircled{\scriptsize{4}}}
\DeclareUnicodeCharacter{2464}{\textcircled{\scriptsize{5}}}
\DeclareUnicodeCharacter{2465}{\textcircled{\scriptsize{6}}}
\DeclareUnicodeCharacter{2466}{\textcircled{\scriptsize{7}}}
\DeclareUnicodeCharacter{2467}{\textcircled{\scriptsize{8}}}
\DeclareUnicodeCharacter{2468}{\textcircled{\scriptsize{9}}}
\DeclareUnicodeCharacter{2070}{\textsuperscript{0}}
\DeclareUnicodeCharacter{2074}{\textsuperscript{4}}
\DeclareUnicodeCharacter{2075}{\textsuperscript{5}}
\DeclareUnicodeCharacter{2076}{\textsuperscript{6}}
\DeclareUnicodeCharacter{2077}{\textsuperscript{7}}
\DeclareUnicodeCharacter{2078}{\textsuperscript{8}}
\DeclareUnicodeCharacter{2079}{\textsuperscript{9}}
\DeclareUnicodeCharacter{02C2}{<}
\DeclareUnicodeCharacter{2033}{\relax\ifmmode '' \else $''$\fi}
\DeclareUnicodeCharacter{2034}{\relax\ifmmode ''' \else $'''$\fi}
\DeclareUnicodeCharacter{2026}{\relax\ifmmode … \else $\ldots$\fi}
\DeclareUnicodeCharacter{0229}{\c{e}}
\DeclareUnicodeCharacter{016F}{\r{u}}
\DeclareUnicodeCharacter{127}{\relax\ifmmode\rm\hbar\else $\rm\hbar$\fi}
\DeclareUnicodeCharacter{3AC}{\relax\ifmmode\acute{\alpha}\else $\acute{\alpha}$\fi}
\DeclareUnicodeCharacter{3AD}{\relax\ifmmode\acute{\varepsilon}\else $\acute{\varepsilon}$\fi}
\DeclareUnicodeCharacter{3AE}{\relax\ifmmode\acute{\eta}\else $\acute{\eta}$\fi}
\DeclareUnicodeCharacter{3AF}{\relax\ifmmode\acute{\iota}\else $\acute{\iota}$\fi}
\DeclareUnicodeCharacter{3CC}{\relax\ifmmode\acute{o}\else $\acute{o}$\fi}
\DeclareUnicodeCharacter{3CD}{\relax\ifmmode\acute{\upsilon}\else $\acute{\upsilon}$\fi}
\DeclareUnicodeCharacter{3CE}{\relax\ifmmode\acute{\omega}\else $\acute{\omega}$\fi}
\DeclareUnicodeCharacter{391}{A}
\DeclareUnicodeCharacter{392}{B}
\DeclareUnicodeCharacter{395}{E}
\DeclareUnicodeCharacter{396}{Z}
\DeclareUnicodeCharacter{397}{H}
\DeclareUnicodeCharacter{399}{I}
\DeclareUnicodeCharacter{39A}{K}
\DeclareUnicodeCharacter{39C}{M}
\DeclareUnicodeCharacter{39D}{N}
\DeclareUnicodeCharacter{39F}{O}
\DeclareUnicodeCharacter{3A1}{P}
\DeclareUnicodeCharacter{3A4}{T}
\DeclareUnicodeCharacter{3A7}{X}

\DeclareUnicodeCharacter{27E6}{\relax\ifmmode \llbracket \else $\llbracket$\fi}
\DeclareUnicodeCharacter{27E7}{\relax\ifmmode \rrbracket \else $\rrbracket$\fi}
\DeclareUnicodeCharacter{00B5}{\relax\ifmmode \mu{} \else $\mu$\fi}
\DeclareUnicodeCharacter{1D434}{\relax\ifmmode A \else $A$\fi}
\DeclareUnicodeCharacter{1D435}{\relax\ifmmode B \else $B$\fi}
\DeclareUnicodeCharacter{1D436}{\relax\ifmmode C \else $C$\fi}
\DeclareUnicodeCharacter{1D437}{\relax\ifmmode D \else $D$\fi}
\DeclareUnicodeCharacter{1D438}{\relax\ifmmode E \else $E$\fi}
\DeclareUnicodeCharacter{1D439}{\relax\ifmmode F \else $F$\fi}
\DeclareUnicodeCharacter{1D43A}{\relax\ifmmode G \else $G$\fi}
\DeclareUnicodeCharacter{1D43B}{\relax\ifmmode H \else $H$\fi}
\DeclareUnicodeCharacter{1D43C}{\relax\ifmmode I \else $I$\fi}
\DeclareUnicodeCharacter{1D43D}{\relax\ifmmode J \else $J$\fi}
\DeclareUnicodeCharacter{1D43E}{\relax\ifmmode K \else $K$\fi}
\DeclareUnicodeCharacter{1D43F}{\relax\ifmmode L \else $L$\fi}
\DeclareUnicodeCharacter{1D440}{\relax\ifmmode M \else $M$\fi}
\DeclareUnicodeCharacter{1D441}{\relax\ifmmode N \else $N$\fi}
\DeclareUnicodeCharacter{1D442}{\relax\ifmmode O \else $O$\fi}
\DeclareUnicodeCharacter{1D443}{\relax\ifmmode P \else $P$\fi}
\DeclareUnicodeCharacter{1D444}{\relax\ifmmode Q \else $Q$\fi}
\DeclareUnicodeCharacter{1D445}{\relax\ifmmode R \else $R$\fi}
\DeclareUnicodeCharacter{1D446}{\relax\ifmmode S \else $S$\fi}
\DeclareUnicodeCharacter{1D447}{\relax\ifmmode T \else $T$\fi}
\DeclareUnicodeCharacter{1D448}{\relax\ifmmode U \else $U$\fi}
\DeclareUnicodeCharacter{1D449}{\relax\ifmmode V \else $V$\fi}
\DeclareUnicodeCharacter{1D44A}{\relax\ifmmode W \else $W$\fi}
\DeclareUnicodeCharacter{1D44B}{\relax\ifmmode X \else $X$\fi}
\DeclareUnicodeCharacter{1D44C}{\relax\ifmmode Y \else $Y$\fi}
\DeclareUnicodeCharacter{1D44D}{\relax\ifmmode Z \else $Z$\fi}
\DeclareUnicodeCharacter{1D44E}{\relax\ifmmode a \else $a$\fi}
\DeclareUnicodeCharacter{1D44F}{\relax\ifmmode b \else $b$\fi}
\DeclareUnicodeCharacter{1D450}{\relax\ifmmode c \else $c$\fi}
\DeclareUnicodeCharacter{1D451}{\relax\ifmmode d \else $d$\fi}
\DeclareUnicodeCharacter{1D452}{\relax\ifmmode e \else $e$\fi}
\DeclareUnicodeCharacter{1D453}{\relax\ifmmode f \else $f$\fi}
\DeclareUnicodeCharacter{1D454}{\relax\ifmmode g \else $g$\fi}
\DeclareUnicodeCharacter{1D456}{\relax\ifmmode i \else $i$\fi}
\DeclareUnicodeCharacter{1D457}{\relax\ifmmode j \else $j$\fi}
\DeclareUnicodeCharacter{1D458}{\relax\ifmmode k \else $k$\fi}
\DeclareUnicodeCharacter{1D459}{\relax\ifmmode l \else $l$\fi}
\DeclareUnicodeCharacter{1D45A}{\relax\ifmmode m \else $m$\fi}
\DeclareUnicodeCharacter{1D45B}{\relax\ifmmode n \else $n$\fi}
\DeclareUnicodeCharacter{1D45C}{\relax\ifmmode o \else $o$\fi}
\DeclareUnicodeCharacter{1D45D}{\relax\ifmmode p \else $p$\fi}
\DeclareUnicodeCharacter{1D45E}{\relax\ifmmode q \else $q$\fi}
\DeclareUnicodeCharacter{1D45F}{\relax\ifmmode r \else $r$\fi}
\DeclareUnicodeCharacter{1D460}{\relax\ifmmode s \else $s$\fi}
\DeclareUnicodeCharacter{1D461}{\relax\ifmmode t \else $t$\fi}
\DeclareUnicodeCharacter{1D462}{\relax\ifmmode u \else $u$\fi}
\DeclareUnicodeCharacter{1D463}{\relax\ifmmode v \else $v$\fi}
\DeclareUnicodeCharacter{1D464}{\relax\ifmmode w \else $w$\fi}
\DeclareUnicodeCharacter{1D465}{\relax\ifmmode x \else $x$\fi}
\DeclareUnicodeCharacter{1D466}{\relax\ifmmode y \else $y$\fi}
\DeclareUnicodeCharacter{1D467}{\relax\ifmmode z \else $z$\fi}

\DeclareUnicodeCharacter{1E67}{\.{\v s}}
\DeclareUnicodeCharacter{1E11}{\relax\ifmmode \c{d} \else $\c{d}$\fi}
\DeclareUnicodeCharacter{1ECB}{\relax\ifmmode \d{i} \else $\d{i}$\fi}
\DeclareUnicodeCharacter{1D8D}{\relax\ifmmode \textlhookx \else $\textlhookx$\fi}
\DeclareUnicodeCharacter{104}{\relax\ifmmode \k{A} \else $\k{A}$\fi}
\DeclareUnicodeCharacter{211E}{\relax\ifmmode \textrecipe \else $\textrecipe$\fi}
\DeclareUnicodeCharacter{29D}{\relax\ifmmode \textctj \else $\textctj$\fi}

\DeclareUnicodeCharacter{1E2E}{\'{\"I}}
\DeclareUnicodeCharacter{23F}{\textrts}

\DeclareUnicodeCharacter{2C73}{\varw}

\DeclareUnicodeCharacter{2127}{\mho}

\DeclareUnicodeCharacter{28C}{\textturnv}
\DeclareUnicodeCharacter{252}{\textturnscripta}
\DeclareUnicodeCharacter{259}{\schwa}
\DeclareUnicodeCharacter{25B}{\m{e}}
\DeclareUnicodeCharacter{266}{\m{h}}
\DeclareUnicodeCharacter{127}{\B{h}}
\DeclareUnicodeCharacter{27E}{\textfishhookr}
\DeclareUnicodeCharacter{281}{\textinvscr}

\usepackage{xifthen}
\usepackage{makecell}
\usetikzlibrary{calc,positioning,arrows.meta}

\definecolor{revcolor}{RGB}{0,0,0}

\titlespacing{\section}{0pt}{12pt}{3pt}
\titlespacing{\subsection}{0pt}{12pt}{3pt}
\titlespacing{\subsubsection}{0pt}{12pt}{3pt}
\titlespacing{\paragraph}{0pt}{12pt}{3pt}
\continuouspages{}
\firstpage{1}
\pubvolume{0}
\issuenum{0}
\elocationid{0}
\articlenumber{0}
\pubyear{2026}
\doiyear{2026}
\copyrightyear{2026}
\datereceived{Day Month Year}
\dateaccepted{Day Month Year}
\dateonlinefirst{}
\datepublished{Day Month Year}

\Title{A Survey of Generative Categories and Techniques in Multimodal Generative Models}

\Author{Longzhen Han\textsuperscript{1}, Awes Mubarak\textsuperscript{1}, Almas Baimagambetov\textsuperscript{1}, Nikolaos Polatidis\textsuperscript{1,*} and Thar Baker\textsuperscript{2}}

\AuthorNames{Longzhen Han, Awes Mubarak, Almas Baimagambetov, Nikolaos Polatidis and Thar Baker}

\address{%
\textsuperscript{1}School of Architecture, Technology and Engineering, University of Brighton, Brighton, BN2 4GJ, United Kingdom

\textsuperscript{2}University of Khorfakkan, Sharjah, United Arab Emirates
}%

\corres{Corresponding Author: Nikolaos Polatidis. Email: N.Polatidis@brighton.ac.uk}

\abstract{Multimodal generative models (MGMs) have rapidly expanded beyond text generation to encompass images, music, video, human motion, and 3D objects. This survey categorises six primary text-conditioned generative modalities and examines how four techniques, namely self-supervised learning, mixture of experts, reinforcement learning from human feedback, and chain-of-thought prompting, are adapted across them, distinguishing shared mechanisms from experimentally supported benefits. We organise evaluation around faithfulness, compositionality, and robustness, and interpret evidence within each study's protocol. Selected within-study comparisons reported in the reviewed literature support four findings: preference gains need not improve text relevance, and reward optimisation can reduce within-prompt diversity; intermediate plans require tests of their realisation in final outputs; expert granularity changes the mechanism and its storage, computation, and runtime costs; and gains against an optimisation score need confirmation outside that scoring path. We further analyse trustworthiness risks, including bias, privacy leakage, and misuse of generated media, and examine the evidence for mitigation strategies.}

\keyword{multimodal generative models; large language models; generative AI; self-supervised learning; mixture of experts; reinforcement learning from human feedback; chain-of-thought; AI safety and evaluation}

\begin{document}

\section{Introduction}\label{sec:introduction}

Since the emergence of GPT-3 in 2020 \cite{brown_language_2020}, text-based large language models (LLMs) have become central to AI research and everyday applications. Their evolution has driven innovations in natural language processing, human-computer interaction, and multimodal applications, supporting integration across diverse fields. Over time, these models have evolved from simple text generation to in-context learning \cite{brown_language_2020, touvron_llama_2023-1}, instruction following \cite{ouyang_training_2022} and multi-step reasoning \cite{deepseek-ai_deepseek-r1_2025}. These developments are changing how people interact with computers, accomplish tasks, and create digital content.

Human perception and understanding draw on multiple modalities, including text, vision, audio, and motion. Advances in hardware enable machines to process, interpret, and generate these diverse data streams, supporting the development of multimodal approaches. Consequently, advanced models not only excel in understanding and generating text but also pair text with vision \cite{radford_learning_2021} or integrate it with audio \cite{elizalde_clap_2023}. Their outputs are also becoming increasingly multimodal and generalised. Today's models often take mixed data types as input. This multimodal integration supports the development of AI systems that can represent and interact with aspects of the physical world.

While text remains a foundational element at the core of their processing, the generative capabilities of these advanced models now span a broad spectrum of output modalities. To better understand this diversity, we categorise the primary generative outputs of multimodal generative models (MGMs) into six areas (Figure~\ref{fig:taxonomy}):

\begin{itemize}
      \item \textbf{Text-to-Text (T2T)},
            which is fundamental for all language-based tasks and natural language processing.
            It serves as the backbone for information retrieval, summarisation, translation, and conversational agents.
      \item \textbf{Text-to-Image (T2I)},
            for visual content generation and analysis, forming the foundation for other generative vision tasks.
      \item \textbf{Text-to-Music (T2M)},
            which supports musical creation. The diversity of instruments and expressive structures makes music complex to model.
      \item \textbf{Text-to-Video (T2V)},
            which combines temporal and visual information for dynamic scene generation.
            Video generation involves modelling aspects of real-world physics.
      \item \textbf{Text-to-Human-Motion (T2HM)},
            vital for applications in animation, robotics, and virtual avatars.
            This modality introduces potential new methods for intuitive human-machine interaction.
      \item  \textbf{Text-to-3D-Objects (T2-3D)},
            important for virtual reality, gaming, and design.
            This capability supports the creation and exploration of objects, and interaction with them, in virtual environments.
\end{itemize}

\begin{figure}[H]
      \centering
      \begin{tikzpicture}[
                  hd/.style={draw, rounded corners=3pt, minimum width=4.3cm, minimum height=0.75cm, align=center, font=\fontfamily{phv}\selectfont\fontsize{9}{11}\selectfont\bfseries},
                  bd/.style={draw, rounded corners=3pt, minimum width=4.3cm, minimum height=0.95cm, align=center, font=\fontfamily{phv}\selectfont\fontsize{9}{11}\selectfont, fill=white},
                  lnk/.style={thick, gray}
            ]
            \node[draw, rounded corners=3pt, fill=yellow!10, minimum width=2.8cm, minimum height=1.1cm, font=\fontfamily{phv}\selectfont\fontsize{9}{11}\selectfont\bfseries] (in) at (0,0) {Text input};
            \foreach \x/\y/\name/\apps/\col/\id in {
                        -4.2/3.1/{Text-to-text (T2T)}/{Summarisation\\Translation}/blue!10/tt,
                        4.2/3.1/{Text-to-image (T2I)}/{Illustration\\Art generation}/blue!25/ti,
                        -4.9/0/{Text-to-music (T2M)}/{Composition\\Sound design}/orange!20/tm,
                        4.9/0/{Text-to-video (T2V)}/{Animation\\Video synthesis}/red!25/tv,
                        -4.2/-3.1/{Text-to-human motion (T2HM)}/{Action sequence\\Gesture}/green!20/th,
                        4.2/-3.1/{Text-to-3D objects (T2-3D)}/{Design\\Virtual reality}/violet!25/t3} {
                        \node[hd, fill=\col] (h\id) at (\x,\y+0.42) {\name};
                        \node[bd, anchor=north] (b\id) at (h\id.south) {\apps};
                  }
            \draw[lnk] (in.north west) -- (btt.south east);
            \draw[lnk] (in.north east) -- (bti.south west);
            \draw[lnk] (in.west) -- (btm.east);
            \draw[lnk] (in.east) -- (btv.west);
            \draw[lnk] (in.south west) -- (hth.north east);
            \draw[lnk] (in.south east) -- (ht3.north west);
      \end{tikzpicture}
      \caption{Taxonomy of the six generative categories of multimodal generative models (MGMs) covered in this survey, with example applications. Each category label defines its abbreviation.}\label{fig:taxonomy}
\end{figure}

These six categories (see Figure~\ref{fig:taxonomy}) represent the primary modalities in which current generative models operate,
with each encompassing a distinct form of data output and a unique set of applications.
We use Text-to-Music rather than the broader Text-to-Audio category because speech, although closely tied to language, is an adjacent task with distinct requirements:
speech synthesis must model prosody, speaker identity and acoustic conditions, and has its own evaluation practice.
Music (T2M) is treated separately because its patterns, structures, and creative elements, including harmony, rhythm, and instrumentation, differ from the linguistic structure of speech.
Thus, while general audio might sometimes be grouped with text or other sound events,
categorising music generation separately highlights
its distinct creative and structural complexities.
Accordingly, the survey focuses on text-conditioned generation; speech and general audio, scene-level 3D generation, and image-to-X or any-to-any mappings are outside detailed coverage (Section~\ref{sec:scope}).
By delineating these capabilities,
we highlight the diverse range of outputs that generative models can produce,
each serving distinct use cases and presenting unique technical challenges.

These categories share concerns without being interchangeable. Images and video share visual representations, and video generators are often initialised from or factorised through image models~\cite{singer_make--video_2022,blattmann_align_2023,girdhar_factorizing_2025}; video nevertheless requires temporal tests that image evaluation cannot supply. Human motion and 3D objects both describe structure in three-dimensional space: motion specifies how an articulated body moves over time, whereas an object asset specifies static geometry and appearance, and a plausible rendering verifies neither physical validity nor complete geometry. Text and music can both be modelled as token sequences, and symbolic notation can serve as an intermediate musical plan~\cite{copet_simple_2023,feng_stepaudio3_2026}, while musical structure and sound quality still require auditory evaluation. These relationships explain why techniques and evaluators are transferred between categories, and why each transfer requires task-specific verification.

Generative capabilities across these modalities rely on foundational architectural innovations,
predominantly the transformer \cite{vaswani_attention_2017} with its core attention mechanism \cite{bahdanau_neural_2015},
and, for many generative vision tasks, diffusion models \cite{ho_denoising_2020}.
As the complexity of modalities evolves and AI systems are tasked with increasingly sophisticated challenges,
both model architectures and training methods continue to develop.
This evolution is critical,
as it often unlocks or significantly enhances the emergent abilities of these models.
Among these advancements,
four techniques are particularly relevant to reasoning and generation.
Three of these primarily shape the model during its training phase:
\textbf{Self-Supervised Learning (SSL)} \cite{radford_improving_2018},
\textbf{Mixture of Experts (MoE)} \cite{jacobs_adaptive_1991},
and \textbf{Reinforcement Learning from Human Feedback (RLHF)} \cite{christiano_deep_2017}.
The fourth, \textbf{Chain-of-Thought (CoT)} prompting \cite{wei_chain--thought_2022},
is a crucial inference-time strategy.
These four techniques provide an organising framework rather than an exhaustive list; Section~\ref{sec:foundations} states the criteria for choosing them and how broadly each is interpreted.

\begin{itemize}
      \item \textbf{Self-Supervised Learning (SSL)}:
            This training-time approach enables models to learn from massive unlabeled datasets
            by performing tasks such as predicting hidden parts of the input.
            SSL builds the foundational understanding of language, patterns, and world knowledge essential for any robust reasoning.
      \item \textbf{Mixture of Experts (MoE)}:
            By selectively activating specialised expert subnetworks for given inputs,
            MoEs substantially increase a model's capacity to store diverse knowledge
            and learn complex patterns crucial for advanced reasoning,
            without a proportional rise in computational cost during training or inference, although all expert weights must still be stored.
      \item \textbf{Reinforcement Learning from Human Feedback (RLHF)}:
            This training-time fine-tuning process aligns models more closely with human preferences and desired behaviours.
            By training on human-ranked responses,
            RLHF makes the model's reasoning more coherent,
            reliable, and better at following complex instructions,
            leading to more helpful and contextually appropriate outputs.
      \item \textbf{Chain-of-Thought (CoT) Prompting}:
            Applied at inference time,
            CoT significantly boosts reasoning by guiding the model to generate a sequence of intermediate steps
            before delivering a final answer.
            These explicit intermediate steps support more accurate and transparent handling of complex, multi-step problems.
\end{itemize}

Recent survey literature provides valuable insights into multimodal machine learning,
particularly regarding MGMs. The survey in \cite{baltrusaitis_multimodal_2019} provides foundational insights into the broader context of multimodal machine learning, establishing essential frameworks and identifying core challenges that support current MGMs research.
Specifically, it categorises multimodal machine learning into five core technical challenges:
representation, translation, alignment, fusion, and co-learning, which address data heterogeneity, modality integration, and knowledge transfer across modalities.

The survey in \cite{caffagni_revolution_2024} reviews recent vision-based MGMs,
examining their architectures, strategies for aligning modalities, and diverse applications,
including visual grounding and image generation.
The survey in \cite{yu_aligning_2025} addresses alignment algorithms for MGMs,
emphasising human preference alignment across multiple modalities,
while the survey in \cite{dang_explainable_2024} offers an extensive examination of interpretability and explainability within MGMs,
a critical factor for trustworthy AI.
The survey in \cite{yin_survey_2024} offers a detailed account of how MGMs are evolving to support finer granularity,
broader modality and language coverage, and diverse application scenarios,
while also advancing methods such as multimodal in-context learning, chain-of-thought reasoning,
and LLM-aided visual understanding.
The survey in \cite{liu_comprehensive_2024} offers a comprehensive overview of MGMs across different modalities,
emphasising their wide-ranging applications and safety concerns.
The survey in \cite{wang_multimodal_2025} provides an in-depth exploration of multimodal Chain-of-Thought (MCoT) reasoning,
showcasing its potential across diverse applications.

More recently, Zhao et al.~\cite{zhao_unified_survey_2025} organise unified understanding--generation models around autoregressive, diffusion and hybrid designs, and Taha~\cite{taha_generative_2026} surveys generative models across text, vision, audio, video and 3D content, aggregating reported results across studies and adding new experiments.
Collectively, these surveys document efforts to improve the functionality, interpretability, and safety of multimodal AI systems.
Table~\ref{tab:related_surveys} positions this survey against them.
Compared with these surveys, this survey makes three contributions: it compares how SSL, MoE, RLHF, and CoT are adapted across six text-conditioned modalities; it interprets evaluation evidence within each study's protocol rather than pooling scores across studies; and it examines the evidence for trustworthiness mitigations.

\begin{table}[H]
\centering
\caption{Positioning against closely related surveys. Descriptions summarise each survey's stated scope; topics not listed are not implied to be absent.}\label{tab:related_surveys}
\footnotesize
\setlength{\tabcolsep}{3pt}
\begin{tabular}{>{\raggedright\arraybackslash}p{0.15\textwidth}>{\raggedright\arraybackslash}p{0.40\textwidth}>{\raggedright\arraybackslash}p{0.41\textwidth}}
\toprule
\textbf{Survey} & \textbf{Organising focus and coverage} & \textbf{Relation to this survey} \\
\midrule
Baltru\v{s}aitis et al., 2019~\cite{baltrusaitis_multimodal_2019} & Multimodal machine learning organised by five challenges: representation, translation, alignment, fusion and co-learning & We focus on text-conditioned generation and compare how specific techniques are adapted across six output modalities \\
Caffagni et al., 2024~\cite{caffagni_revolution_2024} & Visual multimodal LLMs: architectures, modality alignment and applications including visual grounding and image generation & We extend the comparison to music, human motion and 3D objects and organise it by the technique that changes the generator \\
Yu et al., 2025~\cite{yu_aligning_2025} & Alignment algorithms for multimodal LLMs, emphasising human preference alignment & Preference optimisation is one of our four techniques; we relate it to representation learning, expert mixtures and planning, and read reported gains under their evaluation protocols \\
Dang et al., 2024~\cite{dang_explainable_2024} & Interpretability and explainability of multimodal LLMs & We address trustworthiness through evaluation evidence and tested mitigations for misuse, privacy leakage and bias \\
Yin et al., 2024~\cite{yin_survey_2024} & Multimodal LLMs: granularity, modality and language coverage, applications, multimodal in-context learning and chain-of-thought & We examine generators whose outputs are images, music, video, human motion or 3D objects, including generators without a language-model backbone \\
Liu et al., 2024~\cite{liu_comprehensive_2024} & Multimodal LLMs across modalities: concepts, applications and safety & We compare technique adaptation for each output modality and interpret evaluation evidence at the level of each study's protocol \\
Wang et al., 2025~\cite{wang_multimodal_2025} & Multimodal chain-of-thought reasoning and its applications & CoT is one of our four techniques; we distinguish reasoning, planning and search and test whether intermediate structure is realised in generated outputs \\
Zhao et al., 2025~\cite{zhao_unified_survey_2025} & Unified understanding--generation models grouped as diffusion-based, autoregressive and hybrid & We consider unified models only for their text-conditioned generation and compare interventions across specialised generators as well \\
Taha, 2026~\cite{taha_generative_2026} & Generative models for text, vision, audio, video and 3D content; aggregates reported results across studies and adds new experiments & We do not pool scores across studies; comparisons are restricted to within-study contrasts whose conditioning and protocols are recorded \\
This survey & Four techniques compared across six text-conditioned modalities; evaluation evidence interpreted at protocol level; trustworthiness mitigations assessed by harm and evidence & A focused narrative synthesis with selective coverage (Section~\ref{sec:scope}); no new benchmark experiments \\
\bottomrule
\end{tabular}
\end{table}

      {\color{revcolor}
            The remainder of this survey is structured as follows. Section 2 establishes the foundational concepts and background necessary to contextualise the subsequent analysis, then defines the scope and methodology adopted for this review. The main body of the survey begins in Section 3, where we review the historical development of the six generative modalities previously outlined (T2T, T2I, T2M, T2V, T2HM, and T2-3D), closing each with a comparative analysis. Following this, Section 4 reviews the emergence and refinement of the four techniques SSL, MoE, RLHF, and CoT, detailing their mechanisms and impact on model capabilities. Section 5 organises the evaluation of text-to-X generation around faithfulness, compositionality, and robustness, and surveys modality-specific and cross-modal metrics together with the protocols behind representative results. Section 6 examines trustworthiness, safety, and ethics in MGMs, including multimodal bias, misuse risks, privacy leakage, and governance and mitigation strategies.

Section 7 presents a cross-modal discussion of emerging patterns, challenges, and architectural parallels observed across modalities, analysing the transfer of ideas between these modalities and techniques, and derives four findings: preference, relevance and diversity can change differently; intermediate structure must be checked in the final output; expert granularity determines the mechanism and resource comparison; and optimisation-score gains need independent confirmation. Section 8 outlines key future research directions, organised around the four techniques and their extensions to underexplored modalities and evaluation gaps. Section 9 concludes by summarising insights and reflecting on the trajectory toward unified multimodal generative systems. An earlier version of this work is available as a preprint~\cite{han_survey_2025}.
      }

\section{Background and Methodology}\label{sec:scope}

Understanding the evolution of MGMs requires a technical grounding in their foundational architectures
and a clear explanation of how relevant literature is curated and analysed.
This section provides the necessary context and outlines the methodological framework that guides our survey.

\subsection{Background}\label{sec:foundations}

The development of current MGMs builds on text modelling, a data-rich approach to representing human knowledge and communication. Early advances focused on LLMs trained on large text datasets, using SSL techniques to predict masked or subsequent words. These models learned intricate patterns of grammar, acquired factual knowledge, and even developed precursors to complex reasoning, exhibiting emergent abilities through exposure to text.

The transformer architecture \cite{vaswani_attention_2017},
particularly its attention mechanism \cite{bahdanau_neural_2015},
is key to the success of large language models.
This mechanism allows the model to weigh the importance of different words
in a sequence relative to each other regardless of their position.
This ability to capture long-range dependencies proved far more effective
and scalable for sequence modelling than previous architectures,
forming the backbone of most modern large-scale AI models.

Although text provides a rich foundation, vision is another primary source of information for human understanding.
Diffusion Models \cite{ho_denoising_2020} emerged as a powerful generative architecture,
particularly effective for high-fidelity image and video synthesis.
They operate by progressively adding noise to data
and then learning a reverse process to remove noise,
starting from a random signal and conditioned on input,
thereby generating new data samples.
The success of diffusion models extended beyond the visual domain.
Their core principles inspired researchers to adapt and apply similar iterative denoising techniques
for generating other complex data types,
leading to significant advancements in synthesizing realistic audio \cite{schneider_mousai_2024, huang_noise2music_2023},
plausible human motion \cite{zhang_motiondiffuse_2024},
and detailed 3D objects \cite{poole_dreamfusion_2022, lin_magic3d_2023, jun_shap-e_2023, li_instant3d_2024}.
An architectural paradigm also emerged by combining the sequence modelling strengths of Transformers \cite{peebles_scalable_2023},
often used to process conditioning information like text or manage the temporal aspects of generation,
with the high-quality generative capabilities of Diffusion Models.
This combination of Transformers and diffusion models has become central
to many state-of-the-art (SOTA) multimodal generative systems,
enabling more sophisticated control and output quality across diverse tasks.

Building upon these architectural and representational foundations,
the four techniques introduced earlier further extend these models.
SSL provides the rich initial representations across modalities;
MoE architectures support scaling to the large model sizes used
for complex cross-modal reasoning while maintaining computational efficiency;
RLHF aligns the increasingly complex multimodal outputs with human expectations
and safety considerations;
and CoT prompting enhances the explicit reasoning capabilities needed
to solve intricate tasks involving multiple data types.

These four techniques were selected by three criteria: relevance to the survey's questions about conditioning and evaluation, documented application in more than one of the six output modalities, and complementary points of intervention. Representation learning changes training targets or initialisation; expert mixtures change how parameters are used; feedback changes the optimisation target; and CoT-style planning and search change intermediate conditioning or inference. These criteria justify an organising choice; they do not establish that exactly four techniques are foundational. Each technique is interpreted broadly: SSL includes representation transfer, RLHF includes other feedback and preference optimisation, and CoT includes related planning and inference-time search. Other methods, such as instruction tuning, retrieval augmentation and distillation, are discussed only where they bear on a selected comparison, and because the techniques can coexist, evidence for a combined pipeline must not be attributed to each component independently.

Compression, generation and evaluation also introduce separate failure points. A codec can discard details before conditional generation begins; a generator can produce plausible outputs that miss requested attributes; an evaluator can miss those errors. Reconstruction tests probe the first problem, prompt-specific tests the second, and evaluator validation the third, so neither good reconstruction nor a strong pretrained encoder alone establishes conditional generation quality.

Together, sequence modelling architectures such as the Transformer, diffusion models, unified multimodal representations, and advanced training and inference techniques provide the foundation for the text-conditioned generative capabilities surveyed here. They also support the development of more general-purpose multimodal AI systems.

\subsection{Methodology}

To examine the development of MGMs, this survey uses a multi-stage methodology that prioritises transparency, relevance, and breadth. We describe the scope, literature selection, and analytical structure, and acknowledge the main limitations.

This survey focuses on MGMs conditioned on text and producing outputs across the aforementioned six primary modalities.
While the majority of selected works were published after 2021, earlier foundational contributions are also included where they establish critical concepts or techniques. We emphasise models and techniques that generalise beyond narrow tasks and demonstrate capabilities across or within modalities, particularly in the domains of architecture, training strategy, and inference behaviour.
Text plus an image, pose, melody or other control remains eligible, but is distinguished from text-only generation. Speech and general audio are adjacent tasks rather than instances of text or music generation, and our 3D scope concerns objects rather than complete scenes or interactive world models. Image-to-X and any-to-any systems are considered only for their text-conditioned generation capabilities, and non-text-conditioned precursors provide background rather than direct evidence of text-conditioned performance.

The search strategy covered the main lines of work in the field without aiming to identify every relevant study.
Broad and representative keywords related to multimodal learning, generative modelling,
and foundational MGM techniques were used, along with relevant synonyms and acronyms.
Key sources include peer-reviewed conference proceedings (e.g., NeurIPS, CVPR, ACL, ICCV, ICML),
arXiv preprints backed by reputable institutions (e.g., OpenAI, Google DeepMind, Microsoft, Meta, DeepSeek and NVIDIA), as well as demonstrated projects, works under review, or recently submitted manuscripts that are closely aligned with the scope of this survey.
Snowballing techniques include both backward and forward citation tracking,
via Semantic Scholar and Google Scholar, to supplement initial results and uncover influential works.
Papers were selected based on their significant technical contributions to multimodal generative modelling,
such as novel architectures, training strategies, or evaluation frameworks,
as well as empirical rigour demonstrated through benchmark results or compelling qualitative outputs.
Priority was given to peer-reviewed publications and high-impact preprints from reputable institutions,
particularly those accompanied by publicly available code, models, or datasets to support transparency and reproducibility.
Conversely, papers were excluded if they fell outside the defined temporal scope,
were not written in English,
or lacked sufficient technical detail,
especially in the case of proprietary industrial models without publicly documented methodologies.
The reference set was later updated through targeted searches emphasising work released between January 2025 and September 2026, including generation-specific trustworthiness mitigations, and the bibliographic metadata of cited works were checked against publisher, Crossref, arXiv and OpenAlex records.

The selected literature is organised by modality and technique,
with each section structured chronologically to highlight the technological progression and thematic trends.
This chronological structure illustrates how core capabilities, including generalisation, compositionality, and cross-modal alignment, have developed over time.
For the cross-cutting techniques SSL, MoE, RLHF, and CoT,
we examine their implementation across modalities,
focusing on design transferability, scalability, and interoperability.
For the technical comparisons, we summarise study-level information from the cited literature in the comparative tables and accompanying discussion, including conditioning, output representation, the component changed, training or inference objective, dataset and protocol, comparator, and limitations where established. We distinguish whole-system comparisons from ablations that isolate a component. Missing experimental details are treated as unavailable evidence, and an unfilled technique--modality combination reflects this review's selection rather than an absence of research.

No complete historical search or screening log was retained, so the original literature-selection process cannot be fully reconstructed. Recent targeted-search and evidence-checking notes document revision work only.
Despite efforts to ensure breadth and rigour, some limitations remain:
the keyword-based search may miss works with unconventional terminology;
proprietary models are excluded, potentially omitting cutting-edge capabilities;
the review is a focused narrative synthesis rather than an exhaustive systematic review;
and given the field’s rapid evolution,
this survey provides a snapshot that may soon be superseded by further developments.
With this scope and methodology in place, the next section reviews the six categories in turn, building on the architectural foundations above.

\section{Multimodal Generative Models}\label{sec:tasks}

MGMs have developed through advances in architecture, training strategies, and scaling across text, image, audio, video, motion, and object synthesis.
This section reviews the development of MGMs across output types. Within each category, we analyse milestone papers and examine how ideas transfer between fields.
Our review follows a chronological approach, tracing the transition from small,
single-task models to large,
generalised systems capable of few-shot and zero-shot learning.
Each subsection first states the inputs, output representation and constraints of the task, then describes representative approaches, and closes with a comparative analysis of their advantages, limitations and open research questions. The chronological order explains design choices; it does not establish progress on a shared metric.

\subsection{Text-to-Text}

\subsubsection{Inputs, Representation and Constraints}
The input comprises instructions and, where relevant, source text, retrieved evidence or a database schema. Outputs are token sequences, including structured languages such as structured query language (SQL). Fluency, factual support and executable correctness are separate targets; the task definition determines which one is tested.

\subsubsection{Representative Approaches}
An early advance in neural machine translation was reported in \cite{sutskever_sequence_2014}, which introduced an encoder-decoder architecture based on recurrent neural networks (RNNs).
Their model significantly improved upon traditional rule-based and phrase-based methods by mapping input sequences to output sequences.
In their approach, both the encoder and the decoder consisted of four layers of long short-term memory (LSTM) networks,
with each layer executed on a separate GPU to facilitate parallel computation.
The encoder transformed the input sequence into a fixed-dimensional vector representation.
Although this architecture demonstrated that deep neural networks could outperform traditional statistical machine translation systems,
compressing an entire sequence into a single context vector proved limiting, particularly for longer sentences.
To address this challenge, the attention mechanism \cite{bahdanau_neural_2015} was introduced in 2015, retaining the encoder-decoder structure. Unlike LSTMs, which employ gated memory cells to preserve context over time, the attention mechanism allows the decoder to dynamically focus on different parts of the input sequence at each decoding step.
Instead of relying on a single fixed context vector, the model learns to attend to the most relevant portions of the input,
resulting in more context-sensitive and effective output generation.
The next architectural step replaced RNNs entirely.
The Transformer architecture, first introduced in \cite{vaswani_attention_2017},
eliminates recurrence altogether and relies solely on self-attention mechanisms to model dependencies,
leading to improved efficiency and scalability.

The Transformer's multi-head self-attention mechanism helps capture contextual relationships efficiently
by computing multiple attention heads in parallel. Each head processes different aspects of the input, enhancing the model's ability to understand and generate complex patterns. Subsequent breakthroughs emerged with Transformer-based architectures. Bidirectional encoder representations from transformers (BERT) \cite{devlin_bert_2019} introduced a bidirectional Transformer, using masked language modelling (MLM) to predict masked tokens and next sentence prediction (NSP)
to understand sentence relationships. While earlier work favoured an encoder-decoder approach, OpenAI's GPT-1 \cite{radford_improving_2018} employed a 12-layer decoder-only Transformer with masked self-attention heads, 117M parameters, and a context length of 512 tokens. Training purely on left-to-right autoregressive (AR) language modelling focuses on generating the next word given previous words.
This architecture endowed the model with zero-shot behaviour, enabling it to perform tasks without explicit training. It learned patterns and structures during unsupervised learning, achieving significant performance gains. Later, it was fine-tuned with labelled data, with an additional linear output layer.
The OpenAI GPT-2 \cite{radford_language_2019} expanded this approach significantly.
It used a larger decoder-only Transformer with up to 48 layers and up to 1.5B parameters,
with a context length doubled to 1024 tokens.
GPT-2 demonstrated enhanced long-range dependencies and strong zero-shot learning capabilities,
mastering tasks without fine-tuning.
Unlike GPT-1, it required no additional labelled data, relying entirely on unsupervised learning to generalise across natural language processing (NLP) tasks.
OpenAI's GPT-3 \cite{brown_language_2020} uses a Transformer architecture similar to GPT-2's but at a substantially larger scale.
The largest version has 96 layers, a hidden size of 12,288, and 96 attention heads, with a context length of 2048 tokens and dense self-attention.
The model is trained with AR language modelling, predicting the next token given previous tokens,
making it highly capable in zero-shot and few-shot learning tasks.
GPT-3's reasoning ability improved significantly over its predecessors through increases in both model size and training data.
However, it lacks coherence over significantly long passages, which indicates that its reasoning differs from the explicit logic of a rule-based system.
Instead, it mimics reasoning patterns seen in training data.

Deploying large language models introduces substantial safety challenges. Unlike traditional software, their behaviour is shaped by large training datasets, making them less predictable and more prone to generating harmful or biased content. To ensure these models are safe and aligned with human values, their training must be carefully curated. InstructGPT \cite{ouyang_training_2022} addressed this by fine-tuning GPT-3 using reinforcement learning from human feedback (RLHF), aligning model outputs with user instructions rather than mere likelihood maximisation. This marked a crucial shift towards instruction-following behaviour and user alignment.
Another approach focused on efficiency. The GLaM \cite{du_glam_2022} model adopted the MoE structure,
demonstrating that activating only a subset of parameters per inference could significantly reduce computational costs while maintaining performance. Notably, Mixtral \cite{jiang_mixtral_2024} also relied heavily on the MoE design, activating 2 of 8 experts per input vector for efficient scaling.
The LLaMA series exemplifies parallel developments in the open-source community alongside advances in proprietary models. LLaMA 2 \cite{touvron_llama_2023-1} extended this approach by releasing both foundation and fine-tuned chat models, offering stronger performance and safety improvements for real-world deployment.

DeepSeek-R1 \cite{deepseek-ai_deepseek-r1_2025} builds upon DeepSeek-V3, using reinforcement learning based on Group Relative Policy Optimization (GRPO) with rule-based rewards to enhance reasoning. Table~\ref{tab:gpt3_j1} presents a comparative overview of available architectural configurations and training-scale parameters for prominent T2T generation models.

Text-to-text generation also includes structured outputs. Baig et al.~\cite{baig_texttosql_2025} review text-to-SQL generation, where natural-language requests must be grounded in a database schema. Schema linking, cross-domain generalisation, and exact-match versus execution-based evaluation illustrate why successful generation depends on task-specific constraints beyond fluent text. The same distinction between a plausible output and a satisfied constraint recurs in non-text tasks.

\begin{table}[H]
    \centering
    \caption{Models and their hyperparameters for Text-to-Text generation.}\label{tab:gpt3_j1}
    \footnotesize
    \begin{tabular}{lccccccc}
        \toprule
        Model          & $n_{params}$ & $n_{layers}$ & $d_{model}$ & $n_{heads}$ & $d_{head}$ & $n_{vocab}$ & Batch size (tokens) \\
        \midrule
        GPT-3 (6.7B)~\cite{brown_language_2020}   & 6.7B  & 32 & 4,096  & 32 & 128 & 50K & 2M   \\
        GPT-3 (175B)~\cite{brown_language_2020}   & 175B  & 96 & 12,288 & 96 & 128 & 50K & 3.2M \\
        \midrule
        GLaM 0.1B~\cite{du_glam_2022}      & 130M  & 12 & 768   & 12  & 64  & 256K & -- \\
        GLaM 0.1B/64E  & 1.9B  & 12 & 768   & 12  & 64  & 256K & -- \\
        GLaM 1.7B      & 1.7B  & 24 & 2,048 & 16  & 128 & 256K & -- \\
        GLaM 1.7B/32E  & 20B   & 24 & 2,048 & 16  & 128 & 256K & -- \\
        GLaM 1.7B/64E  & 27B   & 24 & 2,048 & 16  & 128 & 256K & -- \\
        GLaM 1.7B/128E & 53B   & 24 & 2,048 & 16  & 128 & 256K & -- \\
        GLaM 1.7B/256E & 105B  & 24 & 2,048 & 16  & 128 & 256K & -- \\
        GLaM 8B        & 8.7B  & 32 & 4,096 & 32  & 128 & 256K & -- \\
        GLaM 8B/64E    & 143B  & 32 & 4,096 & 32  & 128 & 256K & -- \\
        GLaM 137B      & 137B  & 64 & 8,192 & 128 & 128 & 256K & -- \\
        GLaM 64B/64E   & 1.2T  & 64 & 8,192 & 128 & 128 & 256K & -- \\
        \midrule
        Mixtral 8$\times$7B~\cite{jiang_mixtral_2024} & 47B total / 13B active & 32 & 4,096 & 32 & 128 & 32K & -- \\
        \midrule
        LLaMA-2 7B~\cite{touvron_llama_2023-1}  & 7B  & -- & -- & -- & -- & 32K & 4M \\
        LLaMA-2 13B~\cite{touvron_llama_2023-1} & 13B & -- & -- & -- & -- & 32K & 4M \\
        LLaMA-2 70B~\cite{touvron_llama_2023-1} & 70B & -- & -- & -- & -- & 32K & 4M \\
        \bottomrule
    \end{tabular}
    \par\smallskip
    \begin{minipage}{\textwidth}\footnotesize
        $n_{params}$: parameters (for GLaM and Mixtral, all expert parameters); $n_{layers}$: transformer layers; $d_{model}$: hidden size; $n_{heads}$: attention heads; $d_{head}$: dimension per head; $n_{vocab}$: vocabulary size. --: not reported in the cited source. GLaM $x$B/$E$E denotes a sparse model with $E$ experts per MoE layer whose dense counterpart has $x$B parameters. GPT-3 uses the byte-level vocabulary of GPT-2 (50,257 tokens)~\cite{radford_language_2019}. Mixtral routes each token to 2 of 8 experts, so it stores about 47B parameters but uses about 13B per token.
    \end{minipage}
\end{table}

\subsubsection{Comparative Analysis}
\textbf{Intervention and target:} predictive pretraining and human-feedback tuning target different aspects of text generation~\cite{radford_improving_2018,ouyang_training_2022}.
\textbf{Advantage and evidence:} preference evaluation tests whether adaptation improves responses to the sampled instructions.
\textbf{Limitation:} preference does not independently establish factual correctness.
\textbf{Research question:} test whether gains persist on independently checked answers and shifted prompts, alongside inference and annotation costs.

\subsection{Text-to-Image}

\subsubsection{Inputs, Representation and Constraints}
A text description conditions pixels, latent features or image tokens. Layouts, masks and reference images provide additional controls and must be stated explicitly. Object presence, attribute binding and spatial relations require prompt-specific checks beyond perceptual realism.

\subsubsection{Representative Approaches}
Before the adoption of diffusion models and Transformer architectures, generative adversarial networks (GANs) dominated generative modelling.
AttnGAN \cite{xu_attngan_2018} uses the attention mechanism \cite{bahdanau_neural_2015}
in its generator network.
The attentional generative network effectively links words to specific sub-regions in the image,
while a Deep Attentional Multimodal Similarity Model (DAMSM) assesses
how well the generated image aligns with the input text.
Rather than directly comparing text and image similarity,
MirrorGAN \cite{qiao_mirrorgan_2019} uses a semantic text regeneration and alignment module (STREAM) to redescribe the generated image and compare the description with the input text.
It also uses two further modules:
a semantic text embedding module (STEM) and a global-local collaborative attentive module for cascaded image generation (GLAM).

GAN models sometimes struggled with training stability and fine-grained detail. Diffusion models subsequently gained prominence as an alternative.
Inspired by non-equilibrium statistical physics,
the diffusion process gradually disrupts the structure of the data distribution through iterative forward diffusion.
In Denoising Diffusion Probabilistic Models \cite{ho_denoising_2020},
the authors discovered that a reverse diffusion process can effectively restore structure in data,
leading to a versatile and computationally manageable generative framework.
They demonstrated that diffusion models can generate high-quality samples on several datasets (CIFAR-10, LSUN, etc.),
leveraging a Markov chain to incrementally alter one distribution into another.

Before diffusion models dominated image generation, DALL·E \cite{ramesh_zero-shot_2021} took a Transformer-based approach that considers the text and image tokens as a single stream of data, thereby generating the image tokens autoregressively. The model has 12 billion parameters and was trained on 250 million image-text pairs collected from the internet. However, DALL·E 2 \cite{ramesh_hierarchical_2022} employs diffusion models in both stages. In the first stage, a decoder-only Transformer with a causal attention mask processes a sequence comprising the encoded text, CLIP \cite{radford_learning_2021} text embedding, diffusion timestep embedding, noised CLIP image embedding, and a final embedding that the Transformer uses to predict the denoised CLIP image embedding.
The second stage uses diffusion models to generate images conditioned on CLIP image embeddings, with optional text captions.

Diffusion-based methods have become widely adopted for refined and controllable image generation.
Improving image quality increases GPU demand when the model operates directly in pixel space.
Latent Diffusion Models \cite{rombach_high-resolution_2022} address this issue by
applying the diffusion process in the latent space of pretrained autoencoders,
which balances computational complexity and image fidelity.
Training proceeds in two phases: first, an autoencoder is trained to learn a compact and perceptually equivalent latent space from the original data, and then the diffusion model is trained within that space.
To support additional conditioning controls in image generation,
ControlNet \cite{zhang_adding_2023} uses a pretrained Stable Diffusion model as its backbone,
locking its parameters while making a trainable copy of its encoding layers.
These trainable layers are connected to the locked model with a zero convolution layer,
which incrementally increases the parameters from zero while preventing any detrimental noise from affecting the fine-tuning process.
The model accepts one or more control conditions, such as edges, depth, segmentation, and human pose.
DALL·E 3 \cite{betker_improving_2023}, although the training and implementation details are not disclosed,
is inspired by DALL·E 2 \cite{ramesh_hierarchical_2022}.
The researchers seek to address the “prompt following” challenge by training on highly descriptive image captions.
To achieve this, they first develop a robust image captioner capable of generating detailed and accurate image descriptions.
This captioner is then applied to their dataset to enhance caption quality.
Finally, they train T2I models using the improved dataset.

Building upon preceding advancements,
recent studies have extended the flexibility, control, and multimodal capabilities of image generation.
Rectified Flow Transformers \cite{esser_scaling_2024} introduce a novel transformer-based architecture
that establishes a bidirectional flow of information between image and text tokens.
This design is theoretically simpler than mainstream diffusion models.
The authors conducted a scaling study demonstrating that this approach shows no signs of saturation,
highlighting its potential as a more effective alternative.

As model capabilities increase, a further goal is to handle more complex tasks involving multimodal inputs and outputs.
DeepSeek's Janus-Pro \cite{chen_janus-pro_2025} handles both multimodal understanding and generation tasks. It decouples visual encoding into two distinct pathways: a high-dimensional semantic encoder (SigLIP) for multimodal understanding, and a lower-dimensional discrete tokenizer (VQ tokenizer) for fine-grained visual generation.
Both encoders feed into a unified AR transformer, trained in three stages: adaptors and the image head are trained first, all components except the encoders are then pretrained jointly, and supervised fine-tuning finally unlocks the understanding encoder.
Janus-Pro incorporates three major improvements over its predecessor: an optimised training strategy, expanded training data, and a larger model. It makes two improvements to the training process:
\begin{itemize}
    \item Training in Stage I is extended to improve image understanding. \item In Stage II, labelled image data are removed, allowing the model to train directly on Text-to-Image data with dense descriptions, improving efficiency and performance.
\end{itemize}
Janus-Pro provides a flexible model for understanding and generating both visual and textual content.

\subsubsection{Comparative Analysis}
\textbf{Intervention and target:} Diffusion-DPO, a diffusion adaptation of direct preference optimisation (DPO), changes the model using preference pairs, whereas verifier-guided search changes inference selection~\cite{wallace_diffusion_2024,ma_inference-time_2025}.
\textbf{Advantage and evidence:} human comparisons test the former's appeal and alignment; search comparisons examine gains at different sampling budgets.
\textbf{Limitation:} repeated selection may exploit verifier bias and reduce diversity; extra annotations (such as descriptive captions) and extra inference controls (such as ControlNet conditions) are different interventions from text-only generation.
\textbf{Research question:} compare both approaches with independent judgements at matched total cost, including training and verifier calls.

\subsection{Text-to-Music}

\subsubsection{Inputs, Representation and Constraints}
Text descriptions may specify instrumentation, genre, mood or musical structure; melody or audio references add conditioning. Outputs may be waveforms, compressed audio tokens or intermediate symbolic scores. Acoustic quality, text relevance and long-range musical structure require separate tests.

\subsubsection{Representative Approaches}
DeepMind introduced WaveNet \cite{oord_wavenet_2016} in 2016,
a dilated convolutional AR model,
capable of generating raw waveforms at a high resolution.
By learning from raw waveform data, WaveNet effectively captures fine acoustic nuances,
producing realistic speech. It achieved promising results in both audio modelling and speech recognition.
One limitation of WaveNet is its inability to enforce long-range consistency; it instead relies on external conditioning such as linguistic or musical input.
In contrast, NSynth \cite{engel_neural_2017} introduced a WaveNet-style autoencoder that learns temporal hidden codes, improving consistency in long-term structure. Beyond the model, it also introduced the NSynth dataset, a large-scale resource for neural audio synthesis of musical notes.
While Variational Autoencoders effectively produce semantically meaningful latent representations for natural data, they also struggle to model sequences with long-term structure.
To address this, Google MusicVAE \cite{roberts_hierarchical_2018} introduces a novel sequential autoencoder with a hierarchical recurrent decoder. This model encodes an entire sequence into a single latent vector, first generating embeddings for subsequences and then using these embeddings to reconstruct each segment independently. This hierarchical approach mitigates the issue of posterior collapse by encouraging the model to utilise its latent code effectively.
These models provide background for audio and musical representations; they are not text-conditioned music generators.

With the success of unsupervised learning and Transformer architectures in text generation (e.g., GPT-2),
OpenAI Jukebox \cite{dhariwal_jukebox_2020} combined the VQ-VAE and AR transformers to generate raw audio instead of symbolic MIDI representations.
By learning compressed audio representations through VQ-VAE and modelling long-term dependencies with Transformers, Jukebox achieved high-fidelity music synthesis across various genres and artists' styles.

Diffusion models, widely used in T2I generation, have also been applied successfully to T2M generation. Moûsai \cite{schneider_mousai_2024}, a cascading two-stage latent diffusion model, efficiently generates high-quality music lasting several minutes from textual descriptions. To support this model, a dataset called TEXT2MUSIC has been released. It comprises 50,000 text-music pairs, totalling 2,500 hours of music. Designed for real-time inference, Moûsai was trained for approximately one week per stage on a single A100 GPU and runs on a single consumer GPU at a practical inference speed. The first stage trains a music encoder using Diffusion Magnitude-Autoencoding (DMAE), achieving an audio signal compression rate of 64x. The second stage employs a lightweight, specialised 1D U-Net architecture for efficient music synthesis. MuseCoco \cite{lu_musecoco_2023} introduces a structured approach to Text-to-Music generation, combining text-to-attribute understanding with attribute-to-music synthesis. The first stage is built on a BERT backbone model for text-to-attribute understanding, and the second uses a Transformer-based architecture for attribute-to-music generation. The stages are trained independently. In a similar vein, MusicGen \cite{copet_simple_2023} proposes a streamlined architecture focused on explicit and user-friendly control over musical elements. It uses a single-stage Transformer language model with efficient token interleaving, avoiding complex cascaded setups. As a result, it generates high-quality mono and stereo music conditioned on text or melody, enabling fine-grained control.

Google's MusicLM \cite{agostinelli_musiclm_2023} supports high-fidelity music generation from complex text prompts. Using a hierarchical sequence-to-sequence modelling framework, it captures nuanced musical characteristics such as instrumentation, genre, and mood. Trained on 280,000 hours of music, MusicLM can also generate music based on hummed or whistled melodies.
Noise2Music \cite{huang_noise2music_2023} further extends diffusion-based generation.
It uses a two-stage pipeline: a generator creates an intermediate audio representation from text, and a cascade upsamples it into high-fidelity output. A notable innovation was the use of language models to generate synthetic text-audio training pairs.
FluxMusic \cite{fei_flux_2024} extends rectified-flow transformers to text-to-music generation, focusing on musicality and text alignment.
Finally, Stable Audio Open \cite{evans_stable_2025}, developed by Stability AI, represents an open-source alternative in high-fidelity music generation. It offers accessible tools and models trained to produce coherent, multi-style music from textual prompts, supporting reproducibility and community-driven research. Its training data are restricted to Creative Commons recordings, a choice whose consequences are discussed in Section~\ref{sec:trustworthiness}.

\subsubsection{Comparative Analysis}
\textbf{Intervention and target:} compressed audio representations support conditional synthesis, while preference tuning explicitly changes the objective~\cite{schneider_mousai_2024,ziv_mrflowdpo_2025}.
\textbf{Advantage and evidence:} listening comparisons can distinguish musical quality from text relevance; melody-conditioned variants receive information unavailable to a text-only model.
\textbf{Limitation:} overall preference gains need not imply relevance gains, and a reused training reward is not independent validation.
\textbf{Research question:} evaluate these properties separately on held-out prompts and report which improvements are confirmed by independent listeners.

\subsection{Text-to-Video}

\subsubsection{Inputs, Representation and Constraints}
Text conditions a sequence of frames or spatiotemporal latents. An initial frame, camera trajectory or reference clip changes the conditioning task. In addition to object and attribute accuracy, evaluation must consider identity persistence, event order and motion over time.

\subsubsection{Representative Approaches}
Early T2V research used GAN-based models. MoCoGAN \cite{tulyakov_mocogan_2018} introduced a latent space decomposition strategy to separate static elements and motion in videos, allowing for diverse motion trajectories while keeping content fixed. It demonstrated the ability to generate videos of different objects performing the same motion; as an unconditional model, it provides background rather than direct text-to-video evidence. TiVGAN \cite{kim_tivgan_2020} introduced a two-stage generation framework: first, a pretrained T2I GAN generates the first frame. Next, the model gradually learns to generate longer video sequences by extending frames step-by-step.

However, GANs faced challenges such as mode collapse and difficulties in maintaining temporal consistency.
With the success of diffusion models and transformer architectures in T2I generation,
similar concepts have been adapted to the T2V domain, leading to significant advancements in video synthesis.
CogVideo \cite{hong_cogvideo_2022} is a 9-billion-parameter AR transformer-based T2V model built upon the pretrained T2I model CogView2.
It is the largest and first open-source pretrained Transformer for general-domain T2V generation.
The multi-frame-rate hierarchical training strategy enhances text-clip alignment,
significantly improving generation accuracy and enabling precise control over movement intensity.
CogVideo relies on the T2I model CogView2 to reduce computational costs,
highlighting the deep connection between T2I and T2V generation.

Meta’s Make-A-Video \cite{singer_make--video_2022} extends this approach by using large-scale T2I datasets,
eliminating the need for paired text-video datasets.
It integrates knowledge from T2I models with unsupervised video learning,
learning motion by capturing temporal dynamics from unlabelled video data.
This makes the approach highly scalable. The model first learns text-image relationships from T2I models and then learns motion patterns from unlabelled video data.
NVIDIA’s VideoLDM \cite{blattmann_align_2023} builds on latent diffusion by aligning latent representations to ensure temporally consistent, high-resolution outputs with detailed spatial fidelity. Meta’s Emu Video \cite{girdhar_factorizing_2025} adopts a two-stage approach, first generating an image from text, then synthesizing video conditioned on that image, effectively decoupling content and motion to enhance controllability and quality.

Advanced video generation models now demonstrate zero-shot capabilities, improved motion coherence, and longer high-fidelity outputs. Sequential frame generation makes global temporal consistency difficult to maintain. Lumiere \cite{bar-tal_lumiere_2024} introduces a spacetime U-Net architecture that generates an entire video in a single pass. Instead of relying on keyframes and temporal interpolation, Lumiere processes video holistically, ensuring smoother motion and coherence. By using both spatial and temporal information and a pretrained T2I diffusion model, the framework learns to generate full-frame-rate, low-resolution videos at multiple spacetime scales. This initial low-resolution output is then upscaled to achieve high-fidelity results.

Meta's Movie Gen \cite{polyak_movie_2025} trains a 30-billion-parameter transformer for text-to-video generation and adds models for personalisation, editing and audio, enabling coherent video synthesis from abstract prompts or character-centric descriptions.
Despite advances in visual fidelity and narrative control, these models often struggle to represent real-world physical dynamics.

\subsubsection{Comparative Analysis}
\textbf{Intervention and target:} VideoREPA transfers encoder relations, whereas VADER optimises differentiable rewards~\cite{zhang_videorepa_2025,prabhudesai_video_2024}.
\textbf{Advantage and evidence:} their within-study comparisons test auxiliary representation targets and reward adaptation, respectively.
\textbf{Limitation:} frame quality, semantic adherence and physical plausibility are distinct; a frame-level reward does not directly evaluate event structure.
\textbf{Research question:} test temporal and semantic gains together with fixed prompts, video duration and independent evaluators.

\subsection{Text-to-Human-Motion}

\subsubsection{Inputs, Representation and Constraints}
Language conditions joint trajectories, pose parameters or learned motion codes; trajectories, keyframes or body-part constraints supply extra controls. The output describes movement rather than rendered appearance. Text--motion retrieval, movement diversity and kinematic or physical validity test distinct properties.

\subsubsection{Representative Approaches}
Building on advances in language modelling with transformers, TEMOS \cite{petrovich_temos_2022} introduces a variational autoencoder (VAE)-based probabilistic approach to T2HM generation, in contrast to most earlier deterministic models that produce only one motion per text description. Using a text encoder that outputs distribution parameters compatible with the VAE latent space, TEMOS enables the generation of multiple plausible motion sequences for the same input description. The architecture consists of Transformer-based encoders for motion and text. The motion encoder processes motion sequences, while the text encoder, based on DistilBERT, extracts semantic features from textual descriptions. These components work together to generate smooth, realistic human motion.
While TEMOS introduces a probabilistic approach (Transformer-based VAE) to T2HM generation
using a continuous latent space, T2M-GPT \cite{zhang_generating_2023} continues this trajectory by introducing discrete motion representations. Inspired by NLP-style tokenisation, it uses a VQ-VAE to convert motion into codebooks, allowing a GPT-style decoder to model motion as a sequence prediction task. This approach excels at long-range temporal coherence, especially in freeform generation tasks guided by rich language prompts.
In contrast, TM2T \cite{guo_tm2t_2022} takes a different approach by tokenising motion sequences into discrete representations and using a Transformer-based neural machine translation (NMT) model for T2HM and motion-to-text conversion.
This reciprocal learning approach uses motion-to-text supervision to help improve T2HM accuracy by enforcing alignment between generated motions and their textual descriptions. This inverse alignment mechanism ensures that motions generated from text remain semantically meaningful, while also improving motion-text retrieval performance.
While TEMOS and TM2T rely on Transformer-based architectures for T2HM generation, they often struggle with capturing the intricate temporal dynamics and diversity of human motion. To address these limitations, MotionDiffuse \cite{zhang_motiondiffuse_2024} adopts a diffusion-based approach. Its denoising probabilistic model iteratively refines noisy motion samples into realistic human motion, guided by a Transformer-based text encoder and a cross-modality linear Transformer denoiser. Moreover, MotionDiffuse allows multi-level manipulation, responding to fine-grained body-part control and supporting arbitrary-length motion synthesis with time-varying text prompts. Fg-T2M \cite{wang_fg-t2m_2023} builds upon the diffusion paradigm with fine-grained control mechanisms, focusing on disentangling local and global motion cues. It introduces hierarchical diffusion steps that allow better semantic fidelity between text prompts and motion segments, capturing subtle motion variations critical to realism.

Most motion generation models are task-specific and rely solely on text input,
lacking world knowledge for diverse motion synthesis.
MotionGPT \cite{zhang_motiongpt_2024}  introduces a large motion-language model (LMLM)
that fine-tunes LLMs, such as LLaMA,
for motion generation using multimodal control signals (text, keyframes, initial poses)
and a vector-quantised variational autoencoder (VQ-VAE) to discretise motion into tokens for integration with an LLM.
The VQ-VAE translates continuous motion data into a sequence of discrete codes,
making them compatible with the token-based processing of LLMs.
The fine-tuning approach, low-rank adaptation (LoRA), adjusts only a small fraction
of the LLM's parameters, enabling efficient adaptation to the motion generation task.
Light-T2M \cite{zeng_light-t2m_2025} introduces a compact, low-latency model optimised for deployment. It streamlines both encoding and decoding to achieve near real-time synthesis while maintaining competitive generation quality. Recent work further separates motion-prior learning from conditional adaptation \cite{zhang_large_2025, liang_omg_2024}, and adds step-by-step action plans before motion generation \cite{ouyang_motion-r1_2025}; these techniques are compared in Section~\ref{sec:techniques}.

\subsubsection{Comparative Analysis}
\textbf{Intervention and target:} motion-prior learning and preference adaptation address representation reuse and output alignment, respectively~\cite{zhang_large_2025,sheng_exploring_2024}.
\textbf{Advantage and evidence:} augmentation ablations and preference-method comparisons test different mechanisms.
\textbf{Limitation:} reward overfitting can disrupt adaptation; retrieval and motion-distribution measures do not establish physical executability.
\textbf{Research question:} evaluate semantic matching, motion quality and execution constraints separately, without mixing action classification with text retrieval.

\subsection{Text-to-3D-Objects}

\subsubsection{Inputs, Representation and Constraints}
The target is an object representation, such as an implicit field, point representation or mesh, generated from text with any image inputs identified separately. Appearance under rendered views and the underlying geometry are different evaluation targets. Scene layout and world simulation lie outside this object-focused comparison.

\subsubsection{Representative Approaches}
Early T2-3D generation relied on optimisation-based approaches,
in which each 3D object was generated by iteratively refining a neural representation guided by text-conditioned 2D priors.
Following the breakthrough of diffusion models in T2I generation,
one of the primary limitations in applying this approach to 3D synthesis is the lack of large-scale text-3D datasets.
To address this limitation,
DreamFusion \cite{poole_dreamfusion_2022} leverages a pretrained 2D T2I diffusion model as a prior for 3D generation,
eliminating the need for explicit 3D supervision.
The core innovation of DreamFusion is a loss function based on probability density distillation,
which enables the optimisation of a Neural Radiance Field (NeRF) \cite{mildenhall_nerf_2021}
using guidance from the 2D diffusion model.
However, the method has two inherent limitations:
extremely slow optimisation of NeRF;
low-resolution image space supervision on NeRF, leading to low-quality 3D models with a long processing time.
NVIDIA's Magic3D \cite{lin_magic3d_2023} addresses these limitations
by using a coarse-to-fine optimisation strategy that combines both low- and high-resolution diffusion priors to learn a high-fidelity 3D representation of the target content.
First, it generates a coarse 3D model using a low-resolution diffusion prior while accelerating optimisation
with a sparse 3D hash grid for efficient computation.
Then, using this coarse representation as initialisation, Magic3D refines the 3D model into a textured mesh,
optimising it with a high-resolution latent diffusion model.
As a result, Magic3D can produce high-quality 3D meshes in 40 minutes, making it twice as fast as DreamFusion,
which requires approximately 1.5 hours per generation.
While DreamFusion produces NeRFs and Magic3D produces textured meshes,
OpenAI's Point-E \cite{nichol_point-e_2022} proposes a fast, point cloud-based generation framework that bypasses mesh or NeRF representations. It first generates a synthetic view from text with a text-to-image diffusion model, then produces a 3D point cloud with a second diffusion model conditioned on that view, enabling rapid, low-compute 3D synthesis.
OpenAI's Shap-E \cite{jun_shap-e_2023} introduces a conditional generative model
that directly outputs implicit functions representing both geometry and texture.
Training proceeds in two stages: first, an encoder is trained to map explicit 3D assets into implicit function parameters,
and then a diffusion model is trained to generate these parameters directly from text or images.
Unlike DreamFusion and Magic3D, which require gradient-based optimisation for each text prompt,
Shap-E eliminates per-instance optimisation by using a conditional diffusion model,
producing NeRF-based renderings and textured meshes without per-prompt optimisation, although its diffusion sampling remains iterative.
This approach significantly improves efficiency while maintaining high shape diversity and rendering flexibility,
making it a major step towards real-time T2-3D synthesis.
Building on the shift towards real-time T2-3D generation, Instant3D \cite{li_instant3d_2024} introduces a fast, feedforward approach to T2-3D generation, eliminating the need for per-prompt iterative optimisation. The core innovation of Instant3D lies in its effective text conditioning strategy, enabling 3D object synthesis in a single forward pass. Using text embeddings from CLIP, it integrates three mechanisms: cross-attention aligns text descriptions with 3D features; style injection incorporates global text features into the 3D representation; and token-to-plane transformation converts token-level text embeddings into a structured triplane feature map.
Recent work also aligns 3D generation with human preferences over rendered views \cite{zhou_dreamdpo_2025, ye_dreamreward_2025, liu_dreamalign_2025}, compared in Section~\ref{sec:techniques}.

\subsubsection{Comparative Analysis}
\textbf{Intervention and target:} Shap-E learns to generate encoded assets conditionally, whereas DreamDPO optimises an asset using render preferences~\cite{jun_shap-e_2023,zhou_dreamdpo_2025}.
\textbf{Advantage and evidence:} these support amortised synthesis and per-asset preference refinement, respectively; they are not a controlled head-to-head comparison.
\textbf{Limitation:} appearance scores from selected views do not certify usable geometry. Image-conditioned systems receive additional information and belong to a separate track.
\textbf{Research question:} compare text adherence, multiview appearance and asset utility with explicit generation budgets.

\subsection{Towards a Unified Multimodal Horizon}

Advances in one modality can inform development in others, extending capabilities across multimodal generation. From early text-based sequence-to-sequence learning to models that simulate aspects of the world, the evolution has followed a text-centric paradigm, with text remaining the anchor as models extend their generative reach across modalities. Architectural advances have also transferred from text-based models. Transformer-based backbones, originally designed for language, now underpin nearly all modalities. Shared encoder architectures like CLIP \cite{radford_learning_2021} and CLAP \cite{elizalde_clap_2023} connect these modalities: CLIP creates a joint latent space for text and images that is widely used in vision tasks, and CLAP fulfils a similar role for text and audio. These shared latent spaces enable consistent conditioning, alignment, and compositionality across modalities, although embedding similarity alone is not a complete test of spatial composition, musical structure or temporal order.

\begin{figure}[H]
  \centering
  \scalebox{0.7}{
    \begin{tikzpicture}
      \clip (-2, 2) rectangle (18, 15);

      \useasboundingbox (0,0) rectangle (10,15);

      \tikzset{nd/.style={align=center,
            % draw=black,
            % fill=white,
            % text width=5cm,
            % font=\sffamily,
            % inner sep=0.5em,
            % outer sep=0.5em,
            % rounded corners=1ex
          }}

      \newcommand{\nodecontent}[5]{
        \begin{minipage}{#1}
          \raggedright
          % \centering
          \ifthenelse{\isempty{#3}}{}{
            \raisebox{-.4\height}{\includegraphics[width=3mm]{images/#3}}
            \ifthenelse{\isempty{#4}}{
              \hspace{-1.4em}
            }{
              \hspace{-1.0em}
            }
          }
          \ifthenelse{\isempty{#4}}{}{
            \raisebox{-.4\height}{\includegraphics[width=3mm]{images/#4}}
            \ifthenelse{\isempty{#5}}{
              \hspace{-1.4em}
            }{}
          }
          \ifthenelse{\isempty{#5}}{}{
            \raisebox{-.4\height}{\includegraphics[width=3mm]{images/#5}}
          }
          \tiny #2
        \end{minipage}
      }

      \node[nd] (transformer) at (7,8) {\nodecontent{2.4cm}{Transformer~\cite{vaswani_attention_2017}}{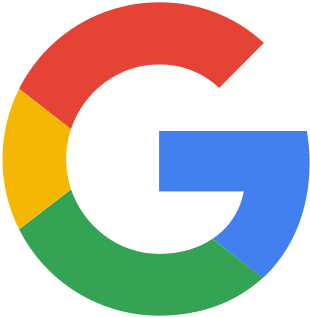}{}{}};
      \node[nd] (diffusion) at ($(transformer.center) + (-90:0.4cm)$) {\nodecontent{2.4cm}{Diffusion}{}{}{}};

      \node[nd] (gpt3) at ($(transformer.center) + (90:1.0cm)$) {\nodecontent{2.4cm}{GPT-3~\cite{brown_language_2020}}{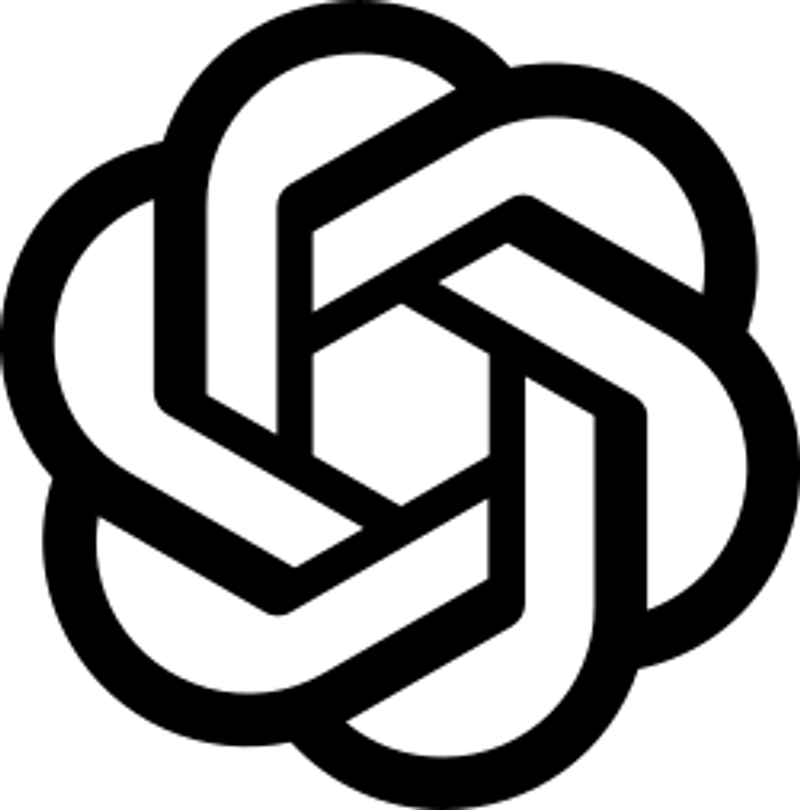}{}{}};

      % 2021-2023 block
      % Jurassic-1 \cite{lieber_jurassic-1_2021} 
      \node[nd] (jurassic) at ($(gpt3.center) + (130:1.6cm)$) {\nodecontent{2.4cm}{Jurassic-1}{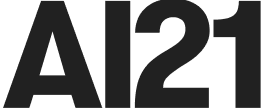}{}{}};
      % MT-NLG \cite{smith_using_2022}
      \node[nd] (mtnlg) at ($(jurassic.center) + (0:2.2cm)$) {\nodecontent{2.4cm}{MT-NLG}{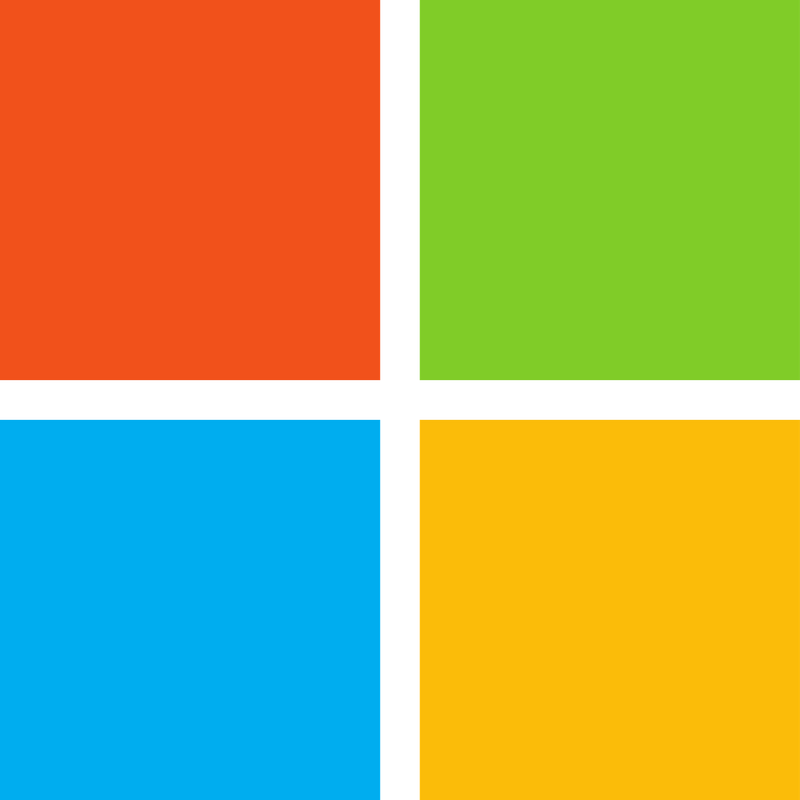}{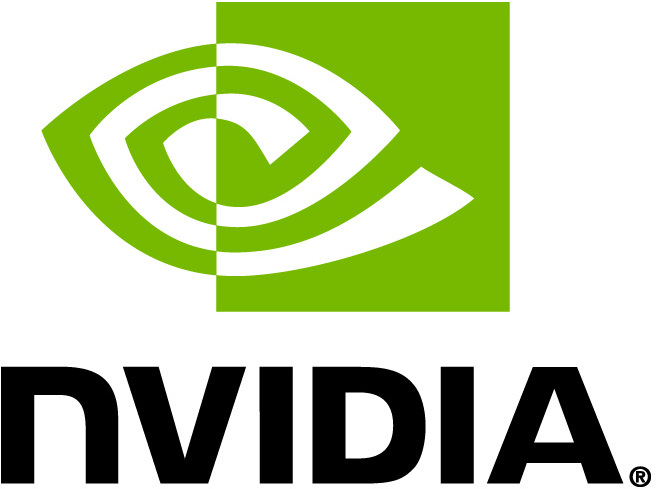}{}};
      % Gopher \cite{rae_scaling_2022}
      \node[nd] (gopher) at ($(jurassic.center) + (90:0.4cm)$) {\nodecontent{2.4cm}{Gopher}{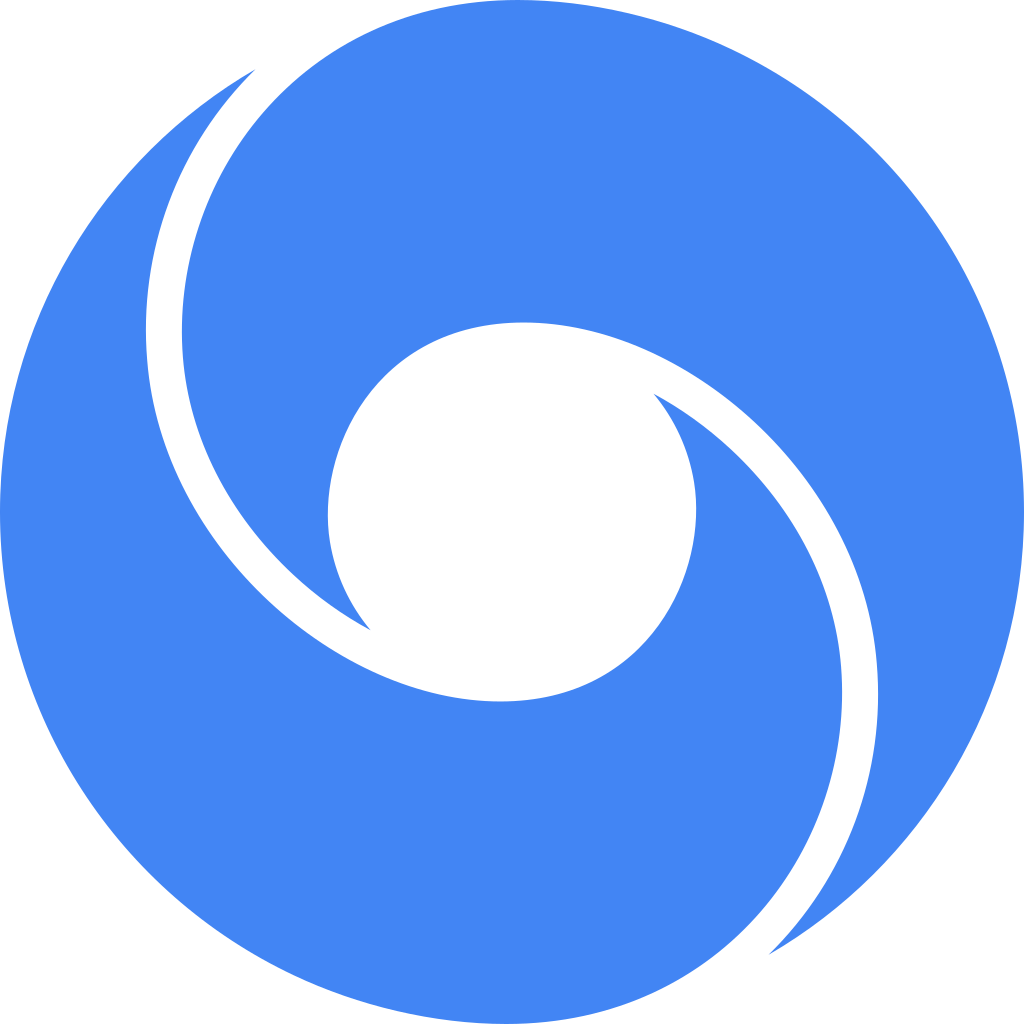}{}{}};
      % InstructGPT, \cite{ouyang_training_2022}
      \node[nd] (instructgpt) at ($(gopher.center) + (0:2.0cm)$) {\nodecontent{2.4cm}{InstructGPT~\cite{ouyang_training_2022}}{o.png}{}{}};
      % GLaM \cite{du_glam_2022}
      \node[nd] (glam) at ($(instructgpt.center) + (0:2.4cm)$) {\nodecontent{2.4cm}{GLaM~\cite{du_glam_2022}}{}{}{}};
      % PaLM 1,2 \cite{chowdhery_palm_2023, anil_palm_2023}
      \node[nd] (palm) at ($(gopher.center) + (90:0.4cm)$) {\nodecontent{2.4cm}{PaLM 1,2}{g.png}{}{}};
      % LLaMA \cite{touvron_llama_2023}
      \node[nd] (llama12) at ($(palm.center) + (0:2.4cm)$) {\nodecontent{2.6cm}{LLaMA 1,2~\cite{touvron_llama_2023-1}}{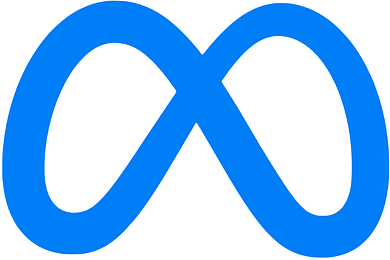}{}{}};

      % 2024 block
      % Llama 3 \cite{grattafiori_llama_2024}
      \node[nd] (llama3) at ($(palm.center) + (90:1.2cm)$) {\nodecontent{2.4cm}{LLaMA 3}{meta.png}{}{}};
      % Gemini \cite{team_gemini_2024}, Gemini 1.5 \cite{team_gemini_2024-1}
      \node[nd] (gemini15) at ($(llama3.center) + (0:2.1cm)$) {\nodecontent{2.6cm}{Gemini}{g.png}{}{}};
      % Mixtral \cite{jiang_mixtral_2024}
      \node[nd] (mixtral) at ($(llama3.center) + (90:0.4cm)$) {\nodecontent{2.4cm}{Mixtral~\cite{jiang_mixtral_2024}}{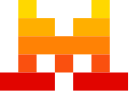}{}{}};
      % GPT-4 \cite{openai_gpt-4_2024}
      \node[nd] (gpt4) at ($(mixtral.center) + (0:1.8cm)$) {\nodecontent{2.4cm}{GPT-4}{o.png}{}{}};
      % DeepSeek-V3 \cite{deepseek-ai_deepseek-v3_2024}
      \node[nd] (deepseek) at ($(gpt4.center) + (0:1.76cm)$) {\nodecontent{2.4cm}{DeepSeek-V3}{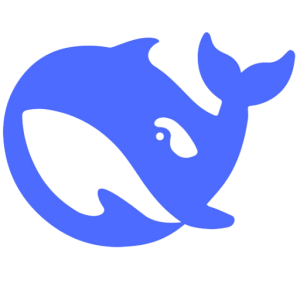}{}{}};

      % 2025 block
      % DeepSeek-R1 \cite{deepseek-ai_deepseek-r1_2025}
      \node[nd] (deepseekr1) at ($(mixtral.center) + (90:1.4cm)$) {\nodecontent{2.4cm}{DeepSeek-R1~\cite{deepseek-ai_deepseek-r1_2025}}{ds.png}{}{}};
      % LLaMA 4 \cite{ai_llama_2025}
      \node[nd] (llama4) at ($(deepseekr1.center) + (0:2.4cm)$) {\nodecontent{2.4cm}{LLaMA 4}{meta.png}{}{}};
      % Gemini 2.5
      \node[nd] (gemini25) at ($(deepseekr1.center) + (90:0.4cm)$) {\nodecontent{2.4cm}{Gemini 2.5}{g.png}{}{}};
      % GPT-4o
      \node[nd] (gpt4o) at ($(gemini25.center) + (0:1.7cm)$) {\nodecontent{2.4cm}{GPT-4o}{o.png}{}{}};

      % t2i
      % DDPM \cite{ho_denoising_2020}
      \node[nd] (ddpm) at ($(diffusion.center) + (20:3.4cm)$) {\nodecontent{2.0cm}{DDPM~\cite{ho_denoising_2020}}{}{}{}};
      % D2C \cite{sinha_d2c_2021}
      \node[nd] (d2c) at ($(ddpm.center) + (0:1.4cm)$) {\nodecontent{2.0cm}{D2C}{}{}{}};
      % DALL·E (OpenAI, Jan 2021): MoE: No (Transformer + dVAE), RLHF: No (Self-supervised on image-text pairs).
      \node[nd] (dalle) at ($(d2c.center) + (0:1.2cm)$) {\nodecontent{2.0cm}{DALL·E~\cite{ramesh_zero-shot_2021}}{o.png}{}{}};

      % 2022 block
      % DALL·E 2 \cite{ramesh_hierarchical_2022}
      \node[nd] (dalle2) at ($(ddpm.center) + (30:2.4cm)$) {\nodecontent{2.0cm}{DALL·E 2~\cite{ramesh_hierarchical_2022}}{o.png}{}{}};
      % GLIDE \cite{nichol_glide_2021}
      \node[nd] (glide) at ($(dalle2.center) + (0:2.2cm)$) {\nodecontent{2.0cm}{GLIDE}{o.png}{}{}};
      % Imagen \cite{saharia_photorealistic_2022}
      \node[nd] (imagen) at ($(dalle2.center) + (90:0.4cm)$) {\nodecontent{2.0cm}{Imagen}{g.png}{}{}};
      % Stable Diffusion \cite{rombach_high-resolution_2022}
      \node[nd] (stable) at ($(imagen.center) + (0:2.2cm)$) {\nodecontent{2.4cm}{Stable Diffusion~\cite{rombach_high-resolution_2022}}{}{}{}};

      % 2023 block
      % SDXL \cite{podell_sdxl_2024}
      \node[nd] (sdxl) at ($(imagen.center) + (30:2.4cm)$) {\nodecontent{2.0cm}{SDXL}{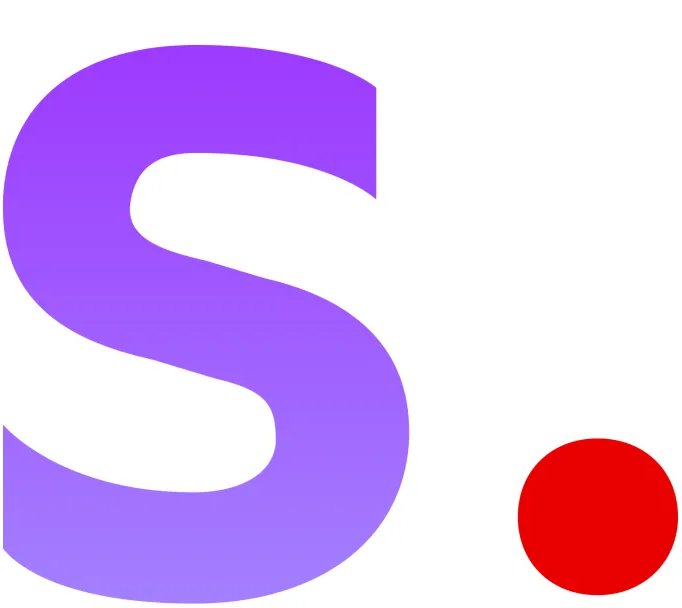}{}{}};
      % Google Muse \cite{chang_muse_2023}
      \node[nd] (muse) at ($(sdxl.center) + (0:2.0cm)$) {\nodecontent{2.4cm}{Google Muse}{g.png}{}{}};
      % ControlNet \cite{zhang_adding_2023}
      \node[nd] (controlnet) at ($(sdxl.center) + (90:0.4cm)$) {\nodecontent{2.4cm}{ControlNet~\cite{zhang_adding_2023}}{}{}{}};
      % DALL·E 3 \cite{betker_improving_2023}
      \node[nd] (dalle3) at ($(controlnet.center) + (0:2.0cm)$) {\nodecontent{2.4cm}{DALL·E 3~\cite{betker_improving_2023}}{o.png}{}{}};

      % Rectified Flow Transformers \cite{esser_scaling_2024}
      \node[nd] (rft) at ($(controlnet.center) + (90:1.4cm)$) {\nodecontent{2.0cm}{RFT~\cite{esser_scaling_2024}}{stability.png}{}{}};
      % EliGen \cite{zhang_eligen_2025}
      \node[nd] (eligen) at ($(rft.center) + (0:1.6cm)$) {\nodecontent{2.0cm}{EliGen}{}{}{}};
      % Janus framework \cite{wu_janus_2024}
      \node[nd] (janus) at ($(rft.center) + (90:0.4cm)$) {\nodecontent{2.4cm}{Janus}{ds.png}{}{}};
      % Janus-Pro \cite{chen_janus-pro_2025}
      \node[nd] (januspro) at ($(janus.center) + (0:1.8cm)$) {\nodecontent{2.4cm}{Janus-Pro~\cite{chen_janus-pro_2025}}{ds.png}{}{}};

      %t2m
      % MuseNet~\cite{payne_christine_musenet_2019}
      \node[nd] (musenet) at ($(transformer.center) + (-20:3.5cm)$) {\nodecontent{2.4cm}{MuseNet}{o.png}{}{}};
      % OpenAI Jukebox \cite{dhariwal_jukebox_2020}
      \node[nd] (jukebox) at ($(musenet.center) + (0:1.8cm)$) {\nodecontent{2.0cm}{Jukebox~\cite{dhariwal_jukebox_2020}}{o.png}{}{}};

      % 2023 block
      % Moûsai \cite{schneider_mousai_2024}
      \node[nd] (mousai) at ($(musenet.center) + (-24:2.4cm)$) {\nodecontent{2.4cm}{Moûsai~\cite{schneider_mousai_2024}}{}{}{}};
      % MusicLM \cite{agostinelli_musiclm_2023}
      \node[nd] (musiclm) at ($(mousai.center) + (0:1.5cm)$) {\nodecontent{2.4cm}{MusicLM~\cite{agostinelli_musiclm_2023}}{g.png}{}{}};
      % Noise2Music (Google, 2023) \cite{huang_noise2music_2023}
      \node[nd] (noise2music) at ($(musiclm.center) + (0:1.9cm)$) {\nodecontent{2.4cm}{Noise2Music~\cite{huang_noise2music_2023}}{g.png}{}{}};
      % MuseCoco \cite{lu_musecoco_2023}
      \node[nd] (musecoco) at ($(mousai.center) + (-90:0.4cm)$) {\nodecontent{2.4cm}{MuseCoco~\cite{lu_musecoco_2023}}{ms.png}{}{}};
      % MusicGen (Meta AI, 2023) \cite{copet_simple_2023}
      \node[nd] (musicgen) at ($(musecoco.center) + (0:2.1cm)$) {\nodecontent{2.4cm}{MusicGen~\cite{copet_simple_2023}}{meta.png}{}{}};

      % 2024 block
      % ETTA (NVIDIA 2024) \cite{lee_etta_2024}
      \node[nd] (etta) at ($(musecoco.center) + (-30:2.4cm)$) {\nodecontent{2.4cm}{ETTA}{n.jpg}{}{}};
      % FLUX \cite{fei_flux_2024}
      \node[nd] (flux) at ($(etta.center) + (0:1.7cm)$) {\nodecontent{2.4cm}{FluxMusic~\cite{fei_flux_2024}}{}{}{}};
      % Stable Audio Open (stability ai 2024) \cite{evans_stable_2025}
      \node[nd] (stableaudio) at ($(etta.center) + (-90:0.4cm)$) {\nodecontent{2.8cm}{Stable Audio Open~\cite{evans_stable_2025}}{stability.png}{}{}};

      % t2v
      % 2022 block
      % CogVideo \cite{hong_cogvideo_2022} 
      \node[nd] (cogvideo) at ($(diffusion.center) + (-120:1.8cm)$) {\nodecontent{2.4cm}{CogVideo~\cite{hong_cogvideo_2022}}{}{}{}};
      % Phenaki (Google, 2022), \cite{villegas_phenaki_2023}
      \node[nd] (phenaki) at ($(cogvideo.center) + (0:1.7cm)$) {\nodecontent{2.4cm}{Phenaki}{g.png}{}{}};
      % NUWA-XL (Microsoft, March 2023), \cite{yin_nuwa-xl_2023}
      \node[nd] (nuwaxl) at ($(cogvideo.center) + (-90:0.4cm)$) {\nodecontent{2.4cm}{NUWA-XL}{ms.png}{}{}};

      % 2023 2024 block
      % VideoLDM (NVIDIA, 2023), \cite{blattmann_align_2023}
      \node[nd] (videoLDM) at ($(nuwaxl.center) + (-90:1.4cm)$) {\nodecontent{2.4cm}{VideoLDM~\cite{blattmann_align_2023}}{n.jpg}{}{}};
      % VideoPoet (Google 2024), \cite{kondratyuk_videopoet_2024}
      \node[nd] (videopoet) at ($(videoLDM.center) + (0:2.1cm)$) {\nodecontent{2.4cm}{VideoPoet}{g.png}{}{}};
      % Emu Video (meta 2023) \cite{girdhar_factorizing_2025}
      \node[nd] (emu) at ($(videopoet.center) + (0:2.1cm)$) {\nodecontent{2.4cm}{Emu Video~\cite{girdhar_factorizing_2025}}{meta.png}{}{}};

      % 2025 block
      % OpenAI Sora (OpenAI, February 2024 - Alpha), \cite{brooks_video_2024}
      \node[nd] (sora) at ($(videoLDM.center) + (-90:1.4cm)$) {\nodecontent{2.4cm}{Sora}{o.png}{}{}};
      % Movie Gen (meta 2024) \cite{polyak_movie_2025}
      \node[nd] (moviegen) at ($(sora.center) + (0:1.5cm)$) {\nodecontent{2.4cm}{MovieGen~\cite{polyak_movie_2025}}{meta.png}{}{}};
      % Cosmos (NVIDIA 2025) \cite{agarwal_cosmos_2025}
      \node[nd] (cosmos) at ($(sora.center) + (-90:0.4cm)$) {\nodecontent{2.4cm}{Cosmos}{n.jpg}{}{}};
      % Veo (Google, 2025) \cite{veo-team_veo_2024}
      \node[nd] (veo) at ($(cosmos.center) + (0:1.7cm)$) {\nodecontent{2.4cm}{Veo}{g.png}{}{}};

      % t2hm

      %   TM2T \cite{guo_tm2t_2022}.
      \node[nd] (tm2t) at ($(diffusion.center) + (-160:2.2cm)$) {\nodecontent{2.4cm}{TM2T~\cite{guo_tm2t_2022}}{}{}{}};
      %   T2M-GPT \cite{zhang_generating_2023}
      \node[nd] (t2mgpt) at ($(tm2t.center) + (180:1.8cm)$) {\nodecontent{2.4cm}{T2M-GPT~\cite{zhang_generating_2023}}{}{}{}};

      %   MotionDiffuse \cite{zhang_motiondiffuse_2024}
      \node[nd] (motiondiffuse) at ($(tm2t.center) + (-160:3.0cm)$) {\nodecontent{2.4cm}{MotionDiffuse~\cite{zhang_motiondiffuse_2024}}{}{}{}};
      %   Fg-T2M \cite{wang_fg-t2m_2023}
      \node[nd] (fgt2m) at ($(motiondiffuse.center) + (180:1.6cm)$) {\nodecontent{2.4cm}{Fg-T2M~\cite{wang_fg-t2m_2023}}{}{}{}};
      %   MotionGPT \cite{zhang_motiongpt_2024}
      \node[nd] (motiongpt) at ($(fgt2m.center) + (-90:0.4cm)$) {\nodecontent{2.4cm}{MotionGPT~\cite{zhang_motiongpt_2024}}{}{}{}};

      %   MotionRL \cite{liu_motionrl_2024}
      \node[nd] (motionrl) at ($(motiongpt.center) + (-90:1.4cm)$) {\nodecontent{2.4cm}{MotionRL}{}{}{}};
      %   MotionGPT-2 \cite{wang_motiongpt-2_2024}
      \node[nd] (motiongpt2) at ($(motionrl.center) + (0:1.7cm)$) {\nodecontent{2.4cm}{MotionGPT-2}{}{}{}};
      %   Light-T2M \cite{zeng_light-t2m_2025}
      \node[nd] (lightt2m) at ($(motionrl.center) + (-90:0.4cm)$) {\nodecontent{2.4cm}{Light-T2M~\cite{zeng_light-t2m_2025}}{}{}{}};

      % t2-3d
      % DreamFusion~\cite{poole_dreamfusion_2022}
      \node[nd] (dreamfusion) at ($(transformer.center) + (170:3.8)$) {\nodecontent{2.4cm}{DreamFusion~\cite{poole_dreamfusion_2022}}{g.png}{}{}};
      % NeRF \cite{mildenhall_nerf_2021}
      \node[nd] (nerf) at ($(dreamfusion.center) + (180:1.4cm)$) {\nodecontent{2.4cm}{NeRF~\cite{mildenhall_nerf_2021}}{}{}{}};

      % Point-E (OpenAI) \cite{nichol_point-e_2022}
      \node[nd] (pointe) at ($(nerf.center) + (90:1.6cm)$) {\nodecontent{2.4cm}{Point-E~\cite{nichol_point-e_2022}}{o.png}{}{}};
      % Shap-E (OpenAI) \cite{jun_shap-e_2023}
      \node[nd] (shape) at ($(pointe.center) + (180:1.8cm)$) {\nodecontent{2.4cm}{Shap-E~\cite{jun_shap-e_2023}}{o.png}{}{}};
      % Magic3D (NVIDIA) \cite{lin_magic3d_2023}
      \node[nd] (magic3d) at ($(pointe.center) + (90:0.4cm)$) {\nodecontent{2.4cm}{Magic3D~\cite{lin_magic3d_2023}}{n.jpg}{}{}};
      % GET3D (NVIDIA) \cite{gao_get3d_2022}
      \node[nd] (get3d) at ($(magic3d.center) + (180:1.8cm)$) {\nodecontent{2.4cm}{GET3D}{n.jpg}{}{}};

      % VolumeDiffusion \cite{tang_volumediffusion_2024}
      \node[nd] (volumediffusion) at ($(get3d.center) + (90:1.8cm)$) {\nodecontent{2.8cm}{VolumeDiffusion}{ms.png}{}{}};
      % Meta 3D Gen \cite{bensadoun_meta_2024}
      \node[nd] (meta3dgen) at ($(volumediffusion.center) + (90:0.4cm)$) {\nodecontent{2.8cm}{Meta 3D Gen}{meta.png}{}{}};
      % Instant3D \cite{li_instant3d_2024}
      \node[nd] (instant3d) at ($(meta3dgen.center) + (0:2.2cm)$) {\nodecontent{2.4cm}{Instant3D~\cite{li_instant3d_2024}}{}{}{}};

      \tikzset{
        textl/.style={
            draw={rgb,255:red,180; green,180; blue,180},
            line width=2pt
          }
      }

      \tikzset{
        imgl/.style={
            draw={rgb,255:red,0; green,191; blue,255},
            line width=2pt
          }
      }

      \tikzset{
        musicl/.style={
            draw={rgb,255:red,128; green,0; blue,128},
            line width=2pt
          }
      }

      \tikzset{
        videol/.style={
            draw={rgb,255:red,255; green,165; blue,0},
            line width=2pt
          }
      }

      \tikzset{
        hml/.style={
            draw={rgb,255:red,34; green,139; blue,34},
            line width=2pt
          }
      }

      \tikzset{
        objl/.style={
            draw={rgb,255:red,0; green,0; blue,139},
            line width=2pt
          }
      }

      \draw[textl] (7,8.2) to[out=80,in=-90] (7,8.8);
      \draw[textl] (7,9.2) to[out=100,in=-90] (7,9.8);
      \draw[textl] (7,11.2) to[out=80,in=-90] (7,12.0);
      \draw[textl] (7,12.8) to[out=100,in=-90] (7,13.8);

      \draw[imgl] (8.2,7.9) to[out=30,in=-130] (11,8.4);
      \draw[imgl] (11.4,9.0) to[out=60,in=-130] (13,9.8);
      \draw[imgl] (13.4,10.6) to[out=60,in=-100] (14.8,11.3);
      \draw[imgl] (14.9,12.2) to[out=80,in=-90] (15.0,13.1);

      \draw[musicl] (8.2,7.8) to[out=-30,in=130] (11,7.1);
      \draw[musicl] (11.2,6.5) to[out=-60,in=110] (13.1,6.1);
      \draw[musicl] (13.3,5.2) to[out=-60,in=90] (14.6,4.5);

      \draw[videol] (7,7.4) to[out=-100,in=-90] (7,6.4);
      \draw[videol] (7,5.4) to[out=-80,in=-90] (7,4.6);
      \draw[videol] (7,3.9) to[out=-100,in=-90] (7,3.2);

      \draw[hml] (5.6,7.7) to[out=210,in=60] (3.2,7.1);
      \draw[hml] (3.0,6.7) to[out=-130,in=60] (1.4,6.0);
      \draw[hml] (1.2,5.3) to[out=-110,in=90] (1.0,4.2);

      \draw[objl] (5.6,7.8) to[out=150,in=-60] (3.2,8.4);
      \draw[objl] (2.6,8.9) to[out=150,in=-60] (1.1,10.0);
      \draw[objl] (0.9,10.9) to[out=110,in=-60] (0.4,12.3);

      \node at (0, 0) {};

    \end{tikzpicture}}
  \caption{\label{sec3:figure_1}This graph depicts the evolution from foundational transformer and diffusion structures, branching into six pathways that highlight milestone models for each modality. Models with a bracketed number are discussed in this survey and listed in the references; the other models are shown for context.}
\end{figure}

Milestone models across modalities build on foundational Transformer and diffusion architectures, with major companies contributing studies in each domain (see Figure~\ref{sec3:figure_1}). The future of multimodal reasoning will not hinge solely on scaling model size, but on fusing modalities through shared semantics, architectures, and spatio-temporal representations. These developments point towards integrated systems for multimodal perception and generation.

\section{Techniques}\label{sec:techniques}

To unlock sophisticated reasoning and generative capabilities across modalities, MGMs increasingly rely on four techniques: SSL, MoE, RLHF, and CoT prompting. These techniques address complementary challenges: learning from unlabelled data, modularising computation, aligning with human intent, and guiding structured inference. MoE selectively activates expert pathways, while CoT structures inference through intermediate reasoning steps.
For each technique, the following subsections trace its adoption across the six modalities and identify what changes, the property targeted, the evaluation used and the limit of the resulting claim.

\subsection{Self-Supervised Learning}
\label{sec:ssl-comparison}
SSL has fundamentally shaped the trajectory of modern generative models, beginning with its pivotal role in LLMs and progressively influencing multimodal domains. The strategy of training models on vast, unlabelled datasets via self-supervised objectives has enabled breakthroughs in scaling, generalisation, and data efficiency. In text-only models, large-scale SSL has proven essential, unlocking emergent reasoning and compositional capabilities with sufficient model and dataset scale. Inspired by this success, researchers in multimodal generation seek to replicate this paradigm across sensory domains, facing additional challenges of modality-specific representations and cross-modal alignment.
This section systematically examines the adoption, adaptation,
and impact of SSL across six generative modalities.
It analyses both the direct benefits within each modality and the broader patterns of technique transfer, synergy, and divergence that emerge when scaling SSL beyond text.

\paragraph{Mechanism and Attribution} The role of SSL in a generator depends on the training stage: it can learn an unconditional prior, a compressed representation, or encoder features that are transferred into conditional training. A representation-transfer objective can be written schematically as $\mathcal{L}=\mathcal{L}_{\mathrm{gen}}+\lambda\mathcal{L}_{\mathrm{align}}$, where $\mathcal{L}_{\mathrm{gen}}$ trains generation, $\mathcal{L}_{\mathrm{align}}$ compares generator features with a pretrained teacher, and $\lambda$ sets their relative weight. Holding the backbone, data and optimisation budget fixed while removing or changing the auxiliary loss is needed to attribute gains to representation transfer. Paired language--output alignment and automatically generated captions supply additional supervision, so a self-supervised component does not make the entire system annotation-free.

\subsubsection{The Foundational Role of SSL in LLMs}
Unsupervised pretraining underpinned the success of GPT-3 \cite{brown_language_2020}.
Unlike traditional supervised learning, which relies on carefully labelled datasets,
SSL allows models to learn from the vast, unstructured,
and unlabelled text readily available on the internet.
This allows them to acquire a deep understanding of grammar, syntax, factual knowledge,
and even rudimentary reasoning capabilities simply by processing massive text corpora \cite{radford_language_2019}.

The core mechanism involves self-supervised objectives, where the model learns by predicting parts of the input data itself. Autoregressive models are well suited to this task, as they generate the next token based on the surrounding context tokens.
This paradigm succeeds largely due to its scalability and independence from fine-grained, labelled data. By leveraging enormous, easily accessible unlabelled datasets such as WebText \cite{radford_language_2019}, researchers have been able to train models with billions or even trillions of parameters, leading to emergent capabilities and state-of-the-art performance across a wide range of natural language tasks. The success of LLMs demonstrates the potential of large-scale unsupervised learning on unlabelled text and provides a possible template for other domains.

Unlike unimodal LLMs, multimodal generation must bridge the gap between distinct data types, typically text and a sensory modality (visual or auditory). This necessitates mechanisms not only for understanding each modality individually but also for learning the complex alignments and correspondences between them. The core challenge lies in learning meaningful, rich representations from potentially vast unlabelled visual or audio datasets in a way that allows them to be effectively controlled and guided by textual input.
This section provides a comparative technical analysis of pretraining strategies across the aforementioned six modalities. It examines the role, prevalence, and effectiveness of unsupervised learning techniques applied to large unlabelled datasets within each modality. It investigates the interplay, cross-pollination, and transferability of these techniques across diverse modalities.

\subsubsection{SSL in Text-to-Text Generation}
OpenAI demonstrated the effectiveness of this approach in GPT-1 \cite{radford_improving_2018}, which was followed by discriminative fine-tuning for tasks such as Natural Language Inference (NLI), paraphrase detection, and story completion. Their use of Transformer networks enabled the model to capture long-range linguistic dependencies, making it highly effective for various NLP applications.
Later LLMs leveraged massive public datasets such as the Common Crawl project\footnote{\href{http://commoncrawl.org}{http://commoncrawl.org}}, which rapidly expanded to 460 TiB of uncompressed content as of January 2025. This abundance of data supported the development of large-scale generative models.

The trajectory of LLM development strongly suggests that unsupervised learning, when combined with massive scale in both model parameters and unlabelled training data, is an exceptionally effective strategy for building powerful foundation models. It established a compelling paradigm: harness the most abundant and cost-effective data source (unlabelled text) with relatively simple, scalable self-supervised objectives to build models with broad capabilities. This approach offers a potential template for developing foundation models in other modalities, contingent upon the availability of comparably large unlabelled datasets and the formulation of equally effective SSL objectives suitable for those data types.
The next-token objective rewards prediction of observed text rather than independently verified factuality or instruction satisfaction, so comparisons with instruction-tuned systems must separate pretraining from subsequent supervision and preference optimisation~\cite{ouyang_training_2022}.

\subsubsection{SSL in Text-to-Image Generation}
The success of T2I models such as the DALL·E series \cite{ramesh_hierarchical_2022, betker_improving_2023}, Stable Diffusion \cite{rombach_high-resolution_2022}, and Contrastive Language-Image Pre-training (CLIP) \cite{radford_learning_2021} largely stems from massive datasets of image-text pairs, such as WebImageText \cite{radford_learning_2021}. However, curating these datasets is resource-intensive and the web-scraped captions often suffer from noise, simplicity, and inherent biases, limiting model performance and fairness. This motivates the exploration of SSL techniques, which offer a pathway to leverage the vast quantities of readily available unlabelled image data.

One direction transfers self-supervised visual representations into the generative model during training. REPA \cite{yu_repa_2025} aligns projected denoising hidden states with clean-image features from a pretrained encoder. Its principal experiments concern class-conditioned ImageNet generation; the separate text-to-image experiment on Microsoft Common Objects in Context (MS-COCO) supplies the relevant evidence for this survey. Class-conditioned improvements in Fr\'echet inception distance (FID) cannot be transferred directly to claims about text adherence, and the encoder's pretraining forms part of the resource budget.

The potential benefits of leveraging unlabelled data via SSL are significant, including access to vastly larger datasets, improved visual representation quality and reduced annotation costs. However, challenges remain, particularly in achieving robust cross-modal alignment between visual concepts learned through self-supervision and specific textual descriptions. This field is rapidly evolving, suggesting a future trend towards hybrid approaches that combine the strengths of large-scale unsupervised visual pretraining with efficient, targeted supervised fine-tuning. This combination may lead to more capable, robust, and potentially fairer T2I generation systems.

\subsubsection{SSL in Text-to-Video Generation}
Compared with T2I, T2V generation introduces the additional dimension of time, requiring models to learn not only visual appearance and text alignment but also temporal consistency and motion dynamics. This necessitates incorporating video data into the pretraining process. A notable early approach, Make-A-Video \cite{singer_make--video_2022}, explicitly separates the learning of appearance and motion. It starts with a pretrained T2I model (trained on paired image-text data to understand appearance and text alignment) and extends it by adding spatiotemporal layers. These new temporal components are then trained exclusively on unlabelled video footage to learn motion patterns and temporal dependencies. This method uses existing T2I capabilities while isolating the learning of motion to the unlabelled video domain.

A crucial element enabling the training of such models with text conditioning is the use of sophisticated data preparation techniques, such as the “recaptioning” technique used in DALL·E 3 \cite{betker_improving_2023}. This involves first training a separate, highly descriptive captioner model and then using this model to automatically generate detailed text captions
for all the videos in the training set. This process effectively converts a large corpus of unlabelled or minimally labelled video
into a pseudo-paired dataset suitable for training a text-conditional generative model. Recaptioning thus creates text supervision and is a different intervention from learning without labels.

VideoREPA \cite{zhang_videorepa_2025} transfers spatial and temporal token relations from a self-supervised video encoder (VideoMAEv2) into CogVideoX fine-tuning. Its diffusion-only, spatial-only and temporal-only ablations test the alignment design, and VideoPhy and VideoPhy2 separately score semantic adherence and physical commonsense: improvement in the latter need not improve the former. These automatic judgements support benchmark-specific plausibility claims, not general physical correctness.

SSL from unlabelled video data is beneficial for enabling T2V models to generate motion, as it provides the necessary temporal signal that static image datasets lack. However, the effectiveness of this pretraining varies.
Although unlabelled video supports learning basic motion, generating complex, long-range, and precisely controllable motion that accurately reflects nuanced text descriptions
remains a significant challenge.
Furthermore, even when motion is learned from unlabelled video,
current state-of-the-art T2V models still typically rely on strong T2I foundations
and powerful text encoders (frequently CLIP-based or large language models)
for understanding the input prompt and guiding the generation process.
This indicates a continued reliance, at least partially,
on principles derived from paired-data training for the crucial text-alignment aspect,
complementing the unsupervised learning of temporal dynamics.

\subsubsection{SSL in Text-to-Music Generation}
T2M generation presents another unique set of challenges and opportunities for SSL. During the SSL phase, models learn fundamental representations of sound, acoustics, and basic musical structures without relying on explicit text labels or annotations. These learned representations encapsulate general knowledge about the audio domain, forming a robust foundation that can subsequently be adapted, typically through fine-tuning, to the specific requirements of the T2M task. Several state-of-the-art T2M models exemplify this reliance on unsupervised pretraining, primarily using Transformer-based or diffusion-based architectures.
\begin{itemize}
      \item{Transformer-Based Architectures}
            These models often treat T2M as a sequence modelling task, operating on discrete audio tokens.
            Google's MusicLM \cite{agostinelli_musiclm_2023}
            uses a hierarchical Transformer architecture,
            conceptualising audio generation as language modelling.
            It relies heavily on pretrained components: semantic tokens from a masked language model (w2v-BERT)
            pretrained on audio capture long-term structure,
            while acoustic tokens from a pretrained SoundStream neural codec ensure high fidelity.
            MusicLM further conditions generation using text and melody embeddings
            from the pretrained MuLan model.
      \item{Diffusion-Based Architectures}
            Moûsai \cite{schneider_mousai_2024} uses a two-stage cascading diffusion process for generating long-form, high-quality music. The first stage involves unsupervised pretraining of a novel Diffusion Magnitude Autoencoder (DMAE) for audio compression. The second stage is a text-conditioned LDM operating in this compressed space, guided by embeddings from a frozen, pretrained Transformer language model. Compression is a component of the pipeline, not evidence that its conditional generator learns without text supervision, and reconstruction quality and prompt adherence require different tests.
\end{itemize}
\noindent{}StepAudio 3 Music uses representations derived from BEST-RQ, a self-supervised speech pretraining method, within a pipeline also containing planning, MoE generation and preference tuning~\cite{feng_stepaudio3_2026}. Its system-level results establish the use of these representations in music generation, but do not isolate their causal contribution.
Self-supervised pretraining on large unlabelled audio datasets supports improvements in generation quality, fidelity, musical coherence, and long-term structure. Masked audio modelling, contrastive learning, and neural audio codecs extract representations from unlabelled audio that can be integrated into Transformer-based language models and diffusion models, improving data efficiency for T2M generation.

\subsubsection{SSL in Text-to-Human-Motion Generation}
Generating realistic 3D human motions from textual prompts involves bridging abstract semantic concepts with complex, continuous motion sequences. The limited availability of high-quality paired datasets motivates leveraging SSL methods to harness vast unlabelled data sources. These methods significantly improve model realism, generalisation, and efficiency by learning fundamental motion priors before applying textual guidance.

The Large Motion Model (LMM) \cite{zhang_large_2025} uses extensive heterogeneous datasets
and applies random masking and frame-rate augmentation strategies during SSL,
creating a unified, generalisable motion prior that significantly boosts performance across various modalities including T2HM generation.
Its fixed-backbone masking and frame-rate ablations probe augmentation choices rather than a no-pretraining comparison; OMG supplies a complementary comparison with and without motion pretraining before text-conditioned controller adaptation~\cite{liang_omg_2024}.

SSL methods in human motion generation are emerging. Compared to other output types, this field lacks large-scale datasets. These methods demonstrate strong potential by harnessing unlabelled data to learn rich and transferable motion representations. They significantly enhance motion realism and adaptability to novel prompts, offering crucial advantages in a data-constrained domain.

\subsubsection{SSL in Text-to-3D-Objects Generation}
Currently, direct SSL on large unlabelled 3D datasets, in the traditional sense of training a backbone model on a pretext task and then fine-tuning it for T2-3D generation, is not the dominant strategy for most state-of-the-art T2-3D systems like Instant3D \cite{li_instant3d_2024} and DreamReward \cite{ye_dreamreward_2025}. The dominant strategies heavily rely on transferring knowledge from the 2D domain \cite{poole_dreamfusion_2022, li_instant3d_2024, mildenhall_nerf_2021} or using available supervised 3D data \cite{jun_shap-e_2023}. The apparent absence of SSL on large unlabelled 3D datasets as a common strategy in the analysed T2-3D models
likely stems from several significant practical and technical challenges, including the difficulty of acquiring large, clean, and standardised unlabelled 3D datasets amid data scarcity, heterogeneity, noise, and the lack of a universally optimal representation.
Developing scalable and powerful self-supervised objectives specifically tailored to capture the complex geometry, topology, and appearance of 3D data remains an open research challenge, since straightforward extensions of successful 2D and NLP techniques may not suffice.

\subsubsection{Cross-Modal Reflections on SSL}
While the fundamental principles of unsupervised learning, self-supervised prediction, contrastive alignment, and masked modelling are broadly shared, the degree to which SSL forms the bedrock of generative capabilities varies significantly across modalities: in T2T, SSL is foundational, as it is the dominant enabler of capability emergence. In T2I, self-supervised visual pretraining is influential yet supplementary. Large-scale paired text-image datasets remain critical for cross-modal alignment. In T2V, motion dynamics crucially benefit from unlabelled video pretraining, but strong foundations from T2I and supervised text conditioning remain essential. In T2M, SSL is highly effective: large unlabelled audio corpora combined with masked and contrastive objectives build strong foundations for T2M generation. In T2HM, SSL is emerging but promising: pretraining on unlabelled motion captures fundamental motor priors, crucial in low-data regimes. In T2-3D, SSL is currently minimal: direct SSL on raw 3D data remains rare due to severe data-related, representational, and computational challenges.

Despite differences, a strong cross-pollination is evident: techniques such as masked modelling (originally from language and vision) and contrastive learning have migrated across modalities, often with adaptations tailored to modality-specific structures. Moreover, modality-bridging models (e.g., CLIP for vision-language, CLAP for audio-text) underscore the growing trend of transferring learned representations and self-supervised methods between domains.

For comparison across these modalities, the useful unit is the intervention and its downstream test: pretrained initialisation versus training from scratch, an auxiliary target versus the same conditional objective, or a learned representation versus an alternative interface. Better reconstruction or representation quality is supporting evidence; generation quality and instruction adherence require their own evaluations.

\subsection{Mixture of Experts}
\label{sec:moe-comparison}

MoE architectures select specialised expert networks for a given input. MoE was first introduced in \cite{jacobs_adaptive_1991} as a method to improve model efficiency by dividing tasks among specialised expert networks, with a gating network determining which expert should be used for a given input, offering conditional computation as a means to increase model capacity without a proportional rise in computational cost. Originally developed to divide learning problems into subtasks handled by specialised networks, MoE was later adopted for scaling language models and has since expanded across diverse modalities. This section systematically traces the adoption and evolution of MoE architectures, highlighting how sparse activation, expert specialisation, and adaptive routing strategies enable greater expressivity, fine-grained control, and efficiency in increasingly complex generative tasks.

\paragraph{Mechanism and Cost} For an input $x$, an output mixture has the form $y=\sum_{k\in S(x)}g_k(x)f_k(x)$, where $f_k$ is expert $k$, $g_k$ its gating weight and $S(x)$ the selected expert set. Sparse routing evaluates a selected subset $S(x)$; dense mixtures evaluate all experts; other designs blend parameters or switch complete denoisers across timesteps. Token-choice routing lets each token select experts, whereas expert-choice routing lets experts select tokens~\cite{jiang_mixtral_2024,sun_ec-dit_2025}. Although only a subset of experts is computed for each token, all expert weights must still be stored and routed across devices, so comparisons should record routing policy, total and active parameters, memory and measured runtime rather than infer efficiency from active parameter counts alone.

\subsubsection{MoE in Text-to-Text Generation}

In 2022, GLaM \cite{du_glam_2022} used the MoE structure to reduce computational costs by activating only a subset of its numerous parameters during each operation. Since then, MoE has been considered a strong candidate for future scaling, with efforts to ensure balanced expert utilisation and prevent overload.
Concurrently, Mistral AI introduced the Mixtral series \cite{jiang_mixtral_2024}, notably the Mixtral 8x7B model, which employs MoE layers with 8 experts and Top-2 routing. Mixtral 8x7B, in particular, is reported to match or outperform a larger dense model (Llama 2 70B) across all evaluated benchmarks while using only about 13B of its 47B parameters per token.

\subsubsection{MoE in Text-to-Image Generation}

ERNIE-ViLG 2.0 \cite{feng_ernie-vilg_2023} introduced the "Mixture of Denoising Experts" (MoDE) concept, employing 10 distinct U-Net based expert denoisers, each responsible for a specific, non-overlapping interval of the diffusion timesteps. The routing mechanism was implicitly determined by the current timestep, enabling expert specialisation across high- and low-noise regimes to enhance text-image alignment and object fidelity. DiffMoE \cite{shi_diffmoe_2025} introduced adaptive two-stage routing, using batch-pool-based routing during training and complexity-adaptive routing during inference to dynamically allocate expert resources based on input difficulty, further optimising generation quality and efficiency. Pushing scalability further, EC-DIT \cite{sun_ec-dit_2025} integrated expert-choice routing within the Diffusion Transformer (DiT) backbone, reaching 97 billion parameters and achieving state-of-the-art performance on the GenEval benchmark,
with routing innovations that balanced computational loads effectively across experts.
GenEval checks prompt-specified objects and attributes~\cite{ghosh_geneval_2023}, whereas FID concerns generated distributions, so improvements on the two are not interchangeable. Most DiffMoE experiments are class-conditional; its text-to-image comparison matches average active parameters while changing routing and total capacity together (Section~\ref{sec:finding-experts}).

\subsubsection{MoE in Text-to-Music Generation}

Explicit MoE use is absent from major T2M models such as Google's \textit{MusicLM} \cite{agostinelli_musiclm_2023} (2023), Meta's \textit{MusicGen} \cite{copet_simple_2023} (2023), and OpenAI's \textit{Jukebox} \cite{dhariwal_jukebox_2020} (2020). These models instead use architectures such as hierarchical sequence-to-sequence transformers, AR transformers with audio tokenisation, and VQ-VAE-based sparse transformers, respectively. MoE is not mainstream in T2M generation because its strengths in efficiency and scalability can conflict with music's requirements for sequential coherence, high-dimensional feature handling, and subtle acoustic control. MoE models struggle with routing complexity: music demands nuanced, continuous feature analysis (pitch, rhythm, harmony), while current MoE routers are better suited for discrete, semantic tokens such as text. Until MoE designs are tailored to music's structural and temporal intricacies, their theoretical benefits remain overshadowed by practical challenges.
A recent exception is StepAudio 3 Music, which uses an autoregressive MoE music generator~\cite{feng_stepaudio3_2026}; because its planning, representation and preference stages change at the same time, its aggregate improvements cannot be attributed to routing alone.

\subsubsection{MoE in Text-to-Video Generation}

CogVideoX \cite{yang_cogvideox_2025} presents a large-scale T2V generation model
built upon a diffusion transformer architecture,
designed to produce 10-second videos at $768\times1360$ resolution and 16 fps.
A key innovation is the use of an “expert transformer with expert adaptive LayerNorm”,
facilitating deep fusion between text and video modalities
and potentially leveraging expert specialisation
to manage different cross-modal relationships and enhance semantic fidelity.
Despite its name, this design normalises text and video hidden states separately rather than routing tokens sparsely among experts.
Alongside this design, CogVideoX incorporates a 3D VAE for spatiotemporal compression,
progressive training strategies, and multi-resolution frame packing,
further boosting efficiency and generation quality.
Wan2.2 A14B \cite{wan_wan22_2025} instead switches between high-noise and low-noise expert denoisers, extending the timestep specialisation of ERNIE-ViLG 2.0 to video; changes in data and training prevent assigning all of its gains to the expert design.
Complementing the generative focus,
MJ-Video \cite{tong_mj-video_2025} addresses evaluation by employing a lightweight (2B parameters) MoE-based reward model.
Built upon the MJ-Bench-Video dataset,
MJ-Video decomposes the complex task of video quality assessment into five dimensions,
Alignment, Safety, Fineness, Coherence \& Consistency, and Bias \& Fairness,
assigning specialised experts to each. This dynamic expert selection enables finer-grained,
more accurate preference predictions compared to traditional monolithic reward models,
demonstrating the use of MoE in both generation and evaluation frameworks for T2V.

\subsubsection{MoE in Text-to-Human-Motion Generation}

A notable application of MoE principles in T2HM generation
is the OMG framework \cite{liang_omg_2024},
which addresses the challenge of generating plausible motions for unseen text prompts using limited paired datasets.
OMG first pretrains a large unconditional diffusion model on over 20 million instances of unlabelled motion data,
then introduces a “Mixture-of-Controllers” (MoC) during fine-tuning.
The MoC block employs text-token-specific experts within a ControlNet architecture,
using cross-attention to route motion features to specialised controllers,
significantly improving zero-shot T2HM alignment.
Complementing this, GenM3 \cite{shi_genm3_2025} tackles multi-source motion data heterogeneity
with a Multi-Expert VQ-VAE and a Multi-path Motion Transformer (MMT).
The MMT uses densely activated modality-specific experts across motion, text, and cross-modal pathways
to better adapt to diverse dataset distributions,
achieving robust representation learning and state-of-the-art performance on HumanML3D.
Together, these frameworks demonstrate MoE's effectiveness in enhancing both generative quality
and cross-modal alignment in T2HM models.
Their evidence concerns conditioning and dense pathways rather than sparse inference savings: GenM3's dense-versus-sparse comparison favours dense activation in the tested configuration.

\subsubsection{MoE in Text-to-3D-Objects Generation}

TripoSG \cite{li_triposg_2025} targets high-fidelity 3D mesh generation from single images
using a Diffusion Transformer (DiT) backbone enhanced with sparse MoE layers.
Following the DiT-MoE design,
it replaces standard FFNs with a structure combining one shared expert and multiple specialised experts,
selecting the top two experts for each token. Operating on latent tokens from a 3D VAE,
this approach enables TripoSG to scale effectively to large 3D datasets,
boosting generation fidelity while maintaining computational efficiency.
Because TripoSG is image-conditioned and its ablations do not measure text faithfulness, it illustrates an architectural transfer rather than direct text-to-3D evidence.

\subsubsection{Cross-Modal Reflections on MoE}

Across modalities, MoE has evolved from a sparsity-driven scaling tool into a modality-sensitive design philosophy,
supporting different forms of specialisation.
Text and vision domains have seen rapid maturation of routing strategies,
while motion, music, and 3D generation are beginning to adapt MoE principles to their structural and temporal complexities.
Innovations such as time- and space-specific expert routing in vision,
token-level expert controllers in motion, and emergent modular experts in video illustrate the transfer of ideas between modalities.
As these fields continue to borrow architectural strategies from one another,
future MoE systems are likely to unify temporal, spatial, and semantic specialisation under adaptive, multimodal frameworks,
further connecting modalities and extending generative capabilities.
Routing granularity, however, determines what is compared: tokens for capacity, denoising stages for noise specialisation, and text-conditioned controllers for local control. Evidence for one granularity cannot justify a general claim that sparse routing improves quality or efficiency across modalities.

\subsection{Reinforcement Learning from Human Feedback}
\label{sec:preference-comparison}

RLHF introduced an important shift in aligning generative models with nuanced human preferences across modalities. Following its success in T2T language models, RLHF developed into a general paradigm comprising supervised fine-tuning (SFT), reward modelling (RM), and reinforcement learning, typically proximal policy optimisation (PPO). Alternatives such as direct preference optimisation (DPO) offer efficiency gains. As generative modelling expanded across modalities, RLHF was adapted to their different requirements, producing new methods, datasets, and fine-tuning strategies.

\paragraph{Optimisation Target} We distinguish the source of feedback (human, automatic metric or model judge) from the optimisation algorithm. Reward-based adaptation seeks outputs with high reward while constraining departure from a reference policy; a representative objective is $\max_\theta\,\mathbb{E}_{c,y\sim p_\theta(\cdot\mid c)}[r(c,y)]-\beta\mathbb{E}_c[D_{\mathrm{KL}}(p_\theta(\cdot\mid c)\|p_{\mathrm{ref}}(\cdot\mid c))]$, where $r$ is the learned reward, $D_{\mathrm{KL}}$ the Kullback--Leibler divergence and $\beta$ controls regularisation~\cite{ouyang_training_2022}. Direct preference fitting instead increases the relative likelihood of a preferred candidate over a rejected one, relative to a reference model. Gains against the training reward need confirmation by separate judgements or task-specific tests.

\subsubsection{RLHF in Text-to-Text Generation}

The researchers at OpenAI and DeepMind first introduced Deep Reinforcement Learning from Human Preferences in 2017 \cite{christiano_deep_2017}, aiming to align AI-generated outputs closely with human preferences,
addressing limitations of traditional RL methods based solely on predefined rewards.
InstructGPT advanced RLHF for large language models \cite{ouyang_training_2022}
through the now-standard three-step process of supervised fine-tuning (SFT), reward modelling (RM), and reinforcement learning (RL) with PPO, demonstrating how alignment could transform LLMs into more helpful, honest, and harmless assistants.
Llama 2 \cite{touvron_llama_2023-1} extended these advances to open-source models,
pioneering the use of separate reward models for helpfulness and safety to manage trade-offs at scale.
However, the analysis in \cite{casper_open_2023} identifies limitations and open problems in RLHF, highlighting the need to address vulnerabilities alongside advances in alignment.
DeepSeek-R1 \cite{deepseek-ai_deepseek-r1_2025} replaces learned preference rewards with rule-based checks for tasks that have verifiable answers: its R1-Zero variant applies group relative policy optimisation (GRPO) to a pretrained base model, rewarding only answer correctness and output format, and average pass@1 on the American Invitational Mathematics Examination (AIME) 2024 rises from 15.6\% to 77.9\% during training. The authors avoid neural reward models for these tasks because of reward hacking, and note that tasks without a reliable verifier remain unresolved.

\subsubsection{RLHF in Text-to-Image Generation}

ImageReward \cite{xu_imagereward_2023} introduced the first large-scale reward model tailored for T2I and proposed Reward Feedback Learning (ReFL) to tune diffusion models more effectively.
The work in \cite{wallace_diffusion_2024} adapts direct preference optimisation (DPO) from LLMs to diffusion models to achieve state-of-the-art prompt adherence without explicit reward modelling.
Diffusion-DPO compares its preference-tuned Stable Diffusion XL (SDXL) with base and refinement variants through human judgements of visual appeal and prompt alignment; avoiding an explicit reward model simplifies one training stage, but does not remove preference bias or establish superior performance for every prompt type.

\subsubsection{RLHF in Text-to-Music Generation}

The work in \cite{cideron_musicrl_2024} reports the first large-scale RLHF effort for T2M generation, using 300k human preferences to train a reward model and fine-tune MusicLM,
achieving significant improvements by blending human and automatic rewards.
MR-FlowDPO \cite{ziv_mrflowdpo_2025} constructs preference pairs using automatic rewards and evaluates audio quality, relevance and musicality separately; its listening comparisons show that improvements in overall preference do not consistently improve human-rated text relevance (Section~\ref{sec:finding-properties}).

\subsubsection{RLHF in Text-to-Video Generation}

T2V alignment has developed through several approaches. DirecT2V \cite{hong_direct2v_2024} introduced an alignment approach using an instruction-tuned LLM as a “director” to dynamically guide frame-by-frame prompts, enhancing coherence without retraining the diffusion model; it changes the conditioning sequence rather than training on feedback, and is revisited as planning in Section~\ref{sec:planning-comparison}. VADER \cite{prabhudesai_video_2024} directly backpropagates reward gradients through diffusion models to obtain sample-efficient updates. LiFT \cite{wang_lift_2025} uses explanatory annotations to refine reward functions: training its reward model with the rationales raises accuracy on held-out ratings from 87.06\% to 93.55\%, although the aligned generator is evaluated with automatic benchmarks rather than new human judgements. These works illustrate different approaches to T2V alignment, from preference data collection to multimodal and flow-aware optimisation.

\subsubsection{RLHF in Text-to-Human-Motion Generation}

Recent work has explored preference learning for T2HM generation. The study in \cite{sheng_exploring_2024} investigates this approach, comparing RLHF and DPO on MotionGPT. The authors showed that DPO, by bypassing reward model overfitting, more effectively improved motion-text alignment in a data-scarce setting. This supports a setting-specific comparison rather than a general claim that DPO avoids overfitting, and a preferred motion is not necessarily physically executable.

\subsubsection{RLHF in Text-to-3D-Objects Generation}

Several studies have developed methods for aligning T2-3D generation with human preferences.
DreamAlign \cite{liu_dreamalign_2025} introduced Direct 3D Preference optimisation (D-3DPO),
a reward-free method leveraging Low-Rank Adaptation (LoRA) to fine-tune multi-view diffusion models,
guided by a newly created HP3D dataset annotated for aesthetic and consistency preferences.
Building on this work, DreamReward \cite{ye_dreamreward_2025} proposed a full-stack approach,
training Reward3D, the first broad T2-3D reward model, on 25,000 expert comparisons,
and integrating its signals via DreamFL to directly refine generation outputs.
Meanwhile, DreamDPO \cite{zhou_dreamdpo_2025} adapted DPO principles for T2-3D optimisation,
favouring relative over absolute preferences and using large multimodal models
for flexible feedback without strict reward calibration.
These render-based preferences support view-dependent comparisons; they do not certify geometry, editability or downstream utility, and DreamReward's automatic comparison reuses its training reward as a metric alongside a separate user study.

\subsubsection{Cross-Modal Reflections on RLHF}

RLHF methods transfer across modalities: text alignment informs adaptations to other outputs, while challenges such as temporal dynamics in video and multi-view consistency in 3D inform broader alignment methods.
Preference optimisation evolved from supervised pairwise comparisons toward multi-objective,
fine-grained, or proxy-guided systems, tailored to the modality's structure and subjective quality axes.
As each modality demands unique evaluation signals,
yet shares fundamental alignment needs, RLHF increasingly provides a shared framework adapted to domain-specific requirements.
For each method, it is therefore useful to record who or what supplies preferences, what is updated, and how evaluation is kept independent of the training reward.

\subsection{Chain-of-Thought}
\label{sec:planning-comparison}

Inference-time scaling has emerged as a crucial axis for advancing generative models,
complementing traditional scaling of model size and training datasets.
At its core, inference-time scaling enables models to allocate more compute dynamically during generation,
enhancing quality, coherence, and reasoning without altering model weights.
A seminal development in this direction is CoT prompting,
initially introduced for large language models to improve multi-step reasoning by encouraging intermediate steps~\cite{wei_chain--thought_2022}.
Its effectiveness, demonstrated in models such as DeepSeek-R1 \cite{deepseek-ai_deepseek-r1_2025},
have inspired widespread adaptations across modalities.
Unlike SSL, MoE, and RLHF, each of which includes a dedicated subsection for the six generative modalities,
this section adopts a different structure, grouping studies into T2T and multimodal CoT due to the relative scarcity of modality-specific studies.
It surveys recent advances that adapt CoT for inference-time reasoning,
including strategies like self-refinement, verifier-guided sampling,
and dynamic compute allocation,
which collectively improve controllability, compositionality, and cross-modal coherence.

\paragraph{Planning versus Selection} A plan-conditioned pipeline can be expressed conceptually as $p(y\mid c)=\sum_z p_\phi(z\mid c)p_\theta(y\mid z,c)$ for discrete plans $z$, where $\phi$ parameterises the planner and $\theta$ the generator. Implementations commonly use a selected plan rather than evaluating this sum. A correct plan can still be rendered incorrectly, so plan accuracy and final realisation require separate tests. Verifier-guided search instead selects among sampled candidates using a score; its comparison must include the cost of candidate generation and verification, and evaluate outputs with a measure independent of the selection score.

\subsubsection{CoT in Text-to-Text Generation}

Inference-time scaling has rapidly evolved to sharpen the reasoning capabilities of LLMs
without altering their core weights.
Simple Test-Time Scaling \cite{muennighoff_s1_2025}
introduced “wait” tokens and “budget forcing” to actively control reasoning length
and quality during inference.
It also fine-tunes the model on reasoning traces, so its gains are not purely training-free, and returns flatten as the budget grows.
DeepSeek-R1-Zero \cite{deepseek-ai_deepseek-r1_2025} instead obtains long reasoning traces through reinforcement learning without supervised reasoning demonstrations, while the full DeepSeek-R1 pipeline adds cold-start demonstrations and a language-consistency reward; a trained reasoning policy and budget control of a fixed model are therefore different interventions.

\subsubsection{CoT in Multimodal Generation}

Several papers either directly leverage the CoT technique or adapt its stepwise, adaptive approach to scale inference in generative models. Video-T1 \cite{liu_video-t1_2025} extends CoT explicitly into video generation, using multi-stage reasoning or refinement during inference. Additionally, Google DeepMind's research on inference-time scaling \cite{ma_inference-time_2025} extends the CoT analogy into search-based adaptive computation: rather than sampling without guidance, it actively searches over initial noise vectors and diffusion paths, guided by verifiers, much like CoT, which explores reasoning paths via intermediate evaluations and allocates extra compute for better outcomes.

\paragraph{Images} In the text-to-image analysis of Ma et al.~\cite{ma_inference-time_2025}, aesthetic and CLIP-based verifier scores can conflict under search, so equal sampling budgets and independent evaluators are needed to decide whether search improves instruction satisfaction rather than only its verifier score. T2I-R1 \cite{jiang_t2i-r1_2025} instead trains a unified autoregressive model, Janus-Pro, to write a semantic plan of the prompt before generating image tokens, optimising both with GRPO and an ensemble of reward models. In its ablation, optimising the image tokens alone lowers output diversity (Vendi score 6.255 versus 6.976 for the base model), whereas jointly optimising the semantic plan and image tokens raises it to 8.203 and improves the WISE cultural-knowledge score from 0.35 to 0.56.

\paragraph{Video} DirecT2V \cite{hong_direct2v_2024} uses an LLM to expand a prompt into frame-level descriptions, with attention modifications supporting temporal consistency; a plausible plan need not be realised consistently by the renderer. Video-T1 treats inference scaling as search: random linear search samples more initial noises, whereas Tree-of-Frames expands and prunes partially generated clips, both ranked by learned video verifiers. VBench scores rise with the search budget for all six tested generators, and Tree-of-Frames reaches comparable scores with about one third of the computation of linear search in the reported configurations.

\paragraph{Music} StepAudio 3 Music can produce ABC notation before music tokens~\cite{feng_stepaudio3_2026}, supplying a symbolic intermediate plan. MusiCoT \cite{lam_analyzable_2025} instead plans in a latent space: the language model first predicts quantised CLAP embeddings for successive ten-second segments and then generates audio tokens. Adding this plan to a MeLoDy-style generator raises the mean opinion score of ten music professionals from 3.35 to 3.72 on the same 100 lyric--prompt pairs; the listeners rated overall quality rather than text relevance.

\paragraph{Motion} Motion-R1 \cite{ouyang_motion-r1_2025} fine-tunes a language model on automatically generated, step-by-step action plans and then applies GRPO with format, semantic-similarity and motion-similarity rewards before emitting motion tokens. On HumanML3D, adding the plans to reward training raises top-3 R-precision from 0.644 to 0.818 and lowers FID from 1.656 to 0.201, although both similarity rewards use the pretrained encoders that also compute these metrics.

\paragraph{3D Objects} HCoG \cite{qin_hcog_2025} uses a language model to decompose a long object description into part blocks ordered from inside to outside by occlusion, then generates the blocks in sequence by extending 3D Gaussians. Its ablation isolates the ordering of the plan: render-based BLIP-VQA scores fall from 0.7295 to 0.5961 with a random order and to 0.5023 with the inverse order.

\subsubsection{Cross-Modal Reflections on CoT}

The adaptation of CoT principles across modalities reflects increasing use of structured, stepwise reasoning and dynamic allocation of computation during inference. In T2I and T2V, CoT-inspired techniques introduce refinement loops, structured sampling, and modular reasoning, improving compositionality, physical plausibility, and narrative coherence. The modular, verifiable, and stepwise structure of CoT offers a shared approach to improving control and efficiency across generative modalities.
Evaluations should nonetheless measure final task success as well as intermediate-plan quality and inference cost, and distinguish a fixed generator with extra search from a model trained to reason.

\subsection{Synergistic Combinations of Techniques}\label{sec:technique-map}

Each technique contributes a distinct strength: data efficiency, modular scalability, alignment, or structured reasoning. Combining them offers further opportunities. Integrating SSL, MoE, RLHF, and CoT can support models that are not only more coherent and controllable but also more capable of generalising across modalities and tasks. This convergence allows MGMs to reason with greater depth, plan with more structure, and respond with finer granularity, supporting more interpretable and adaptive multimodal systems. Figure~\ref{fig:technique_pipeline} illustrates how the four techniques can be combined within a single model, from raw multimodal data to aligned, reasoned output. Notably, the development and adoption of each technique vary considerably across modalities, reflecting differences in data availability, architectural constraints, and task-specific challenges, as summarised in Table~\ref{tab:technique_trends}.
The techniques are not prerequisites for one another, and evidence for a combined pipeline must not be attributed to each component independently.

\begin{figure}[H]
    \centering
\hyphenpenalty=10000\exhyphenpenalty=10000
    \begin{tikzpicture}[
                box/.style={draw, rounded corners, align=center, minimum height=1.3cm, minimum width=2.0cm, font=\fontfamily{phv}\selectfont\fontsize{9}{11}\selectfont},
                lbl/.style={font=\fontfamily{phv}\selectfont\fontsize{8.5}{10}\selectfont, align=center, text width=2.0cm},
                arr/.style={-{Latex[length=2mm]}, thick}
            ]
            \node[box, fill=blue!8]                  (data) {Multimodal\\data};
            \node[box, fill=blue!16, right=0.6cm of data] (ssl)  {SSL\\pretraining};
            \node[box, fill=blue!24, right=0.6cm of ssl]  (moe)  {MoE\\backbone};
            \node[box, fill=blue!32, right=0.6cm of moe]  (rlhf) {RLHF\\alignment};
            \node[box, fill=blue!40, right=0.6cm of rlhf] (cot)  {CoT\\inference};
            \node[box, fill=blue!8,  right=0.6cm of cot]  (out)  {Generated\\output};

            \foreach \a/\b in {data/ssl, ssl/moe, moe/rlhf, rlhf/cot, cot/out}
            \draw[arr] (\a) -- (\b);

            \node[lbl, below=0.3cm of ssl]  {learns from\\unlabelled data};
            \node[lbl, below=0.3cm of moe]  {routes computation\\across experts};
            \node[lbl, below=0.3cm of rlhf] {aligns outputs\\with human\\preferences};
            \node[lbl, below=0.3cm of cot]  {structures reasoning\\at inference};
    \end{tikzpicture}
    \caption{Conceptual pipeline combining self-supervised learning (SSL), mixture of experts (MoE), reinforcement learning from human feedback (RLHF), and chain-of-thought (CoT) prompting within a generic multimodal generative model. The order is illustrative; individual systems use any subset of the four techniques.}\label{fig:technique_pipeline}
\end{figure}

\begin{table}[H]
    \caption{Trends across techniques and modalities. Labels summarise the qualitative assessment of the preceding subsections; the cited studies are representative examples, not performance rankings.}\label{tab:technique_trends}
    \centering
    \footnotesize
    \setlength{\tabcolsep}{3pt}
    \begin{tabular}{>{\raggedright\arraybackslash}p{0.12\textwidth}>{\raggedright\arraybackslash}p{0.2\textwidth}>{\raggedright\arraybackslash}p{0.2\textwidth}>{\raggedright\arraybackslash}p{0.2\textwidth}>{\raggedright\arraybackslash}p{0.2\textwidth}}
        \toprule
        \textbf{Modality}    & \textbf{SSL}            & \textbf{MoE}           & \textbf{RLHF}             & \textbf{CoT}              \\
        \midrule
        T2T   & Core, mature: next-token pretraining~\cite{radford_improving_2018} & Advanced, efficient: GLaM, Mixtral~\cite{du_glam_2022,jiang_mixtral_2024} & Pioneered, refined: InstructGPT~\cite{ouyang_training_2022}; verifiable rewards~\cite{deepseek-ai_deepseek-r1_2025} & Rich, widely used: s1~\cite{muennighoff_s1_2025}; DeepSeek-R1~\cite{deepseek-ai_deepseek-r1_2025} \\
        T2I   & Helpful, limited: REPA~\cite{yu_repa_2025} & Specialised, improving: ERNIE-ViLG 2.0, EC-DiT, DiffMoE~\cite{feng_ernie-vilg_2023,sun_ec-dit_2025,shi_diffmoe_2025} & Active, reward-rich: ImageReward, Diffusion-DPO~\cite{xu_imagereward_2023,wallace_diffusion_2024} & Emerging, refinement: noise search~\cite{ma_inference-time_2025}; T2I-R1~\cite{jiang_t2i-r1_2025} \\
        T2M   & Crucial, effective: MusicLM, Mo\^usai~\cite{agostinelli_musiclm_2023,schneider_mousai_2024} & Rare, routing issues: StepAudio 3 Music~\cite{feng_stepaudio3_2026} & Early, proxy rewards: MusicRL, MR-FlowDPO~\cite{cideron_musicrl_2024,ziv_mrflowdpo_2025} & Sparse, sequentially hard: MusiCoT~\cite{lam_analyzable_2025} \\
        T2V   & Essential for motion: Make-A-Video, VideoREPA~\cite{singer_make--video_2022,zhang_videorepa_2025} & Sparse, task-specific: Wan2.2~\cite{wan_wan22_2025} & Rapid, multimodal rewards: VADER, LiFT~\cite{prabhudesai_video_2024,wang_lift_2025} & Temporal, verifier-based: DirecT2V, Video-T1~\cite{hong_direct2v_2024,liu_video-t1_2025} \\
        T2HM  & Emerging, motion priors: LMM, OMG~\cite{zhang_large_2025,liang_omg_2024} & Growing, control-based: OMG, GenM3~\cite{liang_omg_2024,shi_genm3_2025} & New, preference-driven: Sheng et al.~\cite{sheng_exploring_2024} & Sparse, underexplored: Motion-R1~\cite{ouyang_motion-r1_2025} \\
        T2-3D & Minimal, data-limited   & Rare, few large models: TripoSG (image-conditioned)~\cite{li_triposg_2025} & Recent, viewer-aligned: DreamDPO, DreamReward, DreamAlign~\cite{zhou_dreamdpo_2025,ye_dreamreward_2025,liu_dreamalign_2025} & Nascent, inference search: HCoG~\cite{qin_hcog_2025} \\
        \bottomrule
    \end{tabular}
    \par\smallskip
    \begin{minipage}{\textwidth}\footnotesize
        T2T: text-to-text; T2I: text-to-image; T2M: text-to-music; T2V: text-to-video; T2HM: text-to-human-motion; T2-3D: text-to-3D objects; SSL: self-supervised learning; MoE: mixture of experts; RLHF: reinforcement learning from human feedback; CoT: chain-of-thought; LMM: Large Motion Model.
    \end{minipage}
\end{table}

\section{Evaluation}\label{sec:evaluation}

Having outlined the six text-to-X modalities and four techniques (SSL, MoE, RLHF, and CoT), we organise evaluation around three aspects: \emph{faithfulness} (factual and semantic alignment to the prompt), \emph{compositionality} (the correct binding of multiple concepts, attributes, and relations), and \emph{robustness} (the capacity to maintain human-perceived quality while producing stable, predictable outputs under perturbations of prompt wording, noise, and distribution shift). These axes capture the most recurrent and practically salient failure modes across modalities: hallucinations and misalignments, entangled or incomplete compositions, and brittle behaviour despite high apparent fidelity. Accordingly, we focus on families of widely adopted, conceptually aligned metrics (e.g., contrastive or retrieval-style scores for faithfulness, structured compositional benchmarks for multi-concept binding, and distributional distances, reconstruction metrics, and human or LLM-as-a-judge preferences for quality and robustness), treating them as informative but imperfect proxies that enable cross-modal analysis in the subsequent sections. Because these metrics use different datasets, evaluators and conditioning, we interpret each result within its own study protocol rather than as a point on a longitudinal progress scale.

\subsection{Definitions and Comparison Rules}
Faithfulness asks whether an output respects the prompt, reference information and explicit constraints. Compositionality asks whether it realises combinations of entities, attributes and relations, including held-out combinations when the protocol tests generalisation. Robustness asks whether a specified property persists under a defined perturbation or distribution shift. The three criteria therefore pose increasingly demanding questions: whether requested content is realised at all, whether several requested elements are realised together, and whether either property persists after a specified change. They are used because each applies to all six task families, while each must be operationalized through task-specific tests. Realism, preference and physical validity are related outcomes, but do not substitute for these tests, and for fictional generation adherence to a prompt differs from factual truth about the world.

The three evaluation axes provide a common organising framework, but their operationalisation must remain task-specific; this survey does not establish that they exhaust the relevant evaluation dimensions.

A reported numerical contrast should identify model variant, dataset and split, available conditioning, metric implementation, sample count, uncertainty and comparator. Each row of Tables~\ref{tab:t2t-faithfulness}--\ref{tab:t2-3d-faithfulness} therefore reports a comparison made within a single study. The protocol notes beneath each table list the conditions stated in the cited version, and NE (not established) marks details not recovered from the cited protocol during this review; it does not assert that they are absent from the complete paper or supplement. Values are reproduced as published; point estimates alone do not indicate statistical significance, and results should not be pooled into a common scale. Resource comparisons likewise require measured latency or compute under specified hardware and output settings, and comparisons that add control inputs must be distinguished from text-only comparisons.

\subsection{Faithfulness}

Faithfulness evaluates how well generated content adheres to a prompt’s intent, facts, and semantics, limiting distortions (e.g., hallucinations, compositional or temporal errors) \cite{malin_review_2025}. Across the six modalities, the selected studies use surface-overlap scores, learned-similarity scores and human or model preference judgements, and automated scores are reported to correlate weakly with human perception. Two recurring patterns are clear: metrics built for retrieval often fail to capture generative faithfulness, and key objectives trade off. For example, improving fidelity can reduce controllability, and accelerating inference can undermine temporal coherence.
These approaches now coexist rather than forming a common progression, and reported scores use different datasets, evaluators and conditioning. The tables below therefore present a within-study comparison and its protocol for each study. Scores should not be compared across rows.

\subsubsection{Metrics and What They Establish}
Five families of measures recur in the selected studies. Reference-overlap scores, such as bilingual evaluation understudy (BLEU), compare an output with reference outputs; they are indirect evidence of adherence when many different outputs could be valid. Learned-embedding similarity and retrieval, such as CLIP- or CLAP-based scores and R-precision, estimate global semantic match; they can miss attributes, relations and temporal order, and they lose independence when the evaluator's representation also shaped training.

Structured checks, such as detectors, segmenters, symbolic parsers or execution tests, verify specific constraints, but only those the checker can recognise. Human or model judgements target adherence most directly, yet depend on the question asked, the comparator, rater instructions and the treatment of ties. Distributional measures, including FID, inception score (IS), Fr\'echet audio distance (FAD) and Fr\'echet video distance (FVD), describe the quality or diversity of an output distribution rather than adherence to individual prompts. A faithfulness claim is therefore only as specific as the measure and protocol that support it.

\subsubsection{Modality-Specific Discussions}

\paragraph{T2T} Faithfulness in text generation covers preserving semantics, ensuring factual accuracy and giving behaviourally safe responses. See Table~\ref{tab:t2t-faithfulness}, whose results are reported under their original evaluation protocols.

\begin{table}[H]
    \centering
    \caption{Representative approaches to evaluating faithfulness in text-to-text (T2T) generation.}\label{tab:t2t-faithfulness}
    \footnotesize
    \setlength{\tabcolsep}{3pt}
    \begin{tabular}{>{\raggedright\arraybackslash}p{0.14\textwidth}>{\raggedright\arraybackslash}p{0.19\textwidth}>{\raggedright\arraybackslash}p{0.16\textwidth}>{\raggedright\arraybackslash}p{0.23\textwidth}>{\raggedright\arraybackslash}p{0.19\textwidth}}
        \toprule
        \textbf{Model/study} & \textbf{Task and conditioning} & \textbf{Evaluation target} & \textbf{Metric/protocol and result versus comparator} & \textbf{Interpretation and limitation} \\
        \midrule
        \multicolumn{5}{l}{\textbf{Translation: supporting evidence}} \\
        Attention NMT (2015)~\cite{bahdanau_neural_2015} [S1] & English$\to$French; source sentence; WMT14 news-test-2014, full set & Reference overlap as a proxy for translation adequacy & RNNsearch-50: 26.75 BLEU versus RNNencdec-50: 17.82; source Table 1 & Same-study model comparison; neither factual-response evidence nor an attention-only ablation \\
        Transformer (2017)~\cite{vaswani_attention_2017} [S2] & English$\to$German; source sentence; WMT14 newstest2014 & Reference overlap as a proxy for translation adequacy & Big: 28.4 BLEU versus base: 27.3; source Table 2 & Capacity and training differ; not an isolated component comparison \\
        \midrule
        \multicolumn{5}{l}{\textbf{Grounding and factual response: direct target, evaluator-dependent}} \\
        InstructGPT (2022)~\cite{ouyang_training_2022} [S3] & Closed-domain API instructions with supplied source information & Human-judged unsupported additions & Reported hallucination rates: 21\% versus GPT-3's 41\%; size-pooled evaluation, source Figure 4 & Source-grounding result; neither a universal rate nor a 175B-only comparison \\
        Llama 2 / Llama 2-Chat (2023)~\cite{touvron_llama_2023-1} [S4] & 70B variants; TruthfulQA generation; six-shot QA prompt & Joint truthfulness and informativeness, judged automatically & Chat: 64.14\% versus base: 50.18\%; source Tables 11, 14 and 44 & Joint outcome, not truthfulness alone; full post-training comparison, not an RLHF-only ablation \\
        \bottomrule
    \end{tabular}
\end{table}

{\footnotesize\noindent\textit{Protocol notes.}
\par\nopagebreak[4]\noindent \textbf{S1:} 3,003 test sentences; Moses tokenisation without lowercasing or stemming; 30,000-word shortlists. NE: BLEU implementation, uncertainty.
\par\noindent \textbf{S2:} Beam size 4 and length penalty 0.6; checkpoint averaging over five checkpoints for base and twenty for big. NE: test-sentence count, BLEU implementation, confidence intervals.
\par\noindent \textbf{S3:} Evaluation prompts come from held-out customers. InstructGPT normally denotes PPO-ptx, which mixes pretraining gradients into proximal policy optimisation (PPO); Figure 4 pools model sizes, and the headline percentages are rounded. The plotted error bars are 95\% confidence intervals. NE: closed-domain subset size, numerical interval endpoints.
\par\noindent \textbf{S4:} TruthfulQA~\cite{lin_truthfulqa_2022} contains 817 questions across 38 categories; six random QA examples condition generation, and fine-tuned GPT-3 judges assess truthfulness and informativeness (source Appendix A.4.7). NE: per-model output count, benchmark revision, decoding settings, confidence intervals.
\par\noindent \textbf{Abbreviations and formatting:} NE, not established from the cited protocol in this review; NMT, neural machine translation; BLEU, bilingual evaluation understudy; PPO, proximal policy optimisation; QA, question answering. Bold rows group studies by evaluation target.\par}

Translation overlap does not establish factuality; avoiding unsupported additions to supplied material differs from answering factual questions. Post-training comparisons also combine interventions, so their gains cannot be assigned solely to preference optimisation.

\paragraph{T2I} Faithfulness in text-to-image generation has been assessed through word-region matching, alignment of entities and attributes and human preferences. See Table~\ref{tab:t2i-faithfulness}, whose results are reported under their original evaluation protocols.

\begin{table}[H]
    \centering
    \caption{Representative approaches to evaluating faithfulness in text-to-image (T2I) generation.}\label{tab:t2i-faithfulness}
    \footnotesize
    \setlength{\tabcolsep}{3pt}
    \begin{tabular}{>{\raggedright\arraybackslash}p{0.14\textwidth}>{\raggedright\arraybackslash}p{0.19\textwidth}>{\raggedright\arraybackslash}p{0.16\textwidth}>{\raggedright\arraybackslash}p{0.23\textwidth}>{\raggedright\arraybackslash}p{0.19\textwidth}}
        \toprule
        \textbf{Model/study} & \textbf{Task and conditioning} & \textbf{Evaluation target} & \textbf{Metric/protocol and result versus comparator} & \textbf{Interpretation and limitation} \\
        \midrule
        AttnGAN (2018)~\cite{xu_attngan_2018} [I1] & CUB test descriptions; text-conditioned images & Caption matching: learned retrieval proxy & DAMSM weight 5: AttnGAN2 $67.82\pm4.43\%$ versus AttnGAN1 $58.65\pm5.41\%$ R-precision; source Table 2 & Architecture comparison; evaluator shares the training alignment representation \\
        DALL-E (2021)~\cite{ramesh_zero-shot_2021} [I2] & MS-COCO captions; contrastively reranked samples & Direct human judgement of caption matching & 93.3\% majority preference against DF-GAN; source Figure 7 & Relative preference with candidate selection, not absolute adherence accuracy \\
        DALL-E 2 / unCLIP (2022)~\cite{ramesh_hierarchical_2022} [I3] & MS-COCO validation captions; diffusion-prior unCLIP versus GLIDE & Direct human judgement of caption matching & unCLIP preferred $45.3\%\pm3.0\%$; reported 95\% interval, source Table 1 & Does not show superior caption matching; diversity and photorealism are separate outcomes \\
        ControlNet (2023)~\cite{zhang_adding_2023} [I4] & ADE20K segmentation-conditioned generation; additional spatial input & Supplied-segmentation adherence through a segmenter & Reconstructed IoU: $0.35\pm0.14$ versus ControlNet-lite $0.32\pm0.12$; source Table 2 & Control fidelity, not caption faithfulness; limited by the segmentation evaluator \\
        DALL-E 3 (2023)~\cite{betker_improving_2023} [I5] & Short original MS-COCO 2014 evaluation captions & Caption similarity: learned embedding proxy & CLIP score 32.0 versus DALL-E 2 31.4 and SDXL 30.5; source Table 1 & Full-system comparison; does not isolate recaptioning or check every requested relation \\
        \bottomrule
    \end{tabular}
\end{table}

{\footnotesize\noindent\textit{Protocol notes.}
\par\nopagebreak[4]\noindent \textbf{I1:} Each model generates 30,000 images from unseen descriptions; pretrained DAMSM features rank one matching caption among 99 mismatches. AttnGAN1/2 differ in attention modules and generator stages. NE: meaning of $\pm$, number of independent runs.
\par\noindent \textbf{I2:} 1,000 caption comparisons, five distinct workers per comparison, with one worker excluded. Default sampling reranks 512 candidates at temperature 1. NE: rating count after exclusion, uncertainty.
\par\noindent \textbf{I3:} The protocol uses 1,000 MS-COCO validation captions and permits ``Not sure''; the opposite preference cannot be inferred by subtraction from 100\%. Sampling settings were selected through a photorealism proxy, not matched compute. NE: unique rater count, caption-judgement count.
\par\noindent \textbf{I4:} OneFormer reconstructs the supplied segmentations; real-image IoU is $0.58\pm0.10$. NE: evaluated sample count, prompt policy, segmenter checkpoint, meaning of $\pm$.
\par\noindent \textbf{I5:} CLIP ViT-B/32 evaluates 4,096 captions; scores are mean cosine similarity multiplied by 100, not correctness percentages. Comparators are the DALL-E 2 production version live on 20 September 2023 and SDXL 1.0 with refiner; the source acknowledges possible MS-COCO training overlap. NE: confidence intervals, evaluation-code revision, per-caption sampling details.
\par\noindent \textbf{Abbreviations:} CUB, Caltech-UCSD Birds; DAMSM, deep attentional multimodal similarity model; MS-COCO, Microsoft Common Objects in Context; IoU, intersection over union; CLIP, contrastive language--image pretraining; ViT-B/32, base vision transformer with 32-pixel patches.\par}

These comparisons distinguish learned similarity from direct human judgements and explicit spatial-condition checks. Reusing a training representation for evaluation limits independence. Candidate reranking changes inference cost, while segmentation conditioning changes the task itself.

\paragraph{T2M} Faithfulness in music generation concerns the semantic fit between text and audio, including structure, timbre, and form, assessed through perceptual realism, embedding-space alignment (e.g., MuLan/CLAP \cite{elizalde_clap_2023}) and standardised human preference. See Table~\ref{tab:t2m-faithfulness}, whose results are reported under their original protocols.

\begin{table}[H]
    \centering
    \caption{Representative approaches to evaluating faithfulness in text-to-music (T2M) generation.}\label{tab:t2m-faithfulness}
    \footnotesize
    \setlength{\tabcolsep}{3pt}
    \begin{tabular}{>{\raggedright\arraybackslash}p{0.14\textwidth}>{\raggedright\arraybackslash}p{0.19\textwidth}>{\raggedright\arraybackslash}p{0.16\textwidth}>{\raggedright\arraybackslash}p{0.23\textwidth}>{\raggedright\arraybackslash}p{0.19\textwidth}}
        \toprule
        \textbf{Model/study} & \textbf{Task and conditioning} & \textbf{Evaluation target} & \textbf{Metric/protocol and result versus comparator} & \textbf{Interpretation and limitation} \\
        \midrule
        MusicLM (2023)~\cite{agostinelli_musiclm_2023} [M1] & MusicCaps descriptions; hierarchical audio-token generation & Audio--text alignment proxy & MCC 0.51 versus 0.49 without the semantic modeling stage; source Section 5 & Semantic-stage comparison under this evaluator; MuLan also supplies conditioning \\
        Noise2Music (2023)~\cite{huang_noise2music_2023} [M2] & MusicCaps captions; waveform versus spectrogram diffusion & Audio--text alignment proxy & MuLan audio--text similarity 0.478 versus 0.434; source Table 6 & System-variant comparison; MuLan also contributes training pseudo-labels \\
        MuseCoco (2023)~\cite{lu_musecoco_2023} [M3] & Attribute-to-music stage of a text-to-symbolic-music pipeline; attributes supplied directly & Requested-attribute adherence: rule-based extraction and listener judgements & Prefix control: 67.40\% average attribute control accuracy versus conditional LayerNorm 47.46\% and embedding 36.94\%; source Table 7 & Same-study control-method comparison; bypasses text interpretation, so the full text-to-music pipeline is not tested \\
        MusicGen (2023)~\cite{copet_simple_2023} [M4] & Text conditioning; 300M codebook-pattern ablation on held-out music & Audio--text alignment proxy & Delay: CLAP 0.35, 1,500 steps; flattening: 0.37, 6,000 steps; published Table 4 & Within-study score/sequence-length comparison; steps are not measured wall-clock speedups \\
        Mo\^usai (2024)~\cite{schneider_mousai_2024} [M5] & Text-conditioned audio; study-specific prompts & Audio--text alignment proxy & CLAP 0.13 versus Riffusion 0.06; source Table 4 & Full-system comparison; evaluator checkpoint and scoring set not established \\
        FluxMusic (2024)~\cite{fei_flux_2024} [M6] & Small-model training comparison; held-out MTT prompts & Audio--text alignment proxy & Rectified flow: CLAP 0.312 versus DDIM 0.201; source Table 2 & Evidence for this training comparison, not a universal architecture ranking \\
        \bottomrule
    \end{tabular}
\end{table}

{\footnotesize\noindent\textit{Protocol notes.}
\par\nopagebreak[4]\noindent \textbf{M1:} MCC (MuLan cycle consistency) is the mean cosine similarity between text and generated-music embeddings in MuLan space. MusicCaps contains about 5.5K clips and a balanced 1K subset. NE: ablation sample count and subset, evaluator checkpoint, uncertainty.
\par\noindent \textbf{M2:} Evaluation uses 16 kHz waveforms without the super-resolution cascader and the middle ten seconds of generated MusicCaps clips (about 5.5K captioned clips). NE: evaluated sample count, metric checkpoint, uncertainty.
\par\noindent \textbf{M3:} All three control methods use reduced-size models (a six-layer Linear Transformer). Objective attributes, such as tempo and meter, are checked by rule-based extraction from the generated music; subjective attributes, such as emotion and genre, are judged by 12 listeners who each rate 18 pieces, six per method. Musicality (1--5) is $3.15\pm1.02$, $3.11\pm1.02$ and $2.97\pm0.91$, respectively. NE: sample count for objective attributes, uncertainty of control accuracy, meaning of $\pm$.
\par\noindent \textbf{M4:} The published NeurIPS codebook comparison is Table 4. Ablations use 1,000 thirty-second samples from 528 held-out tracks without artist overlap with training, scored with pretrained LAION CLAP. NE: CLAP checkpoint, uncertainty.
\par\noindent \textbf{M5:} The ACL version reports cosine similarity from pretrained LAION CLAP. MusicGen's separately retrained Mo\^usai is a different evaluated system. NE: prompt-set size and split, evaluator checkpoint, generation settings of the scored samples, uncertainty.
\par\noindent \textbf{M6:} The small variant is trained on ASM/FMA and evaluated on 1,000 randomly selected out-of-domain MTT examples. Both alternatives use batch size 128 and 200K training steps. These results do not describe the 2.1B MusicCaps model. NE: CLAP checkpoint, sampling uncertainty, number of independent runs.
\par\noindent \textbf{Abbreviations:} MCC, MuLan cycle consistency; CLAP, contrastive language--audio pretraining; MTT, MagnaTagATune; ASM, AudioSet music subset; FMA, Free Music Archive; DDIM, denoising diffusion implicit model; LAION, Large-scale Artificial Intelligence Open Network.\par}

Symbolic attribute checks make selected conditions explicit but cover only the attributes represented by the evaluator. Learned audio--text similarity supports scalable comparison while depending on the embedding model; reuse of that representation during training limits evaluation independence. Listener judgements should distinguish relevance from musical quality. Within-study ablations can examine specific design choices, but differences in prompts, sampling and evaluators prevent inferring a MusicLM--MusicGen quality--alignment trade-off from their separate scores. MR-FlowDPO separately evaluates listening quality and relevance~\cite{ziv_mrflowdpo_2025}, providing a within-study basis for distinguishing these outcomes.

\paragraph{T2V} Faithfulness in video generation covers factual grounding and policy adherence, as well as temporal coherence and physically plausible world simulation. See Table~\ref{tab:t2v-faithfulness}, whose results retain their original protocols.

\begin{table}[H]
    \centering
    \caption{Representative approaches to evaluating faithfulness in text-to-video (T2V) generation.}\label{tab:t2v-faithfulness}
    \footnotesize
    \setlength{\tabcolsep}{3pt}
    \begin{tabular}{>{\raggedright\arraybackslash}p{0.14\textwidth}>{\raggedright\arraybackslash}p{0.19\textwidth}>{\raggedright\arraybackslash}p{0.16\textwidth}>{\raggedright\arraybackslash}p{0.23\textwidth}>{\raggedright\arraybackslash}p{0.19\textwidth}}
        \toprule
        \textbf{Model/study} & \textbf{Task and conditioning} & \textbf{Evaluation target} & \textbf{Metric/protocol and result versus comparator} & \textbf{Interpretation and limitation} \\
        \midrule
        TiVGAN (2020)~\cite{kim_tivgan_2020} [V1] & Text conditioning; six-class Kinetics subset; 16 frames & Classifier-based quality/diversity: supporting evidence & Full schedule: IS 5.55 versus 5.13 without initial image-generation stage; source Table 5 & Training-schedule comparison; IS does not check each video's prompt match \\
        CogVideo (2022)~\cite{hong_cogvideo_2022} [V2] & Class names from 30 UCF101 action classes; hierarchical versus one-stage generation & Direct human semantic-relevance ratings & Semantic relevance (1--5): 2.89 versus 2.57 for one-stage sequential generation; source Figure 5 & Same-study design comparison; short class-name prompts test limited semantics \\
        Make-A-Video (2022)~\cite{singer_make--video_2022} [V3] & DrawBench, 200 prompts; English-input CogVideo comparator & Direct human text correspondence & Reported faithfulness preference: 68.75\%; source Table 3 & Full-system comparison; percentages are described per rater although the protocol specifies majority votes \\
        VideoLDM (2023)~\cite{blattmann_align_2023} [V4] & MSR-VTT test captions; SD 2.1 versus SD 1.4 variants & Frame--text embedding proxy & CLIPSIM 0.2929 versus 0.2848; source Table 10 & System variants; frame averaging does not establish action ordering or temporal adherence \\
        Emu Video (2023 preprint; ECCV 2024)~\cite{girdhar_factorizing_2025} [V5] & Text-to-video through an internally generated image; 307 prompts & Direct human spatial/temporal text faithfulness & Factorised versus direct generation: 63.3\% preference; source Table 1a & Matched design comparison; intermediate image is not an additional user-supplied condition \\
        Movie Gen (2024 preprint)~\cite{polyak_movie_2025} [V6] & Movie Gen Video Bench; Runway Gen3 comparator & Direct human text alignment & +10.45 percentage points net win rate; source Table 6 & Win\% minus loss\%, allowing ties; neither win probability nor absolute adherence \\
        \bottomrule
    \end{tabular}
\end{table}

{\footnotesize\noindent\textit{Protocol notes.}
\par\nopagebreak[4]\noindent \textbf{V1:} The ablation holds total training time constant. Kinetics experiments use six classes, 3,032 training videos and a five-layer 3D classifier. NE: evaluation sample count, split, classifier checkpoint, uncertainty.
\par\noindent \textbf{V2:} arXiv v1. Ninety evaluators completed the study, and 30 randomly selected UCF101 class names serve as text. Each video is rated 1--5 for frame texture, motion realism and semantic relevance, and 1--10 for overall quality. The one-stage variant generates all frames sequentially with the first-stage model. Values are the bar labels in Figure 5; the source describes its error bars as standard deviations. NE: videos per condition.
\par\noindent \textbf{V3:} arXiv v1. The human evaluation generates 76-frame, $256\times256$ videos and specifies majority vote from five annotators, judging correspondence separately from quality. Because Table 3 describes percentages of raters, no integer win count or rating denominator is derived. NE: uncertainty, comparator checkpoint, comparison count after filtering.
\par\noindent \textbf{V4:} Appendix G specifies 2,990 MSR-VTT test videos, one randomly selected caption per example, 16 frames at 30 fps and CLIP ViT-B/32 averaged over 47,840 frames. NE: uncertainty, evaluation-code revision.
\par\noindent \textbf{V5:} The ECCV 2024 paper's ablation uses eight-frame generation, 307 prompts and majority vote from five evaluators per comparison; JUICE asks raters to justify choices through spatial and temporal alignment. The pretrained image model, training data, iterations and trainable parameter count are held constant. NE: uncertainty, retained rating count.
\par\noindent \textbf{V6:} arXiv v1, with Runway Gen3 videos collected on 23 September 2024. The benchmark contains 1,003 prompts; text alignment uses majority vote from three annotations, with ties allowed. The 3.74-percentage-point annotation-variance estimate comes from four repetitions on 381 prompts, not a comparator-specific confidence interval. Generation includes prompt rewriting, and videos are spatially matched for comparison. NE: retained rating count.
\par\noindent \textbf{Abbreviations:} IS, inception score; FVD, Fr\'echet video distance; MSR-VTT, Microsoft Research Video to Text; SD, Stable Diffusion; CLIPSIM, mean CLIP similarity between generated frames and text; CLIP, contrastive language--image pretraining.\par}

The comparisons distinguish visual recognisability, frame-level semantic similarity and direct judgements of prompt correspondence. Classifier scores cannot establish whether requested actions occur in the right order, and averaging frame--text similarities does not test temporal relations. Human comparisons target adherence more directly, but depend on the prompt set, comparator, sample selection and voting protocol. VideoREPA distinguishes semantic adherence from physical commonsense~\cite{zhang_videorepa_2025}; neither preference nor reconstruction quality certifies physical understanding.

\paragraph{T2HM} Faithfulness in T2HM generation concerns whether generated motions closely align with textual descriptions in semantics and details, as assessed using metrics such as R-precision, FID, average positional error (APE), and human judgements of alignment. See Table~\ref{tab:t2hm-faithfulness}, whose results retain their original protocols.

\begin{table}[H]
    \centering
    \caption{Representative approaches to evaluating faithfulness in text-to-human-motion (T2HM) generation.}\label{tab:t2hm-faithfulness}
    \footnotesize
    \setlength{\tabcolsep}{3pt}
    \begin{tabular}{>{\raggedright\arraybackslash}p{0.14\textwidth}>{\raggedright\arraybackslash}p{0.19\textwidth}>{\raggedright\arraybackslash}p{0.16\textwidth}>{\raggedright\arraybackslash}p{0.23\textwidth}>{\raggedright\arraybackslash}p{0.19\textwidth}}
        \toprule
        \textbf{Model/study} & \textbf{Task and conditioning} & \textbf{Evaluation target} & \textbf{Metric/protocol and result versus comparator} & \textbf{Interpretation and limitation} \\
        \midrule
        TEMOS (2022)~\cite{petrovich_temos_2022} [H1] & KIT-ML test descriptions and duration; Lin et al. also uses initial pose & Direct human text correspondence & 90.5\% semantic preference over Lin et al.; source Section 4.2/Figure 3 & Unequal conditioning; realism can influence semantic preference \\
        TM2T (2022)~\cite{guo_tm2t_2022} [H2] & HumanML3D test descriptions; generated motion & Learned text--motion retrieval proxy & Top-3 R-precision $0.729\pm0.002$ versus $0.635\pm0.002$ without inverse alignment; source Table 2 & GRU baseline comparison; not a direct check of each requested body-part action \\
        MotionDiffuse (2022 preprint; 2024 journal)~\cite{zhang_motiondiffuse_2024} [H3] & KIT-ML test descriptions with reference motion length & Learned text--motion retrieval proxy & With CLIP present: efficient attention $0.739\pm0.004$ versus conventional attention $0.679\pm0.005$, top-3; source Table 3 & Conditional attention ablation; not a universal benefit of diffusion \\
        T2M-GPT (2023)~\cite{zhang_generating_2023} [H4] & HumanML3D test descriptions; predicted termination & Learned text--motion retrieval proxy & Training corruption $\tau=0.5$: top-3 $0.775\pm0.002$ versus $\tau=0$: $0.685\pm0.003$; source Table 1 & Training ablation; $\tau$ is token-corruption probability, not sampling temperature \\
        Fg-T2M (2023)~\cite{wang_fg-t2m_2023} [H5] & Harder-HumanML3D subset selected for fine-grained wording & Learned text--motion retrieval proxy & Top-3 $0.763\pm0.004$ versus MotionDiffuse $0.725\pm0.005$; source Table 3 & Subset-specific system comparison; retrieval does not certify left/right execution \\
        MotionGPT, Zhang et al. (AAAI 2024)~\cite{zhang_motiongpt_2024} [H6] & HumanML3D validation; 7B model evaluated with text only & Learned text--motion retrieval proxy & Joint multitask training: top-3 0.657 versus separate text-task training 0.577; source Table 4 & Same evaluation condition, different training tasks; not human preference or an isolated LLM benefit \\
        Light-T2M (AAAI 2025)~\cite{zeng_light-t2m_2025} [H7] & HumanML3D; adaptive textual-information injector ablation & Learned text--motion retrieval proxy & With injector: top-3 0.795 versus 0.778 without it; source Table 3 & Architecture and parameter count change together; no universal fidelity ranking \\
        \bottomrule
    \end{tabular}
\end{table}

{\footnotesize\noindent\textit{Protocol notes.}
\par\nopagebreak[4]\noindent \textbf{H1:} The semantic study samples 100 KIT-ML test descriptions, with 2--5 raters per pair (about four on average), majority voting and half credit for ties; catch trials exclude unqualified workers. TEMOS takes text and duration, whereas the comparator also receives the reference initial pose. NE: preference uncertainty, retained rating count.
\par\noindent \textbf{H2:} Source Table 2 evaluates text-to-motion; Table 1's 0.823 concerns motion-to-text generation. The GRU baseline removes inverse alignment. Retrieval ranks one match against 31 mismatches in a learned multimodal space; evaluation repeats 20 times and reports 95\% confidence intervals. Dataset splits are 80\% training, 15\% testing and 5\% validation. NE: evaluated output count, evaluator checkpoint.
\par\noindent \textbf{H3:} Values from the KIT-ML ablation of the 2022 arXiv v1 preprint. The protocol supplies reference motion length and ranks the matching text against 31 test mismatches with a pretrained contrastive evaluator; twenty evaluations and 95\% intervals are specified. Efficient attention behaves differently without CLIP, which limits generalisation of this contrast. NE: output count, evaluator revision.
\par\noindent \textbf{H4:} One matching description is ranked against 31 randomly selected mismatches using pretrained motion/text features; Table 1 reports means and 95\% intervals over 20 evaluations. Both corruption settings use learned termination rather than reference motion length. NE: evaluated output count, evaluator checkpoint.
\par\noindent \textbf{H5:} The harder subset contains 2,582 items selected from 4,382 test items using words such as left/right. NE: meaning of $\pm$ and repetition count, evaluator checkpoint, selected-item identifiers, generated sample count.
\par\noindent \textbf{H6:} This is Zhang et al.'s \textit{Finetuned LLMs Are General-Purpose Motion Generators}, not the separate \textit{Human Motion as a Foreign Language} paper. Both 7B validation settings are evaluated with text only; joint training includes pose-conditioned tasks, but no pose is supplied for this row. NE: uncertainty, repetition and sample counts, retrieval-evaluator version.
\par\noindent \textbf{H7:} Removing the injector reduces parameters from 4.48M to 3.43M and instead concatenates frame and text tokens for the global information layer. The study follows the Guo et al. evaluation pipeline and splits; its Section 4.4 omits ablation confidence intervals, so the main-table intervals do not apply to this contrast. NE: output count, evaluator revision, ablation repetition count.
\par\noindent \textbf{Abbreviations:} KIT-ML, KIT Motion-Language dataset; GRU, gated recurrent unit; CLIP, contrastive language--image pretraining; LLM, large language model; FID, Fr\'echet inception distance.\par}

Motion evaluation must distinguish semantic correspondence from closeness to a single reference trajectory and from distributional quality. Retrieval can reveal changes under a fixed evaluator, but top-3 success does not verify every requested limb, direction or event. Confidence intervals from repeated evaluation reflect evaluation sampling, not variation across independently trained models. Reference length and pose inputs change the available information; validation results must remain distinct from test results. These qualifications also apply to LMM's representation-learning comparisons and motion preference adaptation~\cite{zhang_large_2025,sheng_exploring_2024}. Neither retrieval gains nor lower FID establish physical executability.

\paragraph{T2-3D} This domain emphasises prompt-faithful 3D generation, focusing on geometric and attribute alignment. Assessment ranges from view-based reconstruction measures to perceptual evaluations of semantic consistency. See Table~\ref{tab:t2-3d-faithfulness}, whose results retain their original protocols.

\begin{table}[H]
    \centering
    \caption{Representative approaches to evaluating faithfulness in text-to-3D-object (T2-3D) generation.}\label{tab:t2-3d-faithfulness}
    \footnotesize
    \setlength{\tabcolsep}{3pt}
    \begin{tabular}{>{\raggedright\arraybackslash}p{0.14\textwidth}>{\raggedright\arraybackslash}p{0.19\textwidth}>{\raggedright\arraybackslash}p{0.16\textwidth}>{\raggedright\arraybackslash}p{0.23\textwidth}>{\raggedright\arraybackslash}p{0.19\textwidth}}
        \toprule
        \textbf{Model/study} & \textbf{Task and conditioning} & \textbf{Evaluation target} & \textbf{Metric/protocol and result versus comparator} & \textbf{Interpretation and limitation} \\
        \midrule
        DreamFusion (2022)~\cite{poole_dreamfusion_2022} [D1] & Object-centric MS-COCO validation prompts; text-to-3D optimisation & Text--render retrieval proxy using textureless views & CLIP ViT-L/14 R-precision 58.5\% versus reimplemented Dream Fields 1.4\%; source Table 1 & Different evaluation viewpoints and training objectives; not ground-truth geometric accuracy \\
        Point-E (2022)~\cite{nichol_point-e_2022} [D2] & Text to synthetic image to point cloud; same images across model sizes & Text--render retrieval proxy & CLIP ViT-L/14 R-precision: 300M 45.6\% versus 40M 38.8\%; source Table 1 & Model-size comparison within this pipeline; mesh conversion can lose detail \\
        Magic3D (2023; 2022 preprint protocol)~\cite{lin_magic3d_2023} [D3] & Text prompts from DreamFusion gallery & Supporting evidence: perceived realism and detail & 61.7\% pairwise preference over DreamFusion; preprint Table 1 & Visual-quality question, not explicit prompt adherence \\
        Shap-E (2023)~\cite{jun_shap-e_2023} [D4] & Direct text conditioning; 300M models with expanded 3D data & Text--render retrieval proxy & CLIP ViT-L/14 R-precision 40.9\% versus text-only Point-E 35.5\%; source Table 2 & Base-model/data comparison with different auxiliary pipelines; not an isolated representation effect \\
        Instant3D, Li et al. (2024)~\cite{li_instant3d_2024} [D5] & Text-conditioned assets; multiview videos for human assessment & Direct human text--3D correspondence & 83.62\% consistency selection share among six methods; source Table 1 & Six-way choice, not pairwise win rate; human-study denominator not established \\
        \bottomrule
    \end{tabular}
\end{table}

{\footnotesize\noindent\textit{Protocol notes.}
\par\nopagebreak[4]\noindent \textbf{D1:} arXiv v1, with 153 object-centric MS-COCO validation prompts and two generation seeds. DreamFusion averages azimuths at $30^\circ$ elevation, whereas the baseline uses a held-out $45^\circ$ view, so this is a whole-system comparison rather than a controlled component ablation. NE: azimuth count, distractor pool, uncertainty.
\par\noindent \textbf{D2:} The 300M and 40M variants receive internally generated images, not additional user input. The study fixes 306 synthetic conditioning views and the image-conditioned upsampler across evaluations, using object-centric prompts inherited from Dream Fields. Point-cloud P-FID/P-IS are distinct from semantic adherence. NE: render-view count, distractor pool, evaluator revision, uncertainty.
\par\noindent \textbf{D3:} Protocol and table numbering follow arXiv v1: 397 gallery prompts, three raters per prompt and 1,191 pairwise comparisons. MTurk users choose between videos for realism and detail; DreamFusion outputs come from its gallery. NE: uncertainty.
\par\noindent \textbf{D4:} arXiv v1 Table 2 entries with expanded data and direct text conditioning; 46.4\% instead describes the synthetic-image-conditioned pipeline. Base architecture, size and data are matched, but auxiliary components differ. Evaluation uses MS-COCO validation prompts; Shap-E renders NeRFs at $128\times128$ with random azimuths and $30^\circ$ elevation, using an early-stopped text checkpoint. NE: sample count, guidance setting, distractor pool, evaluator revision, uncertainty.
\par\noindent \textbf{D5:} This is Li et al.'s \textit{Instant Text-to-3D Generation}, not the separate sparse-view generation and large-reconstruction-model paper. The arXiv v2 study compares TextMesh, SJC, DreamFusion, Latent-NeRF, ProlificDreamer and Instant3D, with at least three users per prompt. NE: prompt count, split, retained rating count, uncertainty.
\par\noindent \textbf{Abbreviations:} MS-COCO, Microsoft Common Objects in Context; CLIP, contrastive language--image pretraining; ViT-L/14, large vision transformer with 14-pixel patches; NeRF, neural radiance field; P-FID and P-IS, point-cloud variants of Fr\'echet inception distance and inception score; MTurk, Amazon Mechanical Turk.\par}

Render-based retrieval can test whether an asset appears consistent with a caption under selected views. Textureless rendering reduces the contribution of painted appearance, but neither retrieval nor visual preference verifies complete geometry, topology or downstream usability. Human questions about realism must remain distinct from questions about prompt correspondence. Point-E and Shap-E also show why direct text conditioning and a pipeline with an intermediate generated image need separate descriptions.

\subsubsection{Cross-Modal Synthesis}

Across T2T, T2I, T2V, T2M, T2HM, and T2-3D, evaluation practice has broadened from proxy scores (e.g., BLEU/FID) to embedding-space alignment (CLIP/MuLan/CLAP) and to human preference judgements and learned judges, reflecting the weak correlation of early automated metrics with perceived faithfulness. This arc is explicit in T2T’s BLEU$\rightarrow$truthfulness benchmarks, T2I’s CLIPScore “trap” and compositional tests, T2V’s pivot to large-scale human win rates, and T2M’s adoption of MuLan/CLAP similarity. These approaches now coexist and answer different questions: for example, MR-FlowDPO distinguishes listening preference from text relevance~\cite{ziv_mrflowdpo_2025}, while VideoREPA distinguishes semantic adherence from physical commonsense~\cite{zhang_videorepa_2025}, and neither warrants combining these outcomes into a single faithfulness score.

Methodologically, four levers transfer well across modalities: (i) SSL-style pretraining that yields robust cross-modal embeddings for alignment and retrieval; (ii) planner-style decomposition (e.g., CoT/LLM planning) that improves entity- and relation-level adherence; (iii) preference optimisation (RLHF/RLAIF) that reduces hallucination and aligns to human judgements; and (iv) sparse capacity (MoE) to scale controllably. These appear in T2I/T2V (LLM-guided planning, human-bench optimisation), T2T (RLHF improving TruthfulQA with truthfulness-informativeness trade-offs), and T2HM (preference fine-tuning). Aggregate pipeline gains, however, do not isolate each component's contribution. Comparative claims should therefore specify conditioning and protocols and use within-study ablations where available.

Structure-aware control is a second shared theme: ControlNet-like adapters stabilise where content is placed in T2I; analogous constraints appear as pose/layout priors in T2V, body-part and length controls in T2HM, and geometry priors in T2-3D, consistently improving spatial/temporal fidelity without retraining the full backbone. Because these controls supply information absent from the text, their gains belong to a controlled task rather than to text-only generation.

Common trade-offs recur: fidelity versus controllability; speed versus long-horizon coherence; and, in language, truthfulness versus informativeness under RLHF. A practical cross-modal approach thus combines (1) curated/pretrained representations and retrieval grounding, (2) planner-guided decomposition, (3) lightweight control adapters for structure, and (4) post-training alignment against human preferences with external verifiers or judge models, evaluated primarily by human win rates plus targeted, modality-appropriate diagnostics.

\subsection{Compositionality}\label{sec:compositionality}

Compositionality evaluates whether a model can systematically combine multiple entities, attributes, relations, and constraints expressed in a prompt, and generalise to novel configurations that were rare or absent in training. Across modalities, compositionality is therefore probed by increasing the number of concepts and constraints, designing out-of-distribution splits that recombine known primitives in unseen ways, and disentangling performance on simple versus structurally complex prompts. These tests consistently reveal that global similarity metrics (e.g., BLEU-, CLIP-, or FID-style scores) can remain high even when models drop objects, merge events, or mis-bind roles, motivating specialised benchmarks and diagnostics that explicitly measure role binding, counting, spatiotemporal structure, and logical consistency.

\subsubsection{Metrics and What They Establish}
Compositional metrics score each requested element and report how often all elements are satisfied together. For images, GenEval uses object detection and colour classification to check object co-occurrence, counts, positions and colours~\cite{ghosh_geneval_2023}. Such scores establish satisfaction of the constraints an evaluator can recognise on the benchmark's prompt distribution. They establish compositional generalisation only when the split withholds combinations of seen concepts, and evaluator errors bound their reliability. Other tasks require analogous structured checks, such as symbolic attribute accuracy for music or body-part and timing checks for motion.

\subsubsection{Modality-Specific Discussions}

\paragraph{T2T} In text generation, improving compositionality often involves mechanisms that break complex prompts into manageable parts or allocate model capacity to distinct subtasks. Prompt decomposition is one approach. For example, CoT and least-to-most prompting explicitly teach language models to solve complex instructions step by step by first decomposing them into simpler subproblems and then addressing each sequentially \cite{wei_chain--thought_2022}. This guided reasoning helps models combine multiple concepts into coherent outputs. On the architectural side, an MoE-based language model activates only a few “expert” subnetworks per query \cite{jiang_mixtral_2024}. Whether this separates the concepts or constraints of a prompt, or improves performance on compositional tasks where multiple attributes or facts must be expressed together, requires an explicit routing analysis and held-out combinations rather than aggregate accuracy: Mixtral's own routing analysis found no obvious topic-based specialisation \cite{jiang_mixtral_2024}. For structured generation such as text-to-SQL, schema linking and execution-based evaluation supply task-specific checks~\cite{baig_texttosql_2025}. Together, these techniques, including reasoning-based prompting, modular architectures, and structured generation plans, are proposed to address the compositional generalisation challenge of turning multiple concepts from a single prompt into a coherent, faithful output.

\paragraph{T2I} SOTA T2I models employ several strategies to better compose multiple objects, attributes, and relations described in a prompt into one image: composable or multi-conditioned diffusion, in which the generative model treats different parts of the prompt (e.g. each object or attribute) as separate conditions and combines them, and spatial or object-centric guidance during generation. Explicit spatial conditions, such as those supplied to ControlNet~\cite{zhang_adding_2023}, change the task relative to text-only generation, so the correctness of the supplied or generated layout should be evaluated separately from its realisation in the image.

\paragraph{T2M} Generative models for music have adopted hierarchical and attribute-conditioned strategies to handle compositional prompts (prompts specifying multiple musical concepts like instruments, genre, mood, and melody), breaking the generation process into stages that reflect musical structure. For example, MuseCoco \cite{lu_musecoco_2023} uses a two-stage pipeline: first, a text-to-attributes stage in which the model extracts a set of musical attributes from the prompt (such as tempo, key, instrumentation, style), followed by an attribute-to-music stage that composes a musical piece conditioned on those attributes. The intermediate representation permits separate checks of whether the requested attributes were extracted and whether the output realises them.

\paragraph{T2V} Generative video models face the challenge of compositionality both in space (multiple objects in a scene) and time (multiple actions or events in sequence). To address this, recent models incorporate multi-stage and multi-agent generation frameworks as well as improved temporal architectures. On the architecture side, spatio-temporal compositional diffusion models have been introduced to better control content frame by frame. For instance, VideoTetris \cite{tian_videotetris_2024} is a diffusion-based generator that directly manipulates the cross-attention maps of the denoising process across both space and time. In practice, this means the model can enforce that a certain token (for example, “man”) attends mainly to a consistent region (left side) across all frames, and another token (“dog”) to another region (right side), thereby respecting a prompt such as “a man on the left walking his dog on the right” throughout the video. Such attention composition mechanisms help maintain object permanence and correct spatial relations as the frames progress. Additionally, enhancements like reference-frame guidance and better data augmentation for multi-action scenarios have been used to improve temporal consistency for long or complex prompts \cite{tian_videotetris_2024}.

On 100 generated compositional prompts, VideoTetris raises frame-averaged BLIP-VQA and UniDet scores from 0.4510 to 0.5563 and from 0.1719 to 0.2350 relative to its unmodified backbone; averaging per-frame checks, however, does not test whether objects appear in the requested temporal order. In summary, T2V models improve compositionality through decomposition and modular generation pipelines that treat different concepts separately, and through advanced diffusion or transformer architectures that enforce spatial and temporal arrangements of multiple entities as dictated by the text prompt.

\paragraph{T2HM} Human motion generation from text must frequently combine multiple described actions, body parts, and temporal phases into one fluid motion sequence. Researchers have developed models that either decompose the motion temporally or compose it from multiple generative sources. OMG's text-conditioned controllers and submotion masks, for example, provide a way to direct individual motion components~\cite{liang_omg_2024}; evaluating such control requires checking the requested body parts and time intervals along with the complete movement, since correct individual segments do not guarantee valid transitions.

\paragraph{T2-3D} Within this survey's object-focused scope, compositionality concerns requested parts, attributes and their relationships. Scene-level generation provides adjacent evidence about composition rather than direct evidence for the object-generation comparison. Generating a scene with multiple objects involves ensuring each entity is modelled and placed correctly in three dimensions. To improve compositional synthesis in this modality, pipelines often introduce an explicit scene representation step and optimised composition procedures. A representative example is to use a scene graph or layout as an intermediate: the text prompt is first converted into a structured scene description (identifying the objects and their relations), which then guides the 3D generation. The recent CompGS \cite{ge_compgs_2025} framework follows this idea by representing the scene as a set of 3D Gaussian primitives, essentially one cloud of Gaussians per object, and optimising each object’s shape and detail and the overall scene consistency in turn. By dynamically adjusting each entity’s 3D parameters and also fine-tuning how they compose (overlap, relative scale), the algorithm ensures that even small or complex objects are generated with fidelity and that they fit together correctly (e.g., not colliding incorrectly or floating unnaturally).

In its ablation, removing the entity-level initialisation lowers the automatic T3Bench alignment score for multiple objects from 38.0 to 18.7, a larger drop than removing its staged optimisation (35.2). In this adjacent scene-generation setting, CompGS uses a divide-and-conquer strategy: the system explicitly separates the scene into individual components, uses prior knowledge or auxiliary generative models to give each component a reasonable form and position, and then refines everything jointly with guidance (from vision models or physics rules) to ensure the pieces come together into a plausible, unified 3D scene that matches the prompt’s description. These results support improved scene-level alignment under the reported evaluator; they do not establish that every requested object and relation is correctly realised or directly validate object-level part composition.

\subsubsection{Cross-Modal Synthesis}

Reports across modalities describe similar failures: models under-realise rare concepts, omit or duplicate objects, blur role assignments ("who does what to whom"), or compress multi-step instructions into a single blended event, while global similarity metrics can remain high. Evaluation protocols therefore use stress tests that scale the number of entities and constraints, recombine known concepts in held-out splits, and separate performance on simple versus structurally complex prompts. Models increasingly accept intermediate scaffolds (parse trees, scene graphs, layouts, motion skeletons, chord grids, camera trajectories), either predicted by auxiliary components or supplied by the user; these allow protocols that separately assess whether the scaffold realises the instruction and whether the final output instantiates it without new errors. Stronger structure often reduces sampling diversity, increases inference cost or introduces brittle interfaces between stages. Each of these interventions, however, requires its own component comparison, with constraint types, the number of jointly requested constraints, the construction of held-out combinations and evaluator reliability reported, before gains can be attributed to it.

\subsection{Robustness}\label{sec:robustness}

Robustness evaluates whether a generative model maintains reliable behaviour under perturbations to its inputs, sampling conditions, or deployment context. Whereas faithfulness and compositionality primarily ask whether a single output correctly realises a given prompt, robustness asks whether this behaviour is stable across paraphrased instructions, distribution shifts, longer horizons, and noisy or adversarial conditions. Across the six modalities considered here, robustness is probed by invariance tests (e.g., prompt paraphrasing, seed resampling, domain shifts), stress tests over long sequences (long documents, videos, motions, or musical pieces), and exposure to atypical yet plausible inputs. Below, we first describe how representative state-of-the-art systems in each modality address robustness, and then synthesise the emerging cross-modal patterns.

\subsubsection{Metrics and What They Establish}
Robustness is quantified relative to a baseline: the change in a task score between unperturbed and perturbed conditions, its variation across equivalent prompts or seeds, and the worst-case or subgroup result over a set of perturbations. These quantities establish sensitivity to the perturbations actually tested. They do not certify behaviour under other shifts, and a small change in a poor score does not indicate reliable generation. Good quality on an ordinary test set, diversity across samples and temporal consistency within one output are therefore not sufficient evidence of robustness, and the mechanisms described below are candidate interventions whose contribution requires such a changed-condition test.

\subsubsection{Modality-Specific Discussions}

\paragraph{T2T} In language generation, robustness means the model reliably follows user intent and produces correct, coherent outputs even when prompts are phrased differently. Key approaches include targeted training with human feedback and improved reasoning strategies. InstructGPT \cite{ouyang_training_2022} uses RLHF to better follow instructions and avoid unwanted outputs. This yielded a smaller 1.3B model that outperforms the 175B GPT-3 in adhering to user intent, producing more truthful, helpful answers with far fewer toxic or off-track responses. These preference comparisons, however, do not directly measure consistency across varied phrasing of instructions. The self-consistency decoding strategy \cite{wang_self-consistency_2023} helps LLMs to tackle complex problems. Rather than generating one answer through greedy decoding, the model samples multiple reasoning paths and then chooses the most common answer among them. This majority-vote approach dramatically improved accuracy on mathematical and commonsense problems. It boosts accuracy on a mathematical word-problem benchmark (GSM8K) by 17.9 percentage points, a gain from aggregating sampled reasoning paths rather than a measured invariance to prompt wording.

\paragraph{T2I} Robustness in T2I generation means the output image consistently reflects the prompt’s details (attributes, relationships) despite rewording, and the model remains stable in quality. Diffusion-based generators like Stable Diffusion \cite{rombach_high-resolution_2022} use a guidance technique that interpolates between unconditional and text-conditioned predictions: the model occasionally ignores the text prompt during training (conditioning dropout), and at inference a guidance scale amplifies the influence of the text, which improves prompt adherence at the cost of some diversity. DALL·E 3 \cite{betker_improving_2023} trained on longer, highly descriptive captions written by a dedicated image captioner and, at deployment, GPT-4 expands user prompts similarly; Janus-Pro \cite{chen_janus-pro_2025} mixes real and synthetic data to improve stability in image generation. A robustness protocol would additionally vary meaning-preserving descriptions and report object and attribute satisfaction separately from image quality.

\paragraph{T2M} Robustness in music generation means producing audio that has good acoustic quality and remains faithful to the text prompt’s described style or content, and does so consistently over time. Music is highly structured, so models employ hierarchical generation and alignment mechanisms. MusicLM \cite{agostinelli_musiclm_2023} first generates a coarse sequence of high-level musical “semantics” and then the fine audio waveform details, and it is conditioned on MuLan, a joint text-audio embedding model, whose cycle consistency between the text and the generated audio is used to evaluate alignment. Meta’s MusicGen \cite{copet_simple_2023} instead trains a single Transformer to predict compressed audio tokens from text, using a large high-quality music dataset and a pretrained text encoder. Neither design by itself shows resilience to prompt variations. A stress test should nonetheless specify whether it changes wording, duration or the distribution of musical concepts, and report relevance, structure and acoustic quality separately.

\paragraph{T2V} For video, robustness involves maintaining temporal consistency (no flicker or scene jumps) and staying faithful to the prompt over many frames. VideoLDM \cite{blattmann_align_2023} fine-tunes a pretrained image latent diffusion model for video by inserting temporal alignment layers; only these layers are trained, so the model keeps the per-frame quality of the image backbone, and the decoder is fine-tuned with a 3D convolutional discriminator to reduce flickering. Google's Lumiere \cite{bar-tal_lumiere_2024} introduced a Space-Time U-Net that generates the entire video in one pass instead of generating keyframes and interpolating between them, which can cause inconsistency. Temporal coherence is, however, a property of a clip; robustness additionally requires testing whether it persists when duration, prompt complexity or supplied controls change.

\paragraph{T2HM} In text-driven human motion synthesis, robustness refers to generating accurate, realistic movement that follows the textual description, even for uncommon actions or when controlling specific aspects. MotionDiffuse \cite{zhang_motiondiffuse_2024} was an early diffusion model on motion data; its stochastic generation process can produce diverse yet plausible variations of a described motion, which is one aspect of robustness. MoMADiff \cite{zhang_towards_2025} (“Masked Autoregressive Diffusion for Motion”) explicitly tackles out-of-distribution robustness: earlier T2HM models often used a VQ-VAE to discretise motions into tokens, which limited them to generating combinations of seen motions, whereas MoMADiff operates in a continuous latent motion space and learns to infill randomly masked portions of a motion sequence. When models trained only on HumanML3D generate motions for held-out Motion-X subsets guided by one keyframe per second, MoMADiff retains a top-3 R-precision of 0.907 on the Kungfu subset, whereas the discrete-token MoMask falls to 0.154; because the keyframes supply part of the target motion, this establishes keyframe-guided transfer rather than text-only robustness.

\paragraph{T2-3D} Robust T2-3D generation means that the method can reliably create a 3D asset that matches the prompt from any viewpoint while avoiding common failures such as broken geometry or misaligned views. Magic3D \cite{lin_magic3d_2023} first generates a low-resolution 3D preview with a NeRF supervised at $64\times64$ resolution (accelerated by a sparse hash structure) under a low-resolution diffusion prior, then optimises a high-resolution textured mesh from it with a differentiable renderer and a 2D latent diffusion model at $512\times512$ resolution. In contrast to optimisation-based methods, Shap-E \cite{jun_shap-e_2023} introduced a direct generative approach that learns to output 3D implicit-function (NeRF or SDF) parameters without per-prompt optimisation, so it does not rely on iterative view matching; it achieved comparable or better sample quality than Point-E \cite{nichol_point-e_2022} while operating in a higher-dimensional space of implicit functions. Because it produces an implicit 3D representation, the output can be rendered from any angle, and this amortised approach trades some absolute fidelity for speed. Lighting or camera changes may, however, reveal sensitivity of a render-based evaluator without changing the underlying geometry, so prompt perturbation, asset editing and renderer variation should be reported as distinct tests.

\subsubsection{Cross-Modal Synthesis}

Failures in robustness follow similar patterns across different modalities: small perturbations to prompts or conditions can trigger disproportionate changes, for example flipped answers under paraphrase, artefacts when prompts are lengthened, lost global structure in long pieces of music, or implausible poses and geometry when descriptions deviate from training templates. Evaluation therefore extends protocols focused on faithfulness and compositionality with explicit perturbation regimes: paraphrase and negation tests in T2T; guidance, seed, and conditioning sweeps in T2I, T2V, and T2-3D; duration or tempo variation in T2M; and noise or ablations on control signals in T2HM. Staged generation, diverse training data, strong pretrained encoders and constrained control interfaces are the mechanisms most often credited with stability, at the cost of diversity and inference time. Reports should therefore state the perturbation or shift, its severity, the expected invariant or equivariant property, sample counts and the unperturbed baseline, and distinguish meaning-preserving perturbations from changes that intentionally alter the target.

\subsection{Cross-Modal Summary}

Taken together, these observations indicate that MGMs across modalities exhibit recurring patterns of strengths and blind spots along the three evaluation axes. Table~\ref{tab:eval_summary_axes} summarises representative benchmarks and metrics for each text-to-X setting, highlighting how current practice remains fragmented and often poorly aligned with human judgements, and Figure~\ref{fig:eval_heatmap} condenses these qualitative ratings into a single overview. This motivates the need for more unified, cross-modal evaluation protocols and closer integration of safety- and governance-oriented assessments, which we discuss next. Table~\ref{tab:eval_questions} turns these blind spots into task-specific reporting questions; they are recommendations for interpreting evidence, not claims that every listed test was performed by the selected studies.

\begin{table}[H]
    \centering
    \caption{Illustrative benchmarks and metrics for the three evaluation axes across six modalities. Entries are not exhaustive but highlight representative practices and typical blind spots.}
    \label{tab:eval_summary_axes}
    \footnotesize
    \setlength{\tabcolsep}{3pt}
    \begin{tabular}{>{\raggedright\arraybackslash}p{0.08\textwidth}>{\raggedright\arraybackslash}p{0.28\textwidth}>{\raggedright\arraybackslash}p{0.28\textwidth}>{\raggedright\arraybackslash}p{0.28\textwidth}}
        \toprule
        \textbf{Modality} & \textbf{Faithfulness} & \textbf{Compositionality} & \textbf{Robustness} \\
        \midrule
        T2T & BLEU, ROUGE, EM/F1 on QA and reasoning; judge-LLM preferences; limited sensitivity to hallucinations and long chains & SCAN, COGS, CFQ compositional splits; GSM8K/MATH; performance drops with concept count and nesting depth & Adversarial NLU and paraphrase suites; domain-shift tests; brittle to small lexical or structural perturbations \\
        T2I & CLIPScore, captioning-back metrics, human preferences; high scores can mask object and attribute errors & T2I-CompBench, ConceptMix, relation/layout benchmarks; fidelity-composition divergence at higher realism & FID/IS and CLIP-based scores under corruptions, edits, and style shifts; sensitive to prompt phrasing and noise \\
        T2M & MIR-inspired statistics, genre/mood classifiers, MOS; poor capture of high-level musical form & Multi-tag control of genre, mood, instrumentation, and structure; degradation for complex prompt combinations & Robustness to text perturbations, style interpolation, and length changes; vulnerable to prompt noise and style shift \\
        T2V & CLIP-based video-text alignment, captioning metrics, action recognition; weak coverage of temporal coherence & Multi-entity and multi-event prompts, identity and state tracking over time; long-horizon consistency remains difficult & Robustness to resampling, compression, camera motion, and scene changes; failures accumulate over longer horizons \\
        T2HM & Motion-language retrieval, pose error; limited nuance for rich descriptions & Multi-agent and multi-action prompts, spatial relations, temporal ordering; hard to coordinate interactions and constraints & Robustness to noisy text, viewpoint changes, trajectory perturbations; physical constraints often violated off-distribution \\
        T2-3D & Text-shape retrieval, CLIP-based alignment, part-level IoU; struggles with fine-grained semantics & Multi-object scenes, part-whole relations, spatial layouts; limited support for binding several entities and relations & Robustness to lighting, rendering, and viewpoint changes; geometry may become implausible or non-manifold under edits \\
        \bottomrule
    \end{tabular}
    \par\smallskip
    \begin{minipage}{\textwidth}\footnotesize
        T2T: text-to-text; T2I: text-to-image; T2M: text-to-music; T2V: text-to-video; T2HM: text-to-human-motion; T2-3D: text-to-3D objects; BLEU: bilingual evaluation understudy; ROUGE: recall-oriented understudy for gisting evaluation; EM: exact match; QA: question answering; LLM: large language model; NLU: natural language understanding; CLIP: contrastive language--image pretraining; FID: Fr\'echet inception distance; IS: inception score; MIR: music information retrieval; MOS: mean opinion score; IoU: intersection over union. Benchmark names are given as commonly used in the literature.
    \end{minipage}
\end{table}

\begin{figure}[H]
    \centering
    \begin{tikzpicture}[
            cell/.style={draw, minimum width=3.3cm, minimum height=1.05cm, align=center, font=\fontfamily{phv}\selectfont\fontsize{9}{11}\selectfont},
            rowlbl/.style={font=\fontfamily{phv}\selectfont\fontsize{9}{11}\selectfont\bfseries, align=right, text width=2.1cm},
            collbl/.style={font=\fontfamily{phv}\selectfont\fontsize{9}{11}\selectfont\bfseries, align=center, text width=3.3cm}
        ]
        \node[collbl] at (3.85, 1.0)  {Faithfulness};
        \node[collbl] at (7.15, 1.0)  {Compositionality};
        \node[collbl] at (10.45, 1.0) {Robustness};

        \foreach \y/\name/\fa/\co/\ro in {
                0/T2T/M/M/M,
                -1.2/T2I/M/M/L,
                -2.4/T2M/L/M/L,
                -3.6/T2V/L/L/L,
                -4.8/T2HM/L/L/L,
                -6.0/T2-3D/L/L/L
            }{
                \node[rowlbl] at (0, \y) {\name};
                \foreach \lvl/\xpos in {\fa/3.85, \co/7.15, \ro/10.45}{
                        \ifthenelse{\equal{\lvl}{M}}{\def\ccolor{blue!48}\def\cname{Medium}}{\def\ccolor{blue!18}\def\cname{Low}}
                        \node[cell, fill=\ccolor] at (\xpos, \y) {\cname};
                    }
            }
    \end{tikzpicture}
    \caption{Heatmap view of Table~\ref{tab:eval_summary_axes}: qualitative maturity of evaluation practice per axis and modality. \emph{Medium} denotes established benchmarks/metrics with noted caveats; \emph{Low} denotes practice explicitly described as weak, poor, brittle, or difficult in the surveyed literature. No axis-modality pair reaches a fully mature (\emph{High}) rating. T2T: text-to-text; T2I: text-to-image; T2M: text-to-music; T2V: text-to-video; T2HM: text-to-human-motion; T2-3D: text-to-3D objects.}\label{fig:eval_heatmap}
\end{figure}

\begin{table}[H]
\centering
\caption{Task-specific reporting questions for evaluating adherence, composition and robustness. Robustness requires an explicit baseline and changed condition; the suggested tests are not reported results.}\label{tab:eval_questions}
\footnotesize
\setlength{\tabcolsep}{3pt}
\begin{tabular}{>{\raggedright\arraybackslash}p{0.08\textwidth}>{\raggedright\arraybackslash}p{0.28\textwidth}>{\raggedright\arraybackslash}p{0.28\textwidth}>{\raggedright\arraybackslash}p{0.28\textwidth}}
\toprule
\textbf{Modality} & \textbf{Faithfulness} & \textbf{Compositionality} & \textbf{Robustness} \\
\midrule
T2T & Are source claims and task constraints satisfied? & Are combined constraints jointly satisfied, and are combinations held out? & Does correctness persist across equivalent prompts or a specified domain shift? \\
T2I & Are requested objects and attributes present? & Are attributes bound to the right objects and relations realised? & Does adherence persist under meaning-preserving wording changes across sampled seeds? \\
T2M & Is the description realised, separately from acoustic quality? & Are requested instruments and structural constraints jointly realised? & Does relevance persist under equivalent wording, and are structural constraints met at each tested duration? \\
T2V & Are requested entities and actions realised? & Are event order and identity preserved across frames? & Do adherence and temporal consistency persist under equivalent prompts, and are duration/control changes evaluated against their intended targets? \\
T2HM & Does the movement match the instruction? & Are body-part and temporal constraints jointly satisfied? & Do semantic and movement-validity measures persist under specified language or control perturbations? \\
T2-3D & Does the asset match requested attributes across views? & Are requested parts and their relationships realised? & Does prompt adherence persist under equivalent wording across views, with asset-edit and renderer/evaluator sensitivity assessed separately? \\
\bottomrule
\end{tabular}
\par\smallskip
\begin{minipage}{\textwidth}\footnotesize
T2T: text-to-text; T2I: text-to-image; T2M: text-to-music; T2V: text-to-video; T2HM: text-to-human-motion; T2-3D: text-to-3D objects.
\end{minipage}
\end{table}

\section{Trustworthiness, Safety, and Ethics}\label{sec:trustworthiness}

MGMs extend the risk surface of traditional text-only LLMs from primarily informational errors to perceptual, persuasive, and even real-world harms. High-fidelity synthesis across modalities enables “synthetic consistency”: a deepfake video with a realistic face, cloned voice, and synchronised lip motion is far more convincing than any unimodal fake and can be deployed at scale for misinformation, fraud, and harassment \cite{department_of_industry_science_and_resources_research_2025}. At the same time, these models are trained on internet-scale corpora rich in personal data, copyrighted works, and structural biases, raising profound questions about fairness, privacy, and intellectual property. This survey adopts a five-dimensional lens of truthfulness, safety, robustness, fairness, and privacy that operationalises high-level principles from the NIST AI Risk Management Framework, the OECD AI Principles, and the EU Ethics Guidelines for Trustworthy AI into concrete multimodal failure modes and mitigation strategies \cite{tabassi_artificial_2023, organisation_for_economic_co-operation_and_development_recommendation_2019, high-level_expert_group_on_artificial_intelligence_ethics_2019}. Figure~\ref{fig:trust_wheel} summarises this lens; each dimension is examined in its own subsection below.

\begin{figure}[H]
    \centering
    \begin{tikzpicture}[
            dim/.style={draw, circle, align=center, minimum size=2.1cm, font=\fontfamily{phv}\selectfont\fontsize{9}{11}\selectfont, fill=orange!12},
            hub/.style={draw, circle, align=center, minimum size=1.9cm, font=\fontfamily{phv}\selectfont\fontsize{9}{11}\selectfont\bfseries, fill=orange!30},
            spoke/.style={thick, -{Latex[length=1.8mm]}}
        ]
        \node[hub] (hub) at (0,0) {Trustworthy\\MGMs};
        \foreach \name/\ang/\i in {Truthfulness/90/1, Safety/18/2, Robustness/-54/3, Fairness/-126/4, Privacy/-198/5} {
                \node[dim] (d\i) at (\ang:4.2cm) {\name};
                \draw[spoke] (hub) -- (d\i);
            }
    \end{tikzpicture}
    \caption{The five-dimensional trustworthiness lens for multimodal generative models (MGMs): truthfulness, safety, robustness, fairness, and privacy.}\label{fig:trust_wheel}
\end{figure}

For each mitigation discussed below, we ask which intervention is applied, which harm it targets, what evidence supports it, what limits that evidence, and what capability or quality cost is measured. Direct evidence about the selected generators is distinguished from adjacent work on understanding models, speech and deployment; results for a model answering questions about images do not by themselves establish safer image generation. Direct mitigation evidence is concentrated in text-to-image generation, with one music example of training restricted to licensed data, so transferring these mitigations to other outputs remains a hypothesis to be tested. Table~\ref{tab:trust_mitigations} summarises the mitigations for misuse, privacy leakage and bias.

\begin{table}[H]
\centering
\caption{Selected mitigations for misuse, privacy leakage and bias. Entries summarise the cited evidence and are not comparable effectiveness scores.}\label{tab:trust_mitigations}
\footnotesize
\setlength{\tabcolsep}{2.8pt}
\begin{tabular}{>{\raggedright\arraybackslash}p{0.17\textwidth}>{\raggedright\arraybackslash}p{0.15\textwidth}>{\raggedright\arraybackslash}p{0.22\textwidth}>{\raggedright\arraybackslash}p{0.22\textwidth}>{\raggedright\arraybackslash}p{0.18\textwidth}}
\toprule
\textbf{Mitigation} & \textbf{Targeted harm} & \textbf{Evidence (task)} & \textbf{Limitation} & \textbf{Capability cost} \\
\midrule
\multicolumn{5}{l}{\textbf{Misuse of generated content}} \\
Inference-time safety guidance: SLD~\cite{schramowski_safe_2023} & Inappropriate imagery, such as nudity or violence & Reduced inappropriate content on the I2P prompts, without retraining (T2I) & An inference-time step can be removed when weights are available~\cite{gandikota_erasing_2023} & No adverse effect on overall quality or text alignment reported by the authors \\
Concept erasure in weights: ESD~\cite{gandikota_erasing_2023} & Explicit content; imitation of artistic styles & Comparable to SLD on explicit content; user study of style removal (T2I) & Learned word embeddings recover erased concepts~\cite{pham_circumventing_2024} & Effects on related concepts require separate measurement \\
Prompt and output filters & Unsafe requests and outputs & Deployed as an additional layer & Vulnerable to increasingly sophisticated jailbreaks~\cite{yeo_multimodal_2025} & False refusals of benign prompts require measurement \\
Watermarking: Tree-Ring~\cite{wen_tree-ring_2023} & Undisclosed synthetic media & Detection after common image edits (T2I) & Identifies only the outputs of cooperating providers & Negligible FID change reported \\
\midrule
\multicolumn{5}{l}{\textbf{Privacy leakage, memorisation and copyright}} \\
Licensed training data: Stable Audio Open~\cite{evans_stable_2025} & Reproduction of copyrighted or training audio & No memorisation detected among nearest-neighbour candidates assessed by listening (audio, including music) & Top-50 candidates in one embedding space & Lower music quality than Stable Audio 2.0; data-policy cost not isolated \\
Caption randomisation and augmentation~\cite{somepalli_understanding_2023} & Replication of training images & Reduced replication at training and inference (T2I) & Depends on the tested models and similarity measures & Not reported \\
Attention attenuation: GUARD~\cite{zhao_guard_2026} & Verbatim and template memorisation & Reduced copying on selected prompts (T2I) & Selected configuration; not a formal privacy guarantee & Prompt alignment and image quality evaluated \\
Deduplication, noisy decoding and filtering~\cite{ramakrishnan_assessing_2025} & Leakage of fine-tuning data & Reported reduction in synthetic-secret extraction (T2T) & Single preprint; simple attacks & Utility retention reported; task and aggregation unspecified \\
\midrule
\multicolumn{5}{l}{\textbf{Bias and representation}} \\
Distributional fine-tuning~\cite{shen_finetuning_2024} & Gender, racial and intersectional bias & Reduced bias for occupational prompts (T2I) & Target distributions and attribute classifiers encode normative choices & Explicitly requested attributes must be preserved \\
Reasoning-based prompt revision: FairCoT~\cite{sahili_faircot_2024} & Stereotyped depictions or limited diversity & Improved fairness and diversity across tested T2I systems & Depends on the reasoning model's own biases & No loss of image quality or semantic fidelity reported by the authors \\
\bottomrule
\end{tabular}
\par\smallskip
\begin{minipage}{\textwidth}\footnotesize
Bold rows group mitigations by harm. SLD: safe latent diffusion; I2P: inappropriate image prompts; ESD: Erased Stable Diffusion; T2I: text-to-image; T2T: text-to-text; FID: Fr\'echet inception distance; GUARD: guidance using attractive--repulsive dynamics.
\end{minipage}
\end{table}

\subsection{Truthfulness and Multimodal Hallucination}

For text generation, factual truthfulness concerns whether factual claims are correct and supported by source information. For non-text generators, the closest analogue of hallucination is unfaithful content that is absent from or contradicts the prompt, which Section~\ref{sec:evaluation} treats as a faithfulness failure. This distinction separates factual claims about the world from adherence to a requested fictional or creative output. Misuse of generated media is considered separately in the following subsection.

\subsection{Safety, Misuse, and Deepfakes}

Safety concerns the potential for models to cause harm directly or indirectly. Multimodal systems can be prompted to generate toxic language, graphic or sexual imagery, instructions for wrongdoing, or non-consensual intimate imagery (NCII). These risks are exacerbated by evidence that web-scale training datasets contain NCII and child sexual abuse material (CSAM), especially for image and video domains \cite{al-kfairy_ethical_2024}. Deepfakes constitute an especially acute failure mode: modern pipelines can jointly synthesise realistic faces, body motion, and voices, making it possible to impersonate individuals in live video calls, voice-phishing scams, or political propaganda \cite{salvi_robust_2023}.

A key development is the shift from early detection strategies based on cross-modal inconsistencies (e.g., misaligned lip sync or unnatural prosody) to “synthetic consistency” attacks, in which all modalities are jointly optimised and therefore mutually reinforcing. This undermines traditional deepfake detectors and moves the emphasis towards proactive prevention and provenance: tighter access control and rate-limiting for high-risk capabilities (e.g., face or voice fine-tuning), stronger safety alignment, and universal watermarking of synthetic media \cite{national_telecommunications_and_information_administration_societal_2024, hou_evading_2023}.

Generation-side mitigations intervene at different points. Safe latent diffusion (SLD) changes sampling rather than weights, steering denoising away from inappropriate concepts without additional training and reducing inappropriate content on the inappropriate image prompts (I2P) test set~\cite{schramowski_safe_2023}; because it is an inference step, it can be omitted by anyone who runs the model weights directly. Erased Stable Diffusion (ESD) instead fine-tunes the weights with negative guidance to remove explicit content and selected artistic styles~\cite{gandikota_erasing_2023}, but learned word embeddings can recover erased concepts without altering the sanitised weights~\cite{pham_circumventing_2024}, so erasure reduces exposure to ordinary prompts rather than guaranteeing removal. For provenance, Tree-Ring watermarks embed a pattern in the initial noise and detect it by inverting sampling, with negligible FID change and robustness to common transformations~\cite{wen_tree-ring_2023}; watermarks, however, identify only the outputs of cooperating providers, and provenance metadata such as that of the Coalition for Content Provenance and Authenticity (C2PA) can be lost when content is re-encoded~\cite{coalition_for_content_provenance_and_authenticity_c2pa_2025}. Detection therefore remains necessary for unmarked content, and on deepfakes circulated in 2024, open-source detectors lost much of the accuracy reported on academic benchmarks~\cite{chandra_deepfake-eval-2024_2025}.

Alignment methods centred on RLHF and RL from AI feedback (RLAIF) are being extended to multimodal settings. Frameworks such as Safe RLHF and Safe RLHF-V treat safety as a constrained optimisation problem, using human or AI-generated preference data to encourage refusal of harmful requests while preserving helpfulness on benign multimodal tasks \cite{dai_safe_2024, ji_safe_2025}. Complementary guardrails, including policy filters on both prompts and outputs, act as an additional layer of defence, though they remain vulnerable to increasingly sophisticated jailbreaks \cite{yeo_multimodal_2025}. These refusal-oriented methods govern the responses of assistants rather than image, video, music or motion synthesis, so their evidence is adjacent to generator safety.

\subsection{Robustness and Adversarial Abuse}

In the trustworthiness sense, robustness denotes a model’s predictability and resilience under distribution shift and adversarial conditions \cite{vassilev_adversarial_2024}. Multimodal architectures introduce new attack surfaces beyond text, notably indirect prompt injection via images, audio, or video. In these “Trojan horse” setups, malicious instructions, such as “ignore all safety guidelines”, are encoded as barely perceptible perturbations or steganographic patterns in the non-textual input. When an MGM is asked to describe or reason about the asset, it inadvertently executes the hidden prompt \cite{bagdasaryan_abusing_2023, yeo_multimodal_2025}.

Defensive practice is moving towards a co-evolutionary “red team-blue team” loop, in which automatically generated adversarial prompts and multimodal payloads guide adversarial training and safety fine-tuning. Most documented attacks of this kind target multimodal understanding models; they inform deployed systems that combine understanding and generation, but they are not evidence about the generators compared in this survey, for which the relevant attacks are those on safety filters and concept erasure discussed above.

\subsection{Fairness, Representation, and Multimodal Bias}

Bias in multimodal models manifests in visual, auditory, and embodied outputs as well as in statistical patterns. Large web-scraped datasets encode demographic, cultural, and geographic imbalances that propagate into occupational stereotypes, sexualisation, and Western-centric defaults in image, video, and audio generation \cite{adewumi_fairness_2024}. For example, T2I/T2V models often depict “doctors” as male and “nurses” as female, over-sexualise prompts referring to women, and over-represent Western aesthetics and skin tones, even when prompts are neutral or explicitly specify other cultures \cite{agrawal_evaluating_2025, ghosh_generative_2025}. In music, language models asked to rank or rate musical cultures favour Western traditions \cite{kruspe_musical_2024}, a bias that can propagate into the prompts and plans that language models write for music generation. In T2HM and T2-3D, motion-capture corpora often under-represent diverse body types and mobility patterns, reinforcing narrow norms in generated motion and avatar design \cite{friemert_limitations_2025}.

Mitigation spans both data and alignment. Dataset governance efforts seek to measure and rebalance demographic distributions, reduce explicit stereotypes, and curate targeted data for underrepresented groups, but this remains challenging at internet scale. Alignment-based debiasing is a more recent alternative: Shen et al.~\cite{shen_finetuning_2024} frame fairness as alignment of generated attributes to a user-defined target distribution and fine-tune the diffusion sampling process accordingly, reducing gender, racial and intersectional biases for occupational prompts, while FairCoT revises prompts through iterative chain-of-thought reasoning and reports improved fairness and diversity across tested text-to-image systems without loss of image quality~\cite{sahili_faircot_2024}. Both depend on normative choices about which attributes are balanced and how attributes are inferred from images, and an intervention should diversify attributes that the prompt leaves unspecified while preserving those it explicitly requests. Parallel work advocates for “thick evaluations” that incorporate community perspectives and qualitative assessments of cultural representation, moving beyond aggregate quantitative metrics \cite{qadri_case_2025}.

\subsection{Privacy, Copyright, and Data Governance}

Training on internet-scale corpora exposes multimodal models to large volumes of personal and biometric data. Beyond the leakage of textual personally identifiable information (PII), models may memorise and regenerate unique faces, voices, and gait signatures, enabling cross-modal re-identification and persistent surveillance. Carlini et al.~\cite{carlini_extracting_2023} extract more than a thousand training images, including photographs of individual people, from state-of-the-art diffusion models, and Somepalli et al.~\cite{somepalli_understanding_2023} observe that replication is common for text-conditioned but rare for unconditional diffusion models, implicating caption conditioning as well as duplicated images. In music, MusicLM's analysis finds exact reproduction of training segments to be very rare, whereas approximate matches increase with the length of the audio prompt~\cite{agostinelli_musiclm_2023}.

Copyright and intellectual property (IP) concerns are similarly modality-dependent. In music, legal debates focus on whether training on copyrighted recordings to imitate an artist’s style constitutes infringement. In image and 3D domains, litigation centres on alleged “compressed copies” of training data embedded in model weights and on outputs that are substantially similar to specific copyrighted works. Policy-oriented scholarship has begun to map possible regimes for lawful data use in training (e.g., opt-out mechanisms, collective licensing) and for copyright-aware inference filters that detect near-duplicates and block their generation \cite{sag_copyright_2023}.

Technical mitigations include differential privacy during training to bound the influence of individual data points, targeted data sanitisation and masking, and modality-specific anonymisation for faces, voices, and gait \cite{ramakrishnan_assessing_2025}. For text-to-image generation, randomising and augmenting captions reduces replication at training and inference~\cite{somepalli_understanding_2023}, and guidance using attractive--repulsive dynamics (GUARD) detects prompt-specific attention spikes during denoising and attenuates them, reducing verbatim and template copying while tracking prompt alignment and image quality~\cite{zhao_guard_2026}; such reductions are empirical mitigations rather than formal privacy guarantees. Complementary governance measures such as licensing frameworks, dataset documentation, and explicit “do not train” signals are increasingly recognised as prerequisites for trustworthy multimodal model development \cite{xiang_fairness_2024}. Stable Audio Open adopts this data policy: it trains only on Creative Commons recordings screened for copyrighted music, and a listening check of the 50 nearest training candidates finds no memorisation beyond close renditions of simple sounds~\cite{evans_stable_2025}. The authors suggest that the restriction may limit music quality, but their comparison with Stable Audio 2.0 changes the system and data together and does not isolate that cost.

\subsection{Systemic Oversight and Open Challenges}

Finally, trustworthiness in multimodal generative AI must be embedded in a broader governance ecosystem rather than left to ad hoc technical fixes. A useful abstraction is a three-pillar structure: (i) developer disclosures, (ii) independent audits, and (iii) user-facing labels. Model Cards and Datasheets provide structured documentation of intended use, limitations, and evaluation results. Independent benchmarks such as Trust-videoLLMs \cite{wang_understanding_2025} and Priv-IQ \cite{shahriar_priv-iq_2025} move beyond task accuracy to systematically assess safety, bias, deepfake robustness, and privacy competencies in multimodal LLMs, although they evaluate understanding rather than generation. These are complemented by deepfake-focused datasets and detection challenges that track an evolving adversary ecosystem \cite{chandra_deepfake-eval-2024_2025}. The third pillar, content provenance, is operationalised by standards like C2PA \cite{coalition_for_content_provenance_and_authenticity_c2pa_2025}, which cryptographically bind metadata to assets and enable downstream actors such as platforms, regulators, and end-users to recognise synthetic content.

Many of the benchmarks, audits, and governance mechanisms reviewed in this section can be interpreted as safety-oriented instantiations of the evaluation axes introduced in Section~\ref{sec:evaluation}. They assess whether MGMs remain faithful to ground truth about authenticity and consent, maintain robust behaviour under adversarial prompts and distribution shifts, and satisfy fairness constraints across demographic and cultural groups. Such claims require an explicit threat model for robustness, specified populations and outcomes for fairness, and an empirical or formal guarantee for privacy, reported alongside generation quality and benign-task utility.

\section{Discussion}\label{sec:synthesis}

\subsection{Architectural Paradigms in Text-to-X Generation}

Text-to-X generative models span multiple paradigms, notably GANs, VAEs, AR Transformers, diffusion models, and hybrid systems, each with distinct structural and efficiency characteristics. A high-level comparison of representative models, training objectives, inference regimes, and typical strengths and weaknesses across these paradigms is provided in Table~\ref{tab:discussion}.

\begin{table}[H]
    \centering
    \caption{Comparison of major generative paradigms. Models and numbers are illustrative.}\label{tab:discussion}
    \footnotesize
    \setlength{\tabcolsep}{3pt}
    \begin{tabular}{>{\raggedright\arraybackslash}p{0.1\textwidth}>{\raggedright\arraybackslash}p{0.2\textwidth}>{\raggedright\arraybackslash}p{0.14\textwidth}>{\raggedright\arraybackslash}p{0.13\textwidth}>{\raggedright\arraybackslash}p{0.17\textwidth}>{\raggedright\arraybackslash}p{0.17\textwidth}}
        \toprule
        Paradigm & Examples & Training Loss & Inference & Strengths & Weaknesses \\
        \midrule
        AR (Dense) & GPT-3/4; WaveNet (audio); MotionGPT (motion) & Cross-entropy (MLE) & $T$ sequential steps & High fidelity & High memory $O(T^2)$; slow sampling; high computational cost \\
        AR (MoE) & GLaM (text); Mixtral (text) & Same + sparsity losses & Sparse subnetwork & Larger capacity per flop; efficient inference (sparse) & Complex routing; large total parameter count; training complexity \\
        Diffusion & Stable Diffusion (image); Lumiere (video); Noise2Music (audio) & Denoising score matching & $N$ iterative steps & Stable training; high output quality; inherent stochasticity & Slow sampling; many steps; larger inference cost \\
        GAN & AttnGAN (image); MoCoGAN (video) & Minimax adversarial loss & Single generator pass & Very fast inference; sharp, high-resolution samples & Training instability; mode collapse; lower diversity \\
        VAE & DALL-E1 (discrete VAE tokenizer); MusicVAE (music); TEMOS (motion) & ELBO & Encode-decode pass & Stable training; explicit likelihood bound & Blurry outputs; limited detail; KL trade-off loss \\
        Flow/Score-based & Rectified Flow (image) & Flow matching & Few steps & Straight transport paths & Architectural constraints; numerical stability \\
        \bottomrule
    \end{tabular}
    \par\smallskip
    \begin{minipage}{\textwidth}\footnotesize
        AR: autoregressive; MoE: mixture of experts; GAN: generative adversarial network; VAE: variational autoencoder; MLE: maximum likelihood estimation; ELBO: evidence lower bound; KL: Kullback--Leibler.
    \end{minipage}
\end{table}

\paragraph{GAN-Based Models}
These models \cite{xu_attngan_2018, qiao_mirrorgan_2019, tulyakov_mocogan_2018, kim_tivgan_2020} employ a generator-discriminator game, optimising
\begin{equation}
\min_G\max_D
    \mathbb{E}_{x\sim p_{\text{data}}(x)}[\log D(x)]
    +
    \mathbb{E}_{z\sim p_z(z)}[\log(1-D(G(z)))],
\label{eq:gan-objective}
\end{equation}
where $G$ denotes the generator that maps latent codes $z\sim p_z(z)$ to the data space, $D$ denotes the discriminator that estimates the probability that an input $x$ comes from the real data distribution $p_{\text{data}}(x)$, and the expectations are taken over real data samples $x$ and latent variables $z$. GANs produce samples in a single forward pass (very fast sampling) but suffer from training instability and mode collapse. GANs can yield very sharp images or videos in real time, but with less diversity and weaker likelihood guarantees.

\paragraph{VAE-Based Models}
These models \cite{engel_neural_2017, roberts_hierarchical_2018, ramesh_zero-shot_2021, dhariwal_jukebox_2020, petrovich_temos_2022} encode data into latent codes $z\sim q_\phi(z|x)$ and decode via $p_\theta(x|z)$, optimising the evidence lower bound (ELBO)
\begin{equation}
\mathcal{L}_{\mathrm{ELBO}} = \mathbb{E}_{q_{\phi}(z|x)}[\log p_\theta(x|z)] - \mathrm{KL}\bigl(q_\phi(z|x)\Vert p(z)\bigr),
\label{eq:vae-elbo}
\end{equation}
where $q_\phi(z|x)$ is the variational encoder (approximate posterior), $p_\theta(x|z)$ is the decoder (likelihood model), $p(z)$ is the prior over latent variables, and $\mathrm{KL}(\cdot\Vert\cdot)$ denotes the Kullback-Leibler divergence between two distributions. VAEs train stably and provide a closed-form likelihood lower bound, but often produce blurrier outputs than GANs or diffusion models due to the information bottleneck. VAEs frequently appear as components \cite{blattmann_align_2023} or in audio tokenisation \cite{dhariwal_jukebox_2020}. Hybrid architectures use VAEs as compressors for diffusion or autoregressive sampling, balancing fidelity and efficiency.

\paragraph{Transformer-Based Models} use attention mechanisms to generate text token-by-token,
\begin{equation}
p(x)=\prod_{t=1}^{T} p(x_t\mid x_{<t}),
\label{eq:autoregressive}
\end{equation}
where $x=(x_1,\dots,x_T)$ is a sequence of $T$ tokens and $p(x_t\mid x_{<t})$ is parameterised by a Transformer. Vision Transformers adapt the same blocks for images. Attention computes
\begin{equation}
\mathrm{Attn}(Q,K,V) = \mathrm{softmax}\left(\frac{QK^\top}{\sqrt{d}}\right)V,
\label{eq:attention}
\end{equation}
where $Q,K,V\in\mathbb{R}^{n\times d}$ are the query, key, and value matrices for a sequence of length $n$ and hidden size $d$, $\mathrm{softmax}$ is applied row-wise, and a straightforward implementation incurs $O(n^2 d)$ time complexity and $O(n^2+nd)$ activation storage. Such dense Transformers achieve high representational power and can model long-range dependencies, but inference is inherently sequential (latency $\Omega(T)$ for $T$ tokens) and memory-intensive.

\paragraph{Diffusion-Based Models} perform iterative denoising. A typical forward process adds Gaussian noise.
\begin{equation}
x_t = \sqrt{\alpha_t}\, x_0 + \sqrt{1-\alpha_t}\,\epsilon,\qquad \epsilon\sim\mathcal{N}(0,I),
\label{eq:forward-diffusion}
\end{equation}
for $t=1,\dots,T$, where $x_0$ is a clean data sample, $x_t$ is the noised sample at diffusion step $t$, $\alpha_t\in(0,1]$ controls the signal-to-noise ratio, and $\epsilon$ denotes standard Gaussian noise. The denoising model learns $\epsilon_\theta(x_t,t)$ to predict the noise, with training loss
\begin{equation}
\mathcal{L}_{\text{diff}} =
    \mathbb{E}_{x_0,\epsilon,t}\bigl[\lVert \epsilon - \epsilon_\theta(x_t,t)\rVert_2^2\bigr],
\label{eq:denoising-loss}
\end{equation}
where the expectation is taken over data samples $x_0$, Gaussian noise $\epsilon$, and timesteps $t$ sampled from a predefined schedule, and $\lVert\cdot\rVert_2$ denotes the Euclidean norm. They are mathematically principled (approximating data likelihood) and stable to train, but each generated example incurs repeated forward passes through the denoising model (slow sampling). Latent diffusion variants first encode images into a lower-dimensional VAE space, trading slight quality loss for large speedups.

\paragraph{Hybrid Architectures} combine paradigms to leverage complementary strengths. Examples include \textbf{Diffusion Transformers (DiTs)}, which replace U-Nets with Transformer blocks in diffusion \cite{peebles_scalable_2023}. Unified autoregressive models like Janus-Pro \cite{chen_janus-pro_2025} process multimodal understanding and generation in one transformer, using the LLM to structure generation. \textbf{Cascaded pipelines} use multistage generation: Magic3D \cite{lin_magic3d_2023} first optimises a coarse NeRF on a low-resolution latent ($64\times64$) and then refines a mesh at high resolution. Such cascades divide the computational burden between rapid coarse generation and detailed refinement. \textbf{Masked token modelling} is emerging in motion and audio: MoMADiff \cite{zhang_towards_2025} uses masked sequence prediction (diffusion infilling) instead of fixed VQ tokens, improving handling of novel sequences.

\paragraph{Dense versus Sparse (MoE)} Large Transformers can be scaled by sparsity via MoE. An MoE layer routes each token $x_i$ to a small number of expert feedforward networks: $y_i = \sum_{k} g_{ik} f_k(x_i)$, where gating weights $g_{ik}$ (often top-2 softmax) select expert outputs. This yields enormous model capacity while activating only a fraction of its parameters for each example. For instance, Google’s GLaM \cite{du_glam_2022} has 1.2 trillion total parameters (64 experts per layer) but each token traverses only 96.6B active parameters (8\%). GLaM matches or exceeds GPT-3 (175B) on many tasks, while using about half the inference FLOPs and one-third the training energy of GPT-3. Likewise, the Mixtral \cite{jiang_mixtral_2024} 8x7B model (a sparse variant of Mistral 7B) configures 8 experts per layer and routes each token through 2 experts, giving 47B total capacity but only 13B active per token. It outperforms Llama-2 70B on many benchmarks. MoEs trade off compute (reduced per-token FLOPs) and memory (storing all expert weights) against routing overhead (gating networks, load balancing) and the engineering complexity of sparse distributed training. Such models may also specialise, with different experts focusing on distinct subtasks, although establishing such specialisation requires an explicit routing analysis.

\paragraph{Routing Mechanisms and Compute-Memory Trade-offs} Beyond MoE, routing ideas include adaptive computation (e.g. skipping layers for trivial inputs) and encoder caches, but these are less prevalent in multimodal generation. In MoEs, routers incur little compute relative to experts but require replicating experts across devices (memory overhead). The effective compute cost per sample is lowered (since only a subset of experts run), but the total parameter count and memory footprint remain high. For Transformers, fully dense self-attention demands $O(n^2)$ memory, whereas sparse or linearised attentions reduce that to $O(n)$. Some vision and sequence models use locality (e.g. sliding window or pooling) to reduce cost. Overall, dense models are simpler but costly at scale, while sparse routing (MoE) enables extremely large models with reduced per-token cost.

Beyond these architectural families, modern MGMs are further unified by a set of general-purpose optimisation techniques (SSL, MoE, RLHF, and CoT) which operate largely independently of the output modality. SSL provides shared semantic representations that can be reused across modalities, while MoE increases capacity by routing inputs to sparse expert subsets, enabling specialisation without a linear increase in compute. RLHF and related preference-optimisation schemes align these high-capacity models with desired human behaviours, and CoT-style decompositions expose intermediate reasoning that can guide both decoding and post-hoc verification. Together, these techniques define a common "optimisation backbone" that interacts with the design choices above, shaping the effective trade-offs between fidelity, controllability, and efficiency in text-to-X generation.

\subsection{Technical Trade-offs}

\begin{itemize}
    \item \textbf{Sampling Speed versus Fidelity.} GANs and VAEs generate in one or a few passes, giving very fast sampling: GANs need just a single forward pass through the generator. By contrast, diffusion models typically require hundreds of denoising steps (each a U-Net/Transformer pass). Diffusion’s iterative nature yields high-fidelity, detailed outputs (and a tractable likelihood) but at the expense of latency and compute. Researchers mitigate this via specialised samplers (DDIM, DPM-Solvers) or by distilling into fewer-step models (Rectified Flow Transformers \cite{esser_scaling_2024}). Thus, GANs offer faster sampling but less stable training, whereas diffusion models offer stable training and high output quality at a greater sampling cost.

    \item \textbf{Discretisation Error versus Temporal Consistency.} For modalities with temporal structure (video, music, motion), discrete tokenisation (via VQ-VAEs) allows autoregressive token models but can introduce quantisation noise across time steps. For example, early T2M or T2HM models used a VQ-VAE to convert continuous audio/motion into discrete tokens for autoregressive generation. This discretisation can cause rough transitions or limit long-range coherence. In T2V or motion, explicit temporal architectures (3D convolutions, cross-frame attention) mitigate flicker and abrupt transitions. For instance, VideoLDM \cite{blattmann_align_2023} adds temporal alignment modules at each U-Net layer to encourage frame-by-frame consistency, and Lumiere’s Space-Time U-Net generates the full video jointly (avoiding separate frame conditioning) \cite{bar-tal_lumiere_2024}. Thus, one must trade off the simplicity of discrete tokens against the smoother continuity of continuous latent or frame-level diffusion.

    \item \textbf{Continuous versus Discrete Latent Representations.} Discrete token representations (from VQ-VAEs) support the use of powerful Transformer decoders, but each quantisation incurs approximation error. Continuous latents (as in diffusion or normalising flows) preserve detail but often require denser networks and multiple passes. For example, MotionGPT \cite{zhang_motiongpt_2024} uses a Transformer to model discretised pose sequences, but its quality is bounded by token precision. The newer MoMADiff \cite{zhang_towards_2025} eschews VQ discretisation entirely: it operates in a continuous latent space and trains by randomly masking motion segments and infilling them via diffusion. This avoids quantisation error and better handles out-of-distribution sequences when guided by keyframes, at the cost of requiring sequential denoising. In 3D generation, models like Shap-E \cite{jun_shap-e_2023} directly emit continuous implicit fields (NeRF/SDF parameters), enabling smooth geometry from any viewpoint, whereas discrete voxel/mesh predictions might struggle with fine detail or require heavy post-processing.

    \item \textbf{Conditioning Fidelity versus Controllability.} The choice of conditioning mechanism (and its strength) is key. Classifier-free guidance in diffusion amplifies text influence for better faithfulness, but a higher guidance scale reduces sample diversity and sometimes overfits prompt phrases. Conversely, looser conditioning yields more variety but weaker prompt adherence. Advanced methods improve conditioning: ControlNet \cite{zhang_adding_2023} adds auxiliary inputs (edges, depth maps) to steer spatial content, constraining the denoising path and reducing hallucinations. In language models, complex prompts may be hard to satisfy at once. Systems like DALL·E 3 \cite{betker_improving_2023} address this by “preprocessing” prompts: during training, OpenAI rewrote captions into richer descriptions using a dedicated captioner, and during inference, user prompts are similarly expanded. This effectively teaches the model to expect detailed instructions, boosting output fidelity to nuanced queries. In music, MusicLM \cite{agostinelli_musiclm_2023} conditions generation on high-level semantic tokens using a pretrained joint text-audio embedding (MuLan), ensuring generated audio remains faithful to the text intent. Thus, stronger or multi-stage conditioning (text$\to$attributes$\to$waveform) improves alignment but adds model complexity and sometimes requires more auxiliary data (e.g. annotated attributes or pretrained aligners).
\end{itemize}

\subsection{Cross-Modal Synergies}

Architectural advances often transfer across modalities through shared components or training schemes.

\textbf{Contrastive embeddings} like CLIP (text-image) and MuLan/CLAP (text-audio) provide universal conditioning: models reuse these pretrained encoders for guidance or evaluation. CLIP’s joint embedding underpins many T2I and T2V systems; its scores are standard metrics for prompt alignment. Similarly, MuLan anchors music to text semantics in MusicLM, and CLAP does the same for general audio. These shared embeddings mean that progress in contrastive pretraining benefits multiple modalities.

\textbf{Language models as planners} bridge tasks. GPT-4 and successors, though trained mostly on text, are used to rewrite prompts (DALL·E 3) or decompose complex prompts across domains (coordinating scene layouts in images and videos). Some T2HM works view motion as “pose language”: MotionGPT is a GPT-style model trained on discretised motion tokens, enabling it to accept natural language instructions like a text model. Large LLM backbones therefore increasingly augment non-text generation. For example, an LLM might generate a scene graph, which then conditions a 3D generator.

\textbf{Modular training pipelines} transfer across modalities. Classifier-free guidance transfers across diffusion-based image and video generators; RLHF from language models is adapted to align image and video outputs with preferences; and attention mechanisms refined in vision may inform audio models. Innovations in one modality (e.g. attention or normalisation tricks in Transformers) typically generalise. The overall trajectory is toward multimodal foundation models: e.g. Google’s Gemini and OpenAI’s GPT-4o combine text, image (and potentially other) modalities under one architecture, leveraging cross-domain embeddings and unified training.

However, this shared architecture and optimisation stack still comes with strong modality-specific constraints. Methods originally designed for discrete, symbolic sequences (e.g., token-level language modelling with scalar RLHF rewards) must be adapted to continuous, spatial, or temporally extended outputs. For motion and video, routing in MoE models needs to respect localized or hierarchical semantics (e.g., body parts, spatial regions, or object groups), and CoT-style reasoning often benefits from non-textual intermediate states such as scene graphs, keyframe layouts, or kinematic sketches. Evaluation is further complicated by the fact that perceptual errors (e.g., temporal jitter in motion, phase incoherence in music, or multi-view inconsistency in 3D) are not well captured by token-level or frame-wise metrics. Consequently, cross-modal transfer is inherently asymmetric: techniques developed for text can be reused to structure prompts, conditioning, and high-level planning, but they must be re-engineered to respect the physical, geometric, and temporal structure of each modality.

\subsection{Empirical Evaluation and Trade-offs}

Evaluation studies reveal how these design choices play out. The results reported in Section~\ref{sec:evaluation} come from different datasets, evaluators and protocols, so they illustrate within-study contrasts rather than trends or a common ranking. For text, InstructGPT reports a lower rate of unsupported additions to supplied source material than GPT-3 in the same study (21\% versus 41\%, pooled over model sizes; a source-grounding result, not a universal hallucination rate), at the price of more expensive alignment training. In T2I, AttnGAN's alignment-module ablation and DALL·E 3's CLIP-score comparison with DALL·E 2 and SDXL (Table~\ref{tab:t2i-faithfulness}) answer different questions with different evaluators, so they cannot be arranged as a timeline of improving fidelity.

In music, MusicLM reports a MuLan-based alignment score (MCC$\approx$0.51 against 0.49 without its semantic modelling stage), whereas MusicGen reports CLAP-based scores; because the two use different evaluators and prompts, these numbers cannot by themselves establish that multi-stage systems better match the prompt.

Video models add metrics for temporal consistency (FVD) and CLIP-based prompt match, which measure distinct properties. VideoLDM inserts temporal alignment layers into an image model, and Lumiere's single-shot Space-Time U-Net generates the whole clip jointly instead of frame by frame; both aim to reduce flicker and keep the clip globally coherent, but coherence within a clip does not show that prompt adherence holds, and joint generation can require more compute.

Motion generation metrics probe realism and adherence separately. MotionDiffuse's stochastic generation yields diverse yet plausible variations of a described motion, which is a different property from following the instruction consistently. MoMADiff is motivated by out-of-distribution robustness: by masking inputs, it learns to infill novel motions, addressing the failure mode of VQ-VAE-based systems that merely “stitch together” seen poses; its reported advantage on held-out subsets used keyframe guidance, so it does not establish text-only robustness.

In 3D generation, metrics include multi-view consistency and semantic correctness. Magic3D refines a coarse representation into a textured mesh, whereas Shap-E outputs implicit functions for geometry and texture jointly without per-prompt optimisation and reports comparable or better sample quality than Point-E; these are different pipelines evaluated under different protocols. Evaluation often still relies on proxy tasks (rendering error, human ratings) because automatic 3D metrics are nascent.

These challenges highlight evaluation, alignment, and uncertainty. In high-fidelity regimes, standard distributional metrics (e.g., FID or FVD for images and video, reconstruction scores for motion, or CLIP-based alignment measures) begin to saturate or diverge from human judgements, especially for complex, long-horizon prompts. RLHF and preference-based tuning partially address this by directly optimising for human comparison data, yet they remain expensive to scale across modalities and can overfit to narrow prompt distributions. MoE and modular architectures introduce forms of uncertainty, such as disagreement between experts or unstable routing, that are rarely reflected in existing metrics. CoT and other structured reasoning mechanisms can expose epistemic uncertainty through flawed or inconsistent intermediate steps. Taken together, these observations suggest that explicit mechanisms for robust calibration remain underexplored in non-text domains.

No single paradigm is universally superior; each involves efficiency trade-offs that depend on the use case. Modern systems often combine paradigms, making it important to choose an architecture or hybrid that balances quality, speed, and controllability for the target modality and application. Even though modern architectures such as Transformers and diffusion models currently dominate text-to-X generation, earlier and simpler paradigms retain intrinsic value precisely because of their transparency, inductive biases, and implementation simplicity. Their lightweight components (e.g., autoencoders, GAN-style adversarial losses, factorised temporal models) often offer favourable efficiency and controllability trade-offs, and are increasingly repurposed as modules inside larger systems. Thus, future advanced multimodal generators may hybridise these classical designs with large-scale Transformer/diffusion backbones, rather than fully replacing them.

\subsection{Four Findings for Method Selection and Evaluation}\label{sec:findings}

The preceding comparisons support four findings about choosing and evaluating generation methods. They concern the property improved, the realisation of intermediate structure, the granularity of expert computation and the independence of evaluation. Each finding connects a study-specific result to a research decision; the cases do not form a common performance ranking.

\subsubsection{Preference, Relevance and Diversity Can Change Differently}\label{sec:finding-properties}
Improvement must be specified by outcome. MR-FlowDPO's human study asks separately about overall preference, audio quality, text alignment and musicality~\cite{ziv_mrflowdpo_2025}. For its 1B model against MelodyFlow-1B, it reports net win rates of $16.67\pm10.00$ for overall preference and $1.88\pm9.30$ for text alignment, in percentage points, using 100 pairs per configuration, five annotations per pair and question, and item-level bootstrapping for 95\% uncertainty. The authors find an overall-preference gain but comparable alignment; their 400M comparison improves both outcomes. Motion-R1 supplies a different trade-off: adding similarity rewards to its planned generator improves retrieval but reduces within-prompt variation~\cite{ouyang_motion-r1_2025}, whereas T2I-R1 shows that diversity loss under image-token reward optimisation can be reversed by jointly training semantic planning~\cite{jiang_t2i-r1_2025}. For method selection, quality, adherence and diversity should be reported separately, and higher aggregate preference alone cannot establish that a tuned model is the better choice.

\subsubsection{Intermediate Structure Must Be Checked in the Final Output}\label{sec:finding-realization}
An intermediate plan creates two testable stages: interpreting the request and realising that interpretation. HCoG provides a useful example~\cite{qin_hcog_2025}: its generation-order ablation changes the rendered attribute score, while a separate expert study assesses extracted parts and their ordering, and a documented failure links a reversed order to an incorrectly rendered attribute. MusiCoT offers a complementary, partial check by comparing indicators in its latent musical plan with instrument-track volumes extracted from the generated audio~\cite{lam_analyzable_2025}. Researchers should therefore measure prompt-to-plan correctness, plan-to-output realisation and end-to-end constraint satisfaction separately, under a total inference budget that includes planning and verification.

\subsubsection{Expert Granularity Determines the Mechanism and Cost}\label{sec:finding-experts}
Expert specialisation can change capacity, generation stages or control. Mixtral routes tokens, Wan2.2 switches denoisers by noise regime, and OMG blends controller parameters~\cite{jiang_mixtral_2024,wan_wan22_2025,liang_omg_2024}; these mechanisms do not have interchangeable efficiency implications. DiffMoE makes the capacity distinction concrete~\cite{shi_diffmoe_2025}: its text-to-image comparison uses 1.2B average active parameters for both models but 4.6B total parameters for the expert model versus 1.2B for the dense model, with overall GenEval scores of 0.44 versus 0.38 before supervised fine-tuning and 0.51 versus 0.49 afterwards. This comparison changes both routing and total capacity, and establishes neither a routing-only effect nor equal latency. Method selection should consider which specialisation the task requires and which resource limits deployment, with comparisons of weight storage, active computation, measured latency, and peak memory at the intended settings.

\subsubsection{Optimisation-Score Gains Need Independent Confirmation}\label{sec:finding-independence}
Ma et al.~\cite{ma_inference-time_2025} show for text-to-image search that selecting by an aesthetic verifier can worsen CLIPScore, and selecting by CLIP can worsen the aesthetic score, at a fixed search budget. Evaluator reuse presents a related risk rather than proof of the same failure: Motion-R1 uses shared pretrained representations for similarity rewards and motion evaluation~\cite{ouyang_motion-r1_2025}, and DreamReward reports its training reward as a metric while also supplying a separate human study~\cite{ye_dreamreward_2025}. Studies should record which scorers influence training, checkpoint choice and candidate selection, and then evaluate with held-out tasks and measures outside that optimisation path.

\subsection{Limitations of This Synthesis}
These findings come from selected studies with uneven coverage, including preprints and technical reports. Source versions, conditioning, evaluators and resource budgets differ, and some protocol details were not established in the reviewed material. The comparisons identify useful distinctions and configuration-specific outcomes rather than effect sizes that can be pooled across tasks, and we do not conduct a common reimplementation or new benchmark experiments.

\section{Directions}\label{sec:directions}

As MGMs continue to evolve, future work must not only increase generative capacity but also address the limitations revealed by our evaluation of faithfulness, compositionality, and robustness across text-to-X modalities. Rather than enumerating directions per modality, we organise this section around the four techniques discussed in this survey and highlight how they can be extended or redesigned to address the evaluation gaps identified in Section~\ref{sec:evaluation} and the findings presented in Section~\ref{sec:findings}.

\paragraph{SSL on Video, Motion, and 3D Data}

SSL for video, motion, and 3D data remains fragmented and lacks the coherence seen in text modelling, where next-token prediction has become a standard. While many current video models are pretrained on combinations of text, image, and video data, they often fall short in capturing the deeper physical and causal structures of spatiotemporal domains. Future work should move beyond surface-level frame prediction towards modelling latent dynamics, such as velocity shifts, deformations, and object interactions, enabling models to learn not just what happens next, but why it happens. Integrating 3D human motion and object datasets could foster models with a richer understanding of physical interactions and embodied behaviour. Research should increasingly ground cross-modal alignments in physical laws and structural constraints. SSL in these domains should increasingly represent dynamic and causal relationships rather than surface patterns alone. Proposed objectives should be compared against the same conditional objective with the backbone, data and optimisation budget held fixed, and tested on generation and instruction adherence rather than representation quality alone.

\paragraph{Dataset-Aware and Task-Specialised Experts for Modular Multimodal Reasoning}

As MoE architectures expand into multimodal generation, future work should explore training individual experts on distinct datasets or tasks tailored to specific output types, such as human motion, 3D objects, or temporally grounded scenes. Assigning experts to distinct datasets or task groups can improve performance when handling heterogeneous domains while balancing specialisation and parameter sharing, suggesting that modularity can foster compositional, interpretable reasoning across modalities. Yet, specialisation must be tempered with generalisation: overly narrow training risks brittleness and loss of cross-domain coherence. Curriculum-style or clustered pretraining, followed by joint fine-tuning under adaptive routing, may enable meaningful expert specialisation while preserving collaborative learning across the MoE system. Because expert granularity determines both mechanism and cost (Section~\ref{sec:finding-experts}), such experts should be compared with dense variants under separately matched active-compute and memory limits, measuring throughput, latency and peak memory at realistic batch sizes.

\paragraph{Extending Chain-of-Thought Reasoning to Non-Text Modalities}

CoT prompting, originally designed for stepwise reasoning in language, is increasingly being adapted to modalities where reasoning is spatial, temporal, or structural rather than purely sequential. In image generation, spatial CoT techniques decompose prompts into explicit visual plans before generation, such as object layouts or region-based attributes, and some frameworks generate intermediate visual representations (for example, bounding boxes, sketches) as reasoning artefacts within the model itself. In temporal domains, motion and video CoT decomposes complex activities into sub-goals or frame-level trajectories. Similarly, structural CoT approaches, such as MusiCoT \cite{lam_analyzable_2025}, plan musical progressions or 3D scaffolds before synthesis, showing how reasoning can operate across harmonies or topologies rather than text.

However, applying CoT beyond language presents open challenges. For example, what constitutes a “thought” in image, motion, or 3D space? Visual and structural reasoning may require non-textual intermediate representations, such as layouts, scene graphs, motion paths, or latent templates, that are difficult to define or supervise. The notion of “steps” becomes ambiguous in parallel or geometric tasks like music or mesh construction \cite{wang_multimodal_2025}. Further, CoT-guided generation is prone to error propagation: a flawed early plan can cascade into systemic generation failures unless robustness or self-correction is built in. Future work should therefore couple these modelling advances with stepwise evaluation metrics that quantify the correctness and self-consistency of non-textual reasoning traces, rather than relying solely on final-task performance.

Finally, integration remains a bottleneck: aligning high-level CoT plans from an LLM with the low-level operations of diffusion models or transformers (for instance, how layout constraints guide denoising) is still an open design problem. Addressing these challenges through richer intermediate formats, multimodal-aware reasoning primitives, and bidirectional planner-executor loops will be essential to fully realise CoT in non-text domains. Comparing generated plans with corrected plans under the same generator and budget, and then varying the generator with the plan fixed, would separate errors in interpretation from errors in realisation (Section~\ref{sec:finding-realization}).

\paragraph{Selective RLHF for Expert Modules}

RLHF has traditionally been applied to full model fine-tuning, aligning global behaviour with human preferences. However, as models grow modular, especially under MoE frameworks, future work should explore selective RLHF: applying feedback only to targeted expert modules. For complex modalities such as human motion or 3D generation, targeted adaptation could address constraints related to physics, balance, intent, deformation, and affordances. Rather than fine-tuning the entire model, one approach may be to refine individual experts: one specialising in rigid-body dynamics, another in soft deformation, another in social interaction. Human feedback could be used to train each expert on domain-specific preferences, behaviours or constraints, addressing task complexity while preserving generalisation across the system. Parameter-efficient fine-tuning techniques provide early tools for this modular alignment, which offers a more interpretable, efficient, and scalable pathway for multimodal control. Any such gain should be confirmed under independent tests (Section~\ref{sec:finding-independence}), comparing base and adapted generators on the same held-out prompts with the training reward, a separate evaluator and blinded human judgements.

\paragraph{Evaluation-Centric Benchmarks and Metrics for Text-to-X}

Despite rapid progress, evaluation of multimodal generators remains fragmented and often misaligned with user-perceived quality. Future work should develop benchmarks and metrics that jointly target semantic faithfulness, compositional generalisation, and robustness to distribution shift across modalities. This includes (i) cross-modal preference datasets for pairwise comparison and reward modelling, (ii) calibrated LLM- and MGM-based judges that are explicitly validated against human annotators, and (iii) stress tests for long-horizon consistency, safety, and physical plausibility. Integrating such signals into training, for instance via multi-objective RLHF or differentiable surrogate metrics, could yield models that are optimised directly for the evaluation axes outlined in this survey rather than solely for likelihood-based objectives. The training signal should nonetheless be kept separate from the evaluation metric, and adherence and within-prompt diversity should be reported jointly (Section~\ref{sec:finding-properties}).

\paragraph{Physics-Grounded Generative Modelling}

Future MGMs could incorporate an explicit understanding of physical laws and object affordances to improve the realism of generated outputs. While current models excel at producing visually plausible outputs, they often falter when tasked with modelling real-world physics (for example, object collisions, gravity, and momentum). Integrating differentiable physics engines or using physics-guided self-refinement could enforce physical plausibility during training or inference. This would be especially beneficial for simulation-based applications like robotics, autonomous navigation, and embodied AI.

\section{Conclusion}\label{sec:conclusion}

Compared with existing surveys on multimodal learning, generative modelling, and model alignment, our contribution is threefold. \emph{First,} we adopt a technique-centred taxonomy that organises MGMs around four techniques (SSL, MoE, RLHF, and CoT) and trace their transfer across six text-conditioned modalities, distinguishing shared mechanisms from experimentally supported benefits. \emph{Second,} we organise evaluation around faithfulness, compositionality, and robustness, apply these criteria across text, image, music, motion, and 3D generation, and interpret the evidence at the level of each study's protocol rather than pooling scores across studies. \emph{Third,} we integrate this technical analysis with a dedicated treatment of trustworthiness, safety, and ethics, highlighting how architectural and evaluation choices shape downstream societal risks and which mitigations have been tested.

The discussion connects these findings to shared architectural themes, highlighting transfer across modalities and shared evaluation blind spots. Constraints such as physical plausibility in motion and 3D, and synchronisation in audio-visual generation, motivate richer inductive biases and evaluation beyond generic text or image metrics. Four findings follow from the within-study comparisons: preference gains need not improve relevance, and reward optimisation can reduce within-prompt diversity; intermediate plans require tests of both the plan and its realisation; expert granularity changes the mechanism and the relevant storage, computation and runtime costs; and gains against an optimisation score need confirmation outside that scoring path.

These observations suggest that progress in multimodal generative models will depend more on the co-design of models, data, evaluation, and governance than on a single architectural breakthrough. Efficient architectures that capture cross-modal structure must be accompanied by benchmarks for semantic faithfulness, long-horizon compositionality, and robustness under realistic perturbations. MGMs must also be developed as \emph{trustworthy, safe, and ethical} systems, aligned with user preferences and legal, cultural, and societal constraints. We hope that this framework and synthesis support coordinated progress towards powerful and dependable MGMs.

% CMC requires these six declarations at the end of every article.
\noindent\begin{minipage}{\textwidth}
\acknowledgement{None.}

\funding{The author(s) received no specific funding for this study.}

\authorcontributions{The authors confirm contribution to the paper as follows: study conception and design, all authors; literature search and analysis, all authors; draft manuscript preparation, all authors. All authors reviewed and approved the final version of the manuscript.}

\availabilityofdataandmaterials{This narrative survey generated no new experimental datasets. Study-level information extracted from the cited literature and used in the synthesis appears in the manuscript's comparative tables and discussion. Source publications are listed in the references. Recent literature-search and evidence-checking notes are internal working materials and are not publicly archived; they do not constitute a complete historical search or screening record.}

\ethicsapproval{Not applicable. This study did not involve human or animal subjects.}

\conflictsofinterest{The authors declare no conflicts of interest.}
\end{minipage}

\clearpage
\abbreviations{Abbreviations}{\indent The following abbreviations are used in this manuscript:\\

\noindent
{\footnotesize
\setlength{\tabcolsep}{4pt}
\begin{tabular}{@{}p{0.09\textwidth}p{0.38\textwidth}p{0.09\textwidth}p{0.38\textwidth}@{}}
3D & three-dimensional & MCoT & multimodal chain-of-thought\\
AI & artificial intelligence & MGM & multimodal generative model\\
AIME & American Invitational Mathematics Examination & MIR & music information retrieval\\
APE & average positional error & MLE & maximum likelihood estimation\\
AR & autoregressive & MLM & masked language modelling\\
ASM & AudioSet music subset & MoE & mixture of experts\\
BERT & bidirectional encoder representations from transformers & MOS & mean opinion score\\
BLEU & bilingual evaluation understudy & MS-COCO & Microsoft Common Objects in Context\\
C2PA & Coalition for Content Provenance and Authenticity & MSR-VTT & Microsoft Research Video to Text\\
CLAP & contrastive language--audio pretraining & MTT & MagnaTagATune\\
CLIP & contrastive language--image pretraining & NCII & non-consensual intimate imagery\\
CLIPSIM & mean CLIP similarity between frames and text & NeRF & neural radiance field\\
CoT & chain-of-thought & NIST & National Institute of Standards and Technology\\
CSAM & child sexual abuse material & NLI & natural language inference\\
CUB & Caltech-UCSD Birds & NLP & natural language processing\\
DAMSM & deep attentional multimodal similarity model & NLU & natural language understanding\\
DDIM & denoising diffusion implicit model & NMT & neural machine translation\\
DiT & diffusion transformer & NSP & next sentence prediction\\
DMAE & diffusion magnitude autoencoder & OECD & Organisation for Economic Co-operation and Development\\
DPO & direct preference optimisation & P-FID & point-cloud Fr\'echet inception distance\\
ELBO & evidence lower bound & P-IS & point-cloud inception score\\
EM & exact match & PII & personally identifiable information\\
ESD & Erased Stable Diffusion & PPO & proximal policy optimisation\\
EU & European Union & QA & question answering\\
FAD & Fr\'echet audio distance & RL & reinforcement learning\\
FFN & feed-forward network & RLAIF & reinforcement learning from AI feedback\\
FID & Fr\'echet inception distance & RLHF & reinforcement learning from human feedback\\
FLOPs & floating-point operations & RNN & recurrent neural network\\
FMA & Free Music Archive & ROUGE & recall-oriented understudy for gisting evaluation\\
FVD & Fr\'echet video distance & SD & Stable Diffusion\\
GAN & generative adversarial network & SDF & signed distance function\\
GPT & generative pretrained transformer & SDXL & Stable Diffusion XL\\
GRPO & group relative policy optimisation & SFT & supervised fine-tuning\\
GRU & gated recurrent unit & SLD & safe latent diffusion\\
GUARD & guidance using attractive--repulsive dynamics & SOTA & state of the art\\
I2P & inappropriate image prompts & SQL & structured query language\\
IoU & intersection over union & SSL & self-supervised learning\\
IS & inception score & T2-3D & text-to-3D objects\\
KIT-ML & KIT Motion-Language dataset & T2HM & text-to-human-motion\\
KL & Kullback--Leibler & T2I & text-to-image\\
LAION & Large-scale Artificial Intelligence Open Network & T2M & text-to-music\\
LDM & latent diffusion model & T2T & text-to-text\\
LLM & large language model & T2V & text-to-video\\
LMM & Large Motion Model & VAE & variational autoencoder\\
LoRA & low-rank adaptation & ViT & vision transformer\\
LSTM & long short-term memory & VQ & vector quantisation\\
MCC & MuLan cycle consistency & VQ-VAE & vector-quantised variational autoencoder\\
\end{tabular}}}

\reftitle{References}
\bibliography{sections/references}

\end{document}